%&mn
% MNSAMPLE.TEX
%
% A sample plain TeX single/two column Monthly Notices article.
%
% v1.5  --- released 25th August 1994 (M. Reed)
% v1.4  --- released 22nd February 1994
% v1.3  --- released  8th December 1992
%
% Copyright Cambridge University Press

% The following line automatically loads the mn macros if you are not
% using a format file.
\ifx\mnmacrosloaded\undefined % MN.TEX (Computer Modern version)
%
% plain TeX single / double column macros for the
% Monthly Notices of Royal Astronomical Society
%
% v1.6  (mn.tex)  --- released 18th September 1995 (A. Woollatt)
% v1.5      "     --- released 25th August 1994 (M. Reed)
% v1.4      "     --- released 22nd February 1994
% v1.3  (mnd.tex) --- released 28th November 1992
% v1.26     "     --- released  1st August 1992
% v1.25     "     --- released 25th February 1992
%
% Copyright Cambridge University Press
%
% > Incorporating special symbol code from laa.sty v1.1 (25th Feb 1991)
%   used with the permission of Springer Verlag.
% > Incorporating parts of mssymb.tex (8th July 1987).
% > Incorporating NewFont.sty v ALPHA patchlevel 8 (16th August 1994).
% > Add footlines, add footnotes in double column (18th September
%   1995).

\catcode `\@=11 % @ signs are letters

\def\@version{1.6}
\def\@verdate{18th September 1995}

% Fonts: Computer Modern / Monotype Times (CUP only)
%
% Font family sizes available:
%   8pt, 9pt, 10pt, 11pt, 14pt and 17pt.
%
% Faces available:
%   \rm, math italic, symbol, \it, \bf, \sl, \tt, \sc, \sf, \cal, \em,
%   \mit and \oldstyle.

% define the typeface in use

\newif\ifprod@font

\ifx\@typeface\undefined
  \def\@typeface{Comp. Modern}\prod@fontfalse
\else
  \prod@fonttrue % We want Times
\fi

\def\newfam{\alloc@8\fam\chardef\sixt@@n} % made not outer

\ifprod@font
\font\fiverm=mtr10 at 5pt
\font\fivebf=mtbx10 at 5pt
\font\fiveit=mtti10 at 5pt
\font\fivesl=mtsl10 at 5pt
\font\fivett=cmtt8 at 5pt     \hyphenchar\fivett=-1
\font\fivecsc=mtcsc10 at 5pt
\font\fivesf=mtss10 at 5pt
\font\fivei=mtmi10 at 5pt      \skewchar\fivei='177
\font\fivesy=mtsy10 at 5pt     \skewchar\fivesy='60

\font\sixrm=mtr10 at 6pt
\font\sixbf=mtbx10 at 6pt
\font\sixit=mtti10 at 6pt
\font\sixsl=mtsl10 at 6pt
\font\sixtt=cmtt8 at 6pt      \hyphenchar\sixtt=-1
\font\sixcsc=mtcsc10 at 6pt
\font\sixsf=mtss10 at 6pt
\font\sixi=mtmi10 at 6pt       \skewchar\sixi='177
\font\sixsy=mtsy10 at 6pt      \skewchar\sixsy='60

\font\sevenrm=mtr10 at 7pt
\font\sevenbf=mtbx10 at 7pt
\font\sevenit=mtti10 at 7pt
\font\sevensl=mtsl10 at 7pt
\font\seventt=cmtt8 at 7pt     \hyphenchar\seventt=-1
\font\sevencsc=mtcsc10 at 7pt
\font\sevensf=mtss10 at 7pt
\font\seveni=mtmi10 at 7pt      \skewchar\seveni='177
\font\sevensy=mtsy10 at 7pt     \skewchar\sevensy='60

\font\eightrm=mtr10 at 8pt
\font\eightbf=mtbx10 at 8pt
\font\eightit=mtti10 at 8pt
\font\eighti=mtmi10 at 8pt      \skewchar\eighti='177
\font\eightsy=mtsy10 at 8pt     \skewchar\eightsy='60
\font\eightsl=mtsl10 at 8pt
\font\eighttt=cmtt8             \hyphenchar\eighttt=-1
\font\eightcsc=mtcsc10 at 8pt
\font\eightsf=mtss10 at 8pt

\font\ninerm=mtr10 at 9pt
\font\ninebf=mtbx10 at 9pt
\font\nineit=mtti10 at 9pt
\font\ninei=mtmi10 at 9pt      \skewchar\ninei='177
\font\ninesy=mtsy10 at 9pt     \skewchar\ninesy='60
\font\ninesl=mtsl10 at 9pt
\font\ninett=cmtt9             \hyphenchar\ninett=-1
\font\ninecsc=mtcsc10 at 9pt
\font\ninesf=mtss10 at 9pt

\font\tenrm=mtr10
\font\tenbf=mtbx10
\font\tenit=mtti10
\font\teni=mtmi10		\skewchar\teni='177
\font\tensy=mtsy10		\skewchar\tensy='60
\font\tenex=cmex10
\font\tensl=mtsl10
\font\tentt=cmtt10		\hyphenchar\tentt=-1
\font\tencsc=mtcsc10
\font\tensf=mtss10

\font\elevenrm=mtr10 at 11pt
\font\elevenbf=mtbx10 at 11pt
\font\elevenit=mtti10 at 11pt
\font\eleveni=mtmi10 at 11pt      \skewchar\eleveni='177
\font\elevensy=mtsy10 at 11pt     \skewchar\elevensy='60
\font\elevensl=mtsl10 at 11pt
\font\eleventt=cmtt10 at 11pt     \hyphenchar\eleventt=-1
\font\elevencsc=mtcsc10 at 11pt
\font\elevensf=mtss10 at 11pt

\font\twelverm=mtr10 at 12pt
\font\twelvebf=mtbx10 at 12pt
\font\twelveit=mtti10 at 12pt
\font\twelvesl=mtsl10 at 12pt
\font\twelvett=cmtt12             \hyphenchar\twelvett=-1
\font\twelvecsc=mtcsc10 at 12pt
\font\twelvesf=mtss10 at 12pt
\font\twelvei=mtmi10 at 12pt      \skewchar\twelvei='177
\font\twelvesy=mtsy10 at 12pt     \skewchar\twelvesy='60

\font\fourteenrm=mtr10 at 14pt
\font\fourteenbf=mtbx10 at 14pt
\font\fourteenit=mtti10 at 14pt
\font\fourteeni=mtmi10 at 14pt      \skewchar\fourteeni='177
\font\fourteensy=mtsy10 at 14pt     \skewchar\fourteensy='60
\font\fourteensl=mtsl10 at 14pt
\font\fourteentt=cmtt12 at 14pt     \hyphenchar\fourteentt=-1
\font\fourteencsc=mtcsc10 at 14pt
\font\fourteensf=mtss10 at 14pt

\font\seventeenrm=mtr10 at 17pt
\font\seventeenbf=mtbx10 at 17pt
\font\seventeenit=mtti10 at 17pt
\font\seventeeni=mtmi10 at 17pt      \skewchar\seventeeni='177
\font\seventeensy=mtsy10 at 17pt     \skewchar\seventeensy='60
\font\seventeensl=mtsl10 at 17pt
\font\seventeentt=cmtt12 at 17pt     \hyphenchar\seventeentt=-1
\font\seventeencsc=mtcsc10 at 17pt
\font\seventeensf=mtss10 at 17pt
\else
\font\fiverm=cmr5
\font\fivei=cmmi5             \skewchar\fivei='177
\font\fivesy=cmsy5            \skewchar\fivesy='60
\font\fivebf=cmbx5

\font\sixrm=cmr6
\font\sixi=cmmi6             \skewchar\sixi='177
\font\sixsy=cmsy6            \skewchar\sixsy='60
\font\sixbf=cmbx6

\font\sevenrm=cmr7
\font\sevenit=cmti7
\font\seveni=cmmi7             \skewchar\seveni='177
\font\sevensy=cmsy7            \skewchar\sevensy='60
\font\sevenbf=cmbx7

\font\eightrm=cmr8
\font\eightbf=cmbx8
\font\eightit=cmti8
\font\eighti=cmmi8			\skewchar\eighti='177
\font\eightsy=cmsy8			\skewchar\eightsy='60
\font\eightsl=cmsl8
\font\eighttt=cmtt8			\hyphenchar\eighttt=-1
\font\eightcsc=cmcsc10 at 8pt
\font\eightsf=cmss8

\font\ninerm=cmr9
\font\ninebf=cmbx9
\font\nineit=cmti9
\font\ninei=cmmi9			\skewchar\ninei='177
\font\ninesy=cmsy9			\skewchar\ninesy='60
\font\ninesl=cmsl9
\font\ninett=cmtt9			\hyphenchar\ninett=-1
\font\ninecsc=cmcsc10 at 9pt
\font\ninesf=cmss9

\font\tenrm=cmr10
\font\tenbf=cmbx10
\font\tenit=cmti10
\font\teni=cmmi10		\skewchar\teni='177
\font\tensy=cmsy10		\skewchar\tensy='60
\font\tenex=cmex10
\font\tensl=cmsl10
\font\tentt=cmtt10		\hyphenchar\tentt=-1
\font\tencsc=cmcsc10
\font\tensf=cmss10

\font\elevenrm=cmr10 scaled \magstephalf
\font\elevenbf=cmbx10 scaled \magstephalf
\font\elevenit=cmti10 scaled \magstephalf
\font\eleveni=cmmi10 scaled \magstephalf	\skewchar\eleveni='177
\font\elevensy=cmsy10 scaled \magstephalf	\skewchar\elevensy='60
\font\elevensl=cmsl10 scaled \magstephalf
\font\eleventt=cmtt10 scaled \magstephalf	\hyphenchar\eleventt=-1
\font\elevencsc=cmcsc10 scaled \magstephalf
\font\elevensf=cmss10 scaled \magstephalf

\font\twelverm=cmr10 scaled \magstep1
\font\twelvebf=cmbx10 scaled \magstep1
\font\twelvei=cmmi10 scaled \magstep1      \skewchar\twelvei='177
\font\twelvesy=cmsy10 scaled \magstep1     \skewchar\twelvesy='60

\font\fourteenrm=cmr10 scaled \magstep2
\font\fourteenbf=cmbx10 scaled \magstep2
\font\fourteenit=cmti10 scaled \magstep2
\font\fourteeni=cmmi10 scaled \magstep2		\skewchar\fourteeni='177
\font\fourteensy=cmsy10 scaled \magstep2	\skewchar\fourteensy='60
\font\fourteensl=cmsl10 scaled \magstep2
\font\fourteentt=cmtt10 scaled \magstep2	\hyphenchar\fourteentt=-1
\font\fourteencsc=cmcsc10 scaled \magstep2
\font\fourteensf=cmss10 scaled \magstep2

\font\seventeenrm=cmr10 scaled \magstep3
\font\seventeenbf=cmbx10 scaled \magstep3
\font\seventeenit=cmti10 scaled \magstep3
\font\seventeeni=cmmi10 scaled \magstep3	\skewchar\seventeeni='177
\font\seventeensy=cmsy10 scaled \magstep3	\skewchar\seventeensy='60
\font\seventeensl=cmsl10 scaled \magstep3
\font\seventeentt=cmtt10 scaled \magstep3	\hyphenchar\seventeentt=-1
\font\seventeencsc=cmcsc10 scaled \magstep3
\font\seventeensf=cmss10 scaled \magstep3
\fi

\def\hexnumber#1{\ifcase#1 0\or1\or2\or3\or4\or5\or6\or7\or8\or9\or
  A\or B\or C\or D\or E\or F\fi}

\def\makestrut{%
  \setbox\strutbox=\hbox{%
    \vrule height.7\baselineskip depth.3\baselineskip width \z@}%
}

\def\baselinestretch{1}
\newskip\tmp@bls

\def\b@ls#1{% set baseline using \baselinestretch as a scale factor
  \tmp@bls=#1\relax
  \baselineskip=#1\relax\makestrut
  \normalbaselineskip=\baselinestretch\tmp@bls
  \normalbaselines
}

\def\nostb@ls#1{% set baseline skip ignoring \baselinestretch
  \normalbaselineskip=#1\relax
  \normalbaselines
  \makestrut
}

% families \itfam, \slfam, \bffam, \ttfam defined in PLAIN.
%
% \itfam is \fam4
% \slfam is \fam5
% \bffam is \fam6
% \ttfam is \fam7

\newfam\scfam  % \fam8
\newfam\sffam  % \fam9

\def\mit{\fam\@ne}
\def\cal{\fam\tw@}
\def\em{\ifdim\fontdimen1\font>\z@ \rm\else\it\fi}

\textfont3=\tenex
\scriptfont3=\tenex
\scriptscriptfont3=\tenex

\setbox0=\hbox{\tenex B} \p@renwd=\wd0 % width of the big left (

\def\eightpoint{% 8^6^5 on 10pt
  \def\rm{\fam0\eightrm}%
  \textfont0=\eightrm \scriptfont0=\sixrm \scriptscriptfont0=\fiverm%
  \textfont1=\eighti  \scriptfont1=\sixi  \scriptscriptfont1=\fivei%
  \textfont2=\eightsy \scriptfont2=\sixsy \scriptscriptfont2=\fivesy%
  \textfont\itfam=\eightit\def\it{\fam\itfam\eightit}%
  \ifprod@font
    \scriptfont\itfam=\sixit
      \scriptscriptfont\itfam=\fiveit
  \else
    \scriptfont\itfam=\eightit
      \scriptscriptfont\itfam=\eightit
  \fi
  \textfont\bffam=\eightbf%
    \scriptfont\bffam=\sixbf%
      \scriptscriptfont\bffam=\fivebf%
  \def\bf{\fam\bffam\eightbf}%
  \textfont\slfam=\eightsl\def\sl{\fam\slfam\eightsl}%
  \ifprod@font
    \scriptfont\slfam=\sixsl
      \scriptscriptfont\slfam=\fivesl
  \else
    \scriptfont\slfam=\eightsl
      \scriptscriptfont\slfam=\eightsl
  \fi
  \textfont\ttfam=\eighttt\def\tt{\fam\ttfam\eighttt}%
  \ifprod@font
    \scriptfont\ttfam=\sixtt
      \scriptscriptfont\ttfam=\fivett
  \else
    \scriptfont\ttfam=\eighttt
      \scriptscriptfont\ttfam=\eighttt
  \fi
  \textfont\scfam=\eightcsc\def\sc{\fam\scfam\eightcsc}%
  \ifprod@font
    \scriptfont\scfam=\sixcsc
      \scriptscriptfont\scfam=\fivecsc
  \else
    \scriptfont\scfam=\eightcsc
      \scriptscriptfont\scfam=\eightcsc
  \fi
  \textfont\sffam=\eightsf\def\sf{\fam\sffam\eightsf}%
  \ifprod@font
    \scriptfont\sffam=\sixsf
      \scriptscriptfont\sffam=\fivesf
  \else
    \scriptfont\sffam=\eightsf
      \scriptscriptfont\sffam=\eightsf
  \fi
  \def\oldstyle{\fam\@ne\eighti}%
  \b@ls{10pt}\rm\@viiipt%
}
\def\@viiipt{}

\def\ninepoint{% 9^6^5 on 11pt (two col) / 12 (single col)
  \def\rm{\fam0\ninerm}%
  \textfont0=\ninerm \scriptfont0=\sixrm \scriptscriptfont0=\fiverm%
  \textfont1=\ninei  \scriptfont1=\sixi  \scriptscriptfont1=\fivei%
  \textfont2=\ninesy \scriptfont2=\sixsy \scriptscriptfont2=\fivesy%
  \textfont\itfam=\nineit\def\it{\fam\itfam\nineit}%
  \ifprod@font
    \scriptfont\itfam=\sixit
      \scriptscriptfont\itfam=\fiveit
  \else
    \scriptfont\itfam=\nineit
      \scriptscriptfont\itfam=\nineit
  \fi
  \textfont\bffam=\ninebf%
    \scriptfont\bffam=\sixbf%
      \scriptscriptfont\bffam=\fivebf%
  \def\bf{\fam\bffam\ninebf}%
  \textfont\slfam=\ninesl\def\sl{\fam\slfam\ninesl}%
  \ifprod@font
    \scriptfont\slfam=\sixsl
      \scriptscriptfont\slfam=\fivesl
  \else
    \scriptfont\slfam=\ninesl
      \scriptscriptfont\slfam=\ninesl
  \fi
  \textfont\ttfam=\ninett\def\tt{\fam\ttfam\ninett}%
  \ifprod@font
    \scriptfont\ttfam=\sixtt
      \scriptscriptfont\ttfam=\fivett
  \else
    \scriptfont\ttfam=\ninett
      \scriptscriptfont\ttfam=\ninett
  \fi
  \textfont\scfam=\ninecsc\def\sc{\fam\scfam\ninecsc}%
  \ifprod@font
    \scriptfont\scfam=\sixcsc
      \scriptscriptfont\scfam=\fivecsc
  \else
    \scriptfont\scfam=\ninecsc
      \scriptscriptfont\scfam=\ninecsc
  \fi
  \textfont\sffam=\ninesf\def\sf{\fam\sffam\ninesf}%
  \ifprod@font
    \scriptfont\sffam=\sixsf
      \scriptscriptfont\sffam=\fivesf
  \else
    \scriptfont\sffam=\ninesf
      \scriptscriptfont\sffam=\ninesf
  \fi
  \def\oldstyle{\fam\@ne\ninei}%
  \b@ls{\TextLeading plus \Feathering}\rm\@ixpt%
}
\def\@ixpt{}

\def\tenpoint{% 10^7^5 on 11pt
  \def\rm{\fam0\tenrm}%
  \textfont0=\tenrm \scriptfont0=\sevenrm \scriptscriptfont0=\fiverm%
  \textfont1=\teni  \scriptfont1=\seveni  \scriptscriptfont1=\fivei%
  \textfont2=\tensy \scriptfont2=\sevensy \scriptscriptfont2=\fivesy%
  \textfont\itfam=\tenit\def\it{\fam\itfam\tenit}%
  \ifprod@font
    \scriptfont\itfam=\sevenit
      \scriptscriptfont\itfam=\fiveit
  \else
    \scriptfont\itfam=\tenit
      \scriptscriptfont\itfam=\tenit
  \fi
  \textfont\bffam=\tenbf%
    \scriptfont\bffam=\sevenbf%
      \scriptscriptfont\bffam=\fivebf%
  \def\bf{\fam\bffam\tenbf}%
  \textfont\slfam=\tensl\def\sl{\fam\slfam\tensl}%
  \ifprod@font
    \scriptfont\slfam=\sevensl
      \scriptscriptfont\slfam=\fivesl
  \else
    \scriptfont\slfam=\tensl
      \scriptscriptfont\slfam=\tensl
  \fi
  \textfont\ttfam=\tentt\def\tt{\fam\ttfam\tentt}%
  \ifprod@font
    \scriptfont\ttfam=\seventt
      \scriptscriptfont\ttfam=\fivett
  \else
    \scriptfont\ttfam=\tentt
      \scriptscriptfont\ttfam=\tentt
  \fi
  \textfont\scfam=\tencsc\def\sc{\fam\scfam\tencsc}%
  \ifprod@font
    \scriptfont\scfam=\sevencsc
      \scriptscriptfont\scfam=\fivecsc
  \else
    \scriptfont\scfam=\tencsc
      \scriptscriptfont\scfam=\tencsc
  \fi
  \textfont\sffam=\tensf\def\sf{\fam\sffam\tensf}%
  \ifprod@font
    \scriptfont\sffam=\sevensf
      \scriptscriptfont\sffam=\fivesf
  \else
    \scriptfont\sffam=\tensf
      \scriptscriptfont\sffam=\tensf
  \fi
  \def\oldstyle{\fam\@ne\teni}%
  \b@ls{11pt}\rm\@xpt%
}
\def\@xpt{}

\def\elevenpoint{% 11^8^6 on 13pt
  \def\rm{\fam0\elevenrm}%
  \textfont0=\elevenrm \scriptfont0=\eightrm \scriptscriptfont0=\sixrm%
  \textfont1=\eleveni  \scriptfont1=\eighti  \scriptscriptfont1=\sixi%
  \textfont2=\elevensy \scriptfont2=\eightsy \scriptscriptfont2=\sixsy%
  \textfont\itfam=\elevenit\def\it{\fam\itfam\elevenit}%
  \ifprod@font
    \scriptfont\itfam=\eightit
      \scriptscriptfont\itfam=\sixit
  \else
    \scriptfont\itfam=\elevenit
      \scriptscriptfont\itfam=\elevenit
  \fi
  \textfont\bffam=\elevenbf%
    \scriptfont\bffam=\eightbf%
      \scriptscriptfont\bffam=\sixbf%
  \def\bf{\fam\bffam\elevenbf}%
  \textfont\slfam=\elevensl\def\sl{\fam\slfam\elevensl}%
  \ifprod@font
    \scriptfont\slfam=\eightsl
      \scriptscriptfont\slfam=\sixsl
  \else
    \scriptfont\slfam=\elevensl
      \scriptscriptfont\slfam=\elevensl
  \fi
  \textfont\ttfam=\eleventt\def\tt{\fam\ttfam\eleventt}%
  \ifprod@font
    \scriptfont\ttfam=\eighttt
      \scriptscriptfont\ttfam=\sixtt
  \else
    \scriptfont\ttfam=\eleventt
      \scriptscriptfont\ttfam=\eleventt
  \fi
  \textfont\scfam=\elevencsc\def\sc{\fam\scfam\elevencsc}%
  \ifprod@font
    \scriptfont\scfam=\eightcsc
      \scriptscriptfont\scfam=\sixcsc
  \else
    \scriptfont\scfam=\elevencsc
      \scriptscriptfont\scfam=\elevencsc
  \fi
  \textfont\sffam=\elevensf\def\sf{\fam\sffam\elevensf}%
  \ifprod@font
    \scriptfont\sffam=\eightsf
      \scriptscriptfont\sffam=\sixsf
  \else
    \scriptfont\sffam=\elevensf
      \scriptscriptfont\sffam=\elevensf
  \fi
  \def\oldstyle{\fam\@ne\eleveni}%
  \b@ls{13pt}\rm\@xipt%
}
\def\@xipt{}

\def\fourteenpoint{% 14^10^7 on 17pt
  \def\rm{\fam0\fourteenrm}%
  \textfont0\fourteenrm  \scriptfont0\tenrm  \scriptscriptfont0\sevenrm%
  \textfont1\fourteeni   \scriptfont1\teni   \scriptscriptfont1\seveni%
  \textfont2\fourteensy  \scriptfont2\tensy  \scriptscriptfont2\sevensy%
  \textfont\itfam=\fourteenit\def\it{\fam\itfam\fourteenit}%
  \ifprod@font
    \scriptfont\itfam=\tenit
      \scriptscriptfont\itfam=\sevenit
  \else
    \scriptfont\itfam=\fourteenit
      \scriptscriptfont\itfam=\fourteenit
  \fi
  \textfont\bffam=\fourteenbf%
    \scriptfont\bffam=\tenbf%
      \scriptscriptfont\bffam=\sevenbf%
  \def\bf{\fam\bffam\fourteenbf}%
  \textfont\slfam=\fourteensl\def\sl{\fam\slfam\fourteensl}%
  \ifprod@font
    \scriptfont\slfam=\tensl
      \scriptscriptfont\slfam=\sevensl
  \else
    \scriptfont\slfam=\fourteensl
      \scriptscriptfont\slfam=\fourteensl
  \fi
  \textfont\ttfam=\fourteentt\def\tt{\fam\ttfam\fourteentt}%
  \ifprod@font
    \scriptfont\ttfam=\tentt
      \scriptscriptfont\ttfam=\seventt
  \else
    \scriptfont\ttfam=\fourteentt
      \scriptscriptfont\ttfam=\fourteentt
  \fi
  \textfont\scfam=\fourteencsc\def\sc{\fam\scfam\fourteencsc}%
  \ifprod@font
    \scriptfont\scfam=\tencsc
      \scriptscriptfont\scfam=\sevencsc
  \else
    \scriptfont\scfam=\fourteencsc
      \scriptscriptfont\scfam=\fourteencsc
  \fi
  \textfont\sffam=\fourteensf\def\sf{\fam\sffam\fourteensf}%
  \ifprod@font
    \scriptfont\sffam=\tensf
      \scriptscriptfont\sffam=\sevensf
  \else
    \scriptfont\sffam=\fourteensf
      \scriptscriptfont\sffam=\fourteensf
  \fi
  \def\oldstyle{\fam\@ne\fourteeni}%
  \b@ls{17pt}\rm\@xivpt%
}
\def\@xivpt{}

\def\seventeenpoint{% 17^12^10 on 20pt
  \def\rm{\fam0\seventeenrm}%
  \textfont0\seventeenrm  \scriptfont0\twelverm  \scriptscriptfont0\tenrm%
  \textfont1\seventeeni   \scriptfont1\twelvei   \scriptscriptfont1\teni%
  \textfont2\seventeensy  \scriptfont2\twelvesy  \scriptscriptfont2\tensy%
  \textfont\itfam=\seventeenit\def\it{\fam\itfam\seventeenit}%
  \ifprod@font
    \scriptfont\itfam=\twelveit
      \scriptscriptfont\itfam=\tenit
  \else
    \scriptfont\itfam=\seventeenit
      \scriptscriptfont\itfam=\seventeenit
  \fi
  \textfont\bffam=\seventeenbf%
    \scriptfont\bffam=\twelvebf%
      \scriptscriptfont\bffam=\tenbf%
  \def\bf{\fam\bffam\seventeenbf}%
  \textfont\slfam=\seventeensl\def\sl{\fam\slfam\seventeensl}%
  \ifprod@font
    \scriptfont\slfam=\twelvesl
      \scriptscriptfont\slfam=\tensl
  \else
    \scriptfont\slfam=\seventeensl
      \scriptscriptfont\slfam=\seventeensl
  \fi
  \textfont\ttfam=\seventeentt\def\tt{\fam\ttfam\seventeentt}%
  \ifprod@font
    \scriptfont\ttfam=\twelvett
      \scriptscriptfont\ttfam=\tentt
  \else
    \scriptfont\ttfam=\seventeentt
      \scriptscriptfont\ttfam=\seventeentt
  \fi
  \textfont\scfam=\seventeencsc\def\sc{\fam\scfam\seventeencsc}%
  \ifprod@font
    \scriptfont\scfam=\twelvecsc
      \scriptscriptfont\scfam=\tencsc
  \else
    \scriptfont\scfam=\seventeencsc
      \scriptscriptfont\scfam=\seventeencsc
  \fi
  \textfont\sffam=\seventeensf\def\sf{\fam\sffam\seventeensf}%
  \ifprod@font
    \scriptfont\sffam=\twelvesf
      \scriptscriptfont\sffam=\tensf
  \else
    \scriptfont\sffam=\seventeensf
      \scriptscriptfont\sffam=\seventeensf
  \fi
  \def\oldstyle{\fam\@ne\seventeeni}%
  \b@ls{20pt}\rm\@xviipt%
}
\def\@xviipt{}

\lineskip=1pt      \normallineskip=\lineskip
\lineskiplimit=\z@ \normallineskiplimit=\lineskiplimit

% BOLD MATH SYMBOLS

\def\loadboldmathnames{%
  \def\balpha{{\bmath{\alpha}}}%
  \def\bbeta{{\bmath{\beta}}}%
  \def\bgamma{{\bmath{\gamma}}}%
  \def\bdelta{{\bmath{\delta}}}%
  \def\bepsilon{{\bmath{\epsilon}}}%
  \def\bzeta{{\bmath{\zeta}}}%
  \def\boldeta{{\bmath{\eta}}}%
  \def\btheta{{\bmath{\theta}}}%
  \def\biota{{\bmath{\iota}}}%
  \def\bkappa{{\bmath{\kappa}}}%
  \def\blambda{{\bmath{\lambda}}}%
  \def\bmu{{\bmath{\mu}}}%
  \def\bnu{{\bmath{\nu}}}%
  \def\bxi{{\bmath{\xi}}}%
  \def\bpi{{\bmath{\pi}}}%
  \def\brho{{\bmath{\rho}}}%
  \def\bsigma{{\bmath{\sigma}}}%
  \def\btau{{\bmath{\tau}}}%
  \def\bupsilon{{\bmath{\upsilon}}}%
  \def\bphi{{\bmath{\phi}}}%
  \def\bchi{{\bmath{\chi}}}%
  \def\bpsi{{\bmath{\psi}}}%
  \def\bomega{{\bmath{\omega}}}%
  \def\bvarepsilon{{\bmath{\varepsilon}}}%
  \def\bvartheta{{\bmath{\vartheta}}}%
  \def\bvarpi{{\bmath{\varpi}}}%
  \def\bvarrho{{\bmath{\varrho}}}%
  \def\bvarsigma{{\bmath{\varsigma}}}%
  \def\bvarphi{{\bmath{\varphi}}}%
  \def\baleph{{\bmath{\aleph}}}%
  \def\bimath{{\bmath{\imath}}}%
  \def\bjmath{{\bmath{\jmath}}}%
  \def\bell{{\bmath{\ell}}}%
  \def\bwp{{\bmath{\wp}}}%
  \def\bRe{{\bmath{\Re}}}%
  \def\bIm{{\bmath{\Im}}}%
  \def\bpartial{{\bmath{\partial}}}%
  \def\binfty{{\bmath{\infty}}}%
  \def\bprime{{\bmath{\prime}}}%
  \def\bemptyset{{\bmath{\emptyset}}}%
  \def\bnabla{{\bmath{\nabla}}}%
  \def\btop{{\bmath{\top}}}%
  \def\bbot{{\bmath{\bot}}}%
  \def\btriangle{{\bmath{\triangle}}}%
  \def\bforall{{\bmath{\forall}}}%
  \def\bexists{{\bmath{\exists}}}%
  \def\bneg{{\bmath{\neg}}}%
  \def\bflat{{\bmath{\flat}}}%
  \def\bnatural{{\bmath{\natural}}}%
  \def\bsharp{{\bmath{\sharp}}}%
  \def\bclubsuit{{\bmath{\clubsuit}}}%
  \def\bdiamondsuit{{\bmath{\diamondsuit}}}%
  \def\bheartsuit{{\bmath{\heartsuit}}}%
  \def\bspadesuit{{\bmath{\spadesuit}}}%
  \def\bsmallint{{\bmath{\smallint}}}%
  \def\btriangleleft{{\bmath{\triangleleft}}}%
  \def\btriangleright{{\bmath{\triangleright}}}%
  \def\bbigtriangleup{{\bmath{\bigtriangleup}}}%
  \def\bbigtriangledown{{\bmath{\bigtriangledown}}}%
  \def\bwedge{{\bmath{\wedge}}}%
  \def\bvee{{\bmath{\vee}}}%
  \def\bcap{{\bmath{\cap}}}%
  \def\bcup{{\bmath{\cup}}}%
  \def\bddagger{{\bmath{\ddagger}}}%
  \def\bdagger{{\bmath{\dagger}}}%
  \def\bsqcap{{\bmath{\sqcap}}}%
  \def\bsqcup{{\bmath{\sqcup}}}%
  \def\buplus{{\bmath{\uplus}}}%
  \def\bamalg{{\bmath{\amalg}}}%
  \def\bdiamond{{\bmath{\diamond}}}%
  \def\bbullet{{\bmath{\bullet}}}%
  \def\bwr{{\bmath{\wr}}}%
  \def\bdiv{{\bmath{\div}}}%
  \def\bodot{{\bmath{\odot}}}%
  \def\boslash{{\bmath{\oslash}}}%
  \def\botimes{{\bmath{\otimes}}}%
  \def\bominus{{\bmath{\ominus}}}%
  \def\boplus{{\bmath{\oplus}}}%
  \def\bmp{{\bmath{\mp}}}%
  \def\bpm{{\bmath{\pm}}}%
  \def\bcirc{{\bmath{\circ}}}%
  \def\bbigcirc{{\bmath{\bigcirc}}}%
  \def\bsetminus{{\bmath{\setminus}}}%
  \def\bcdot{{\bmath{\cdot}}}%
  \def\bast{{\bmath{\ast}}}%
  \def\btimes{{\bmath{\times}}}%
  \def\bstar{{\bmath{\star}}}%
  \def\bpropto{{\bmath{\propto}}}%
  \def\bsqsubseteq{{\bmath{\sqsubseteq}}}%
  \def\bsqsupseteq{{\bmath{\sqsupseteq}}}%
  \def\bparallel{{\bmath{\parallel}}}%
  \def\bmid{{\bmath{\mid}}}%
  \def\bdashv{{\bmath{\dashv}}}%
  \def\bvdash{{\bmath{\vdash}}}%
  \def\bnearrow{{\bmath{\nearrow}}}%
  \def\bsearrow{{\bmath{\searrow}}}%
  \def\bnwarrow{{\bmath{\nwarrow}}}%
  \def\bswarrow{{\bmath{\swarrow}}}%
  \def\bLeftrightarrow{{\bmath{\Leftrightarrow}}}%
  \def\bLeftarrow{{\bmath{\Leftarrow}}}%
  \def\bRightarrow{{\bmath{\Rightarrow}}}%
  \def\bleq{{\bmath{\leq}}}%
  \def\bgeq{{\bmath{\geq}}}%
  \def\bsucc{{\bmath{\succ}}}%
  \def\bprec{{\bmath{\prec}}}%
  \def\bapprox{{\bmath{\approx}}}%
  \def\bsucceq{{\bmath{\succeq}}}%
  \def\bpreceq{{\bmath{\preceq}}}%
  \def\bsupset{{\bmath{\supset}}}%
  \def\bsubset{{\bmath{\subset}}}%
  \def\bsupseteq{{\bmath{\supseteq}}}%
  \def\bsubseteq{{\bmath{\subseteq}}}%
  \def\bin{{\bmath{\in}}}%
  \def\bni{{\bmath{\ni}}}%
  \def\bgg{{\bmath{\gg}}}%
  \def\bll{{\bmath{\ll}}}%
  \def\bnot{{\bmath{\not}}}%
  \def\bleftrightarrow{{\bmath{\leftrightarrow}}}%
  \def\bleftarrow{{\bmath{\leftarrow}}}%
  \def\brightarrow{{\bmath{\rightarrow}}}%
  \def\bmapstochar{{\bmath{\mapstochar}}}%
  \def\bsim{{\bmath{\sim}}}%
  \def\bsimeq{{\bmath{\simeq}}}%
  \def\bperp{{\bmath{\perp}}}%
  \def\bequiv{{\bmath{\equiv}}}%
  \def\basymp{{\bmath{\asymp}}}%
  \def\bsmile{{\bmath{\smile}}}%
  \def\bfrown{{\bmath{\frown}}}%
  \def\bleftharpoonup{{\bmath{\leftharpoonup}}}%
  \def\bleftharpoondown{{\bmath{\leftharpoondown}}}%
  \def\brightharpoonup{{\bmath{\rightharpoonup}}}%
  \def\brightharpoondown{{\bmath{\rightharpoondown}}}%
  \def\blhook{{\bmath{\lhook}}}%
  \def\brhook{{\bmath{\rhook}}}%
  \def\bldotp{{\bmath{\ldotp}}}%
  \def\bcdotp{{\bmath{\cdotp}}}%
}

% Make \, work in non-math mode
\def\,{\relax\ifmmode \mskip\thinmuskip\else \thinspace\fi}
\let\protect=\relax

\long\def\@ifundefined#1#2#3{\expandafter\ifx\csname
  #1\endcsname\relax#2\else#3\fi}

%%%%%%%%%%%%%%%%%%%%%%%%%%%%%%%%%%%%%%%%%

% NewFont.sty: ALPHA VERSION patchlevel 8, 16th August 1994, M. Reed

% \addtom@thgroup{math font loading info}
% Adds to internal \math@groups definition, which is executed at the end
% of each size changing command. It is called by \NewSymbolFont.

\newtoks\math@groups \math@groups={}
\def\addtom@thgroup#1#2{#1\expandafter{\the#1#2}} %  \mac={new\the\mac}

% Make TeX change the values of \s@ze, \ss@ze, \sss@ze when \@npt is
% executed. This makes it possible for math characters to be loaded
% at the correct size automatically when the size is changed.

% \addtosizeh@ok{x}{10}{7}{5}

\def\addtosizeh@ok#1#2#3#4{%
  \expandafter\def\csname @#1pt\endcsname{%
    \def\s@ze{#2}\def\ss@ze{#3}\def\sss@ze{#4}\the\math@groups%
  }%
}

% \resetsizehook allows the size parameters to be reset after \addtosizeh@ok
% has been called (it re-defines \@npt).
% e.g JFM which requires \xpt to have 10.5pt instead of 10pt.
% Note: \resetsizehook must be used in the preamble BEFORE any
% \New... commands.

% e.g. \resetsizehook{x}{10.5}{7}{5}

\let\resetsizehook=\addtosizeh@ok

% Standard LaTeX sizes

\ifprod@font
%  \addtosizeh@ok{v}    {5} {5}  {5}
%  \addtosizeh@ok{vi}   {6} {6}  {6}
%  \addtosizeh@ok{vii}  {7} {6}  {5}
  \addtosizeh@ok{viii} {8} {6}  {5}
  \addtosizeh@ok{ix}   {9} {6}  {5}
  \addtosizeh@ok{x}    {10}{7}  {5}
  \addtosizeh@ok{xi}   {11}{8}  {6}
%  \addtosizeh@ok{xii}  {12}{8}  {6}
  \addtosizeh@ok{xiv}  {14}{10} {7}
  \addtosizeh@ok{xvii} {17}{12}{10}
%  \addtosizeh@ok{xx}   {20}{14}{12}
%  \addtosizeh@ok{xxv}  {25}{20}{17}
\else
%  \addtosizeh@ok{v}    {5}     {5}     {5}
%  \addtosizeh@ok{vi}   {6}     {6}     {6}
%  \addtosizeh@ok{vii}  {7}     {6}     {5}
  \addtosizeh@ok{viii} {8}     {6}     {5}
  \addtosizeh@ok{ix}   {9}     {6}     {5}
  \addtosizeh@ok{x}    {10}    {7}     {5}
  \addtosizeh@ok{xi}   {10.95} {8}     {6}
%  \addtosizeh@ok{xii}  {12}    {8}     {6}
  \addtosizeh@ok{xiv}  {14.4}  {10}    {7}
  \addtosizeh@ok{xvii} {17.28} {12}    {10}
%  \addtosizeh@ok{xx}   {20.74} {14.4}  {12}
%  \addtosizeh@ok{xxv}  {24.88} {20.74} {17.28}
\fi

\def\get@font#1#2#3{%
  \edef\fonts@ze{\romannumeral#3}%         10 -> x
  \edef\fontn@me{\fonts@ze#1}%             AMSa -> xAMSa
  \@ifundefined{\fontn@me}%
    {%%\typeout{defining \fontn@me}%
     \global\expandafter\font\csname \fontn@me\endcsname=#2 at #3pt}%
    {}%
}

\def\ass@tfont#1#2{%
  \xdef\fam@name{\csname #1\endcsname}%
  \xdef\font@name{\csname #2\endcsname}%
  \let\textfont@name\font@name
  \textfont\fam@name\textfont@name
}

\def\ass@sfont#1#2{%
  \xdef\fam@name{\csname #1\endcsname}%
  \xdef\font@name{\csname #2\endcsname}%
  \let\textfont@name\font@name
  \scriptfont\fam@name\textfont@name
}

\def\ass@ssfont#1#2{%
  \xdef\fam@name{\csname #1\endcsname}%
  \xdef\font@name{\csname #2\endcsname}%
  \let\textfont@name\font@name
  \scriptscriptfont\fam@name\textfont@name
}

%                fam name  base font  (allocates a \newfam)
% \NewSymbolFont {AMSa}    {mtxm10}

\def\NewSymbolFont#1#2{%
  \expandafter\ifx\csname sym#1fam\endcsname\relax % if not defined
    \expandafter\newfam\csname sym#1fam\endcsname
    \expandafter\edef\csname sym#1fam\endcsname{\the\allocationnumber}%
    \addtom@thgroup\math@groups{%
      \get@font{#1}{#2}{\s@ze}%
      \ass@tfont{sym#1fam}{\fontn@me}%
      \get@font{#1}{#2}{\ss@ze}%
      \ass@sfont{sym#1fam}{\fontn@me}%
      \get@font{#1}{#2}{\sss@ze}%
      \ass@ssfont{sym#1fam}{\fontn@me}%
    }%
  \else
    \errmessage{Family `#1' already defined}%
  \fi
}

%                symbol         type fam    pos (hex)
% \NewMathSymbol {\blacksquare} {0}  {AMSa} {04}

\def\NewMathSymbol#1#2#3#4{%
  \edef\f@mly{\expandafter\hexnumber{\csname sym#3fam\endcsname}}%
  \mathchardef#1="#2\f@mly#4\relax
}

%                  macro name  type  fam1   pos  fam2   pos
% \NewMathDelimiter{\ulcorner} {4}   {AMSa} {70} {AMSb} {70}

\newif\ifd@f

\def\NewMathDelimiter#1#2#3#4#5#6{%
  \d@ftrue
  \expandafter\ifx\csname sym#3fam\endcsname\relax
    \d@ffalse \errmessage{Family `#3' is not defined}%
  \fi
  \expandafter\ifx\csname sym#5fam\endcsname\relax
    \d@ffalse \errmessage{Family `#5' is not defined}%
  \fi
  \ifd@f
    \edef\f@mly{\expandafter\hexnumber{\csname sym#3fam\endcsname}}%
    \edef\f@mlytw@{\expandafter\hexnumber{\csname sym#5fam\endcsname}}%
    \xdef#1{\delimiter"#2\f@mly #4\f@mlytw@ #6\relax}%
  \fi
}

%                  macro name  base font  skewchar setting e.g '60 (octal)
% \NewMathAlphabet {mathbssi}  {mtmisb10} {}

\def\setboxz@h{\setbox\z@\hbox}
\def\wdz@{\wd\z@}
\def\boxz@{\box\z@}
\def\setbox@ne{\setbox\@ne}
\def\wd@ne{\wd\@ne}

\def\math@atom#1#2{%
   \binrel@{#1}\binrel@@{#2}}
\def\binrel@#1{\setboxz@h{\thinmuskip0mu
  \medmuskip\m@ne mu\thickmuskip\@ne mu$#1\m@th$}%
 \setbox@ne\hbox{\thinmuskip0mu\medmuskip\m@ne mu\thickmuskip
  \@ne mu${}#1{}\m@th$}%
 \setbox\tw@\hbox{\hskip\wd@ne\hskip-\wdz@}}
\def\binrel@@#1{\ifdim\wd2<\z@\mathbin{#1}\else\ifdim\wd\tw@>\z@
 \mathrel{#1}\else{#1}\fi\fi}

\def\m@thit{1}

\def\set@skchar#1{\global\expandafter\skewchar
  \csname\fontn@me\endcsname=#1\relax}

\def\NewMathAlphabet#1#2#3{%
  \def\tst{#3}%
  \ifx\tst\empty\else % if a \skewchar setting is present
    \expandafter\gdef\csname #1@sc\endcsname{}%  \def\cmd@sc{..}
  \fi
  \expandafter\def\csname #1\endcsname{%  \def\cmd{\protect\@cmd}
    \protect\csname @#1\endcsname}%
  \expandafter\def\csname @#1\endcsname##1{%  \def\@cmd{..}
    {%
    \begingroup
      \get@font{#1}{#2}{\s@ze}%
      \@ifundefined{#1@sc}{}{\set@skchar{#3}}%
      \ass@tfont{m@thit}{\fontn@me}%
      \get@font{#1}{#2}{\ss@ze}%
      \@ifundefined{#1@sc}{}{\set@skchar{#3}}%
      \ass@sfont{m@thit}{\fontn@me}%
      \get@font{#1}{#2}{\sss@ze}%
      \@ifundefined{#1@sc}{}{\set@skchar{#3}}%
      \ass@ssfont{m@thit}{\fontn@me}%
      \math@atom{##1}{%
      \mathchoice%
        {\hbox{$\m@th\displaystyle##1$}}%
        {\hbox{$\m@th\textstyle##1$}}%
        {\hbox{$\m@th\scriptstyle##1$}}%
        {\hbox{$\m@th\scriptscriptstyle##1$}}}%
    \endgroup
    }%
  }%
}

%                  macro name  base font  hyphenchar setting e.g -1 (off)
% \NewTextAlphabet {textbfit}  {mtbxti10} {}

% save a family if \NewTextAlphabet is not used.
\newif\iffirstta  \firsttatrue

\def\set@hchar#1{\global\expandafter\hyphenchar
  \csname\fontn@me\endcsname=#1\relax}

\def\NewTextAlphabet#1#2#3{%
  \iffirstta
    \global\firsttafalse
    \newfam\scratchfam
    \edef\scrt@fam{\the\allocationnumber}%
  \fi
  \def\tst{#3}%
  \ifx\tst\empty\else % if a \hyphenchar setting is required
    \expandafter\gdef\csname #1@hc\endcsname{}%  \def\cmd@sc{..}
  \fi
  \expandafter\def\csname #1\endcsname{%  \def\cmd{\protect\t@cmd}
    \protect\csname t@#1\endcsname}%
  \long\expandafter\def\csname t@#1\endcsname##1{%  \def\t@cmd{..}
    \ifmmode
      \typeout{Warning: do not use \expandafter\string\csname #1\endcsname
        \space in math mode}\fi%
    {%
      \get@font{#1}{#2}{\s@ze}\let\t@xtfnt=\fontn@me\relax
      \@ifundefined{#1@hc}{}{\set@hchar{#3}}%
      \ass@tfont{scrt@fam}{\fontn@me}%
      \get@font{#1}{#2}{\ss@ze}%
      \@ifundefined{#1@hc}{}{\set@hchar{#3}}%
      \ass@sfont{scrt@fam}{\fontn@me}%
      \get@font{#1}{#2}{\sss@ze}%
      \@ifundefined{#1@hc}{}{\set@hchar{#3}}%
      \ass@ssfont{scrt@fam}{\fontn@me}%
      \fam\scratchfam\csname\t@xtfnt\endcsname
    ##1%
    }%
  }%
  \expandafter\def\csname #1shape%  \def\cmdshape{\protect\@cmdshape}
    \endcsname{\protect\csname @#1shape\endcsname}%
  \expandafter\def\csname @#1shape\endcsname{%  \def\@cmdshape
    \ifmmode
      \typeout{Warning: do not use \expandafter\string\csname
        #1shape\endcsname \space in math mode}\fi
      \get@font{#1}{#2}{\s@ze}\let\t@xtfnt=\fontn@me\relax
      \@ifundefined{#1@hc}{}{\set@hchar{#3}}%
      \ass@tfont{scrt@fam}{\fontn@me}%
      \get@font{#1}{#2}{\ss@ze}%
      \@ifundefined{#1@hc}{}{\set@hchar{#3}}%
      \ass@sfont{scrt@fam}{\fontn@me}%
      \get@font{#1}{#2}{\sss@ze}%
      \@ifundefined{#1@hc}{}{\set@hchar{#3}}%
      \ass@ssfont{scrt@fam}{\fontn@me}%
      \fam\scratchfam\csname\t@xtfnt\endcsname
  }%
}

% \bmath{math text}

\ifprod@font
  \def\math@itfnt{mtmib10}
  \def\math@syfnt{mtbsy10}
\else
  \def\math@itfnt{cmmib10}
  \def\math@syfnt{cmbsy10}
\fi

\def\m@thsy{2}

\def\bmath{\protect\@bmath}
\def\@bmath#1{%
  {%
  \begingroup
    \get@font{mthit}{\math@itfnt}{\s@ze}\set@skchar{'177}%
    \ass@tfont{m@thit}{\fontn@me}%
    \get@font{mthit}{\math@itfnt}{\ss@ze}\set@skchar{'177}%
    \ass@sfont{m@thit}{\fontn@me}%
    \get@font{mthit}{\math@itfnt}{\sss@ze}\set@skchar{'177}%
    \ass@ssfont{m@thit}{\fontn@me}%
    \get@font{mthsy}{\math@syfnt}{\s@ze}\set@skchar{'60}%
    \ass@tfont{m@thsy}{\fontn@me}%
    \get@font{mthsy}{\math@syfnt}{\ss@ze}\set@skchar{'60}%
    \ass@sfont{m@thsy}{\fontn@me}%
    \get@font{mthsy}{\math@syfnt}{\sss@ze}\set@skchar{'60}%
    \ass@ssfont{m@thsy}{\fontn@me}%
    \math@atom{#1}{%
    \mathchoice%
      {\hbox{$\m@th\displaystyle#1$}}%
      {\hbox{$\m@th\textstyle#1$}}%
      {\hbox{$\m@th\scriptstyle#1$}}%
      {\hbox{$\m@th\scriptscriptstyle#1$}}}%
  \endgroup
  }%
}

%%%%%%%%%%%%%%%%%%%%%%%%%%%%%%%%%%%%%%%%%

% Astronomy and Astrophysics symbol macros

\def\sun{\hbox{$\odot$}}

\def\diameter{{\ifmmode\mathchoice
{\ooalign{\hfil\hbox{$\displaystyle/$}\hfil\crcr
{\hbox{$\displaystyle\mathchar"20D$}}}}
{\ooalign{\hfil\hbox{$\textstyle/$}\hfil\crcr
{\hbox{$\textstyle\mathchar"20D$}}}}
{\ooalign{\hfil\hbox{$\scriptstyle/$}\hfil\crcr
{\hbox{$\scriptstyle\mathchar"20D$}}}}
{\ooalign{\hfil\hbox{$\scriptscriptstyle/$}\hfil\crcr
{\hbox{$\scriptscriptstyle\mathchar"20D$}}}}
\else{\ooalign{\hfil/\hfil\crcr\mathhexbox20D}}%
\fi}}

\def\sq{\ifmmode\squareforqed\else{\unskip\nobreak\hfil
\penalty50\hskip1em\null\nobreak\hfil\squareforqed
\parfillskip=0pt\finalhyphendemerits=0\endgraf}\fi}
\def\squareforqed{\hbox{\rlap{$\sqcap$}$\sqcup$}}

% Simulated Blackboard Bold symbols

\def\bbbc{{\mathchoice {\setbox0=\hbox{$\displaystyle\rm C$}\hbox{\hbox
to0pt{\kern0.4\wd0\vrule height0.9\ht0\hss}\box0}}
{\setbox0=\hbox{$\textstyle\rm C$}\hbox{\hbox
to0pt{\kern0.4\wd0\vrule height0.9\ht0\hss}\box0}}
{\setbox0=\hbox{$\scriptstyle\rm C$}\hbox{\hbox
to0pt{\kern0.4\wd0\vrule height0.9\ht0\hss}\box0}}
{\setbox0=\hbox{$\scriptscriptstyle\rm C$}\hbox{\hbox
to0pt{\kern0.4\wd0\vrule height0.9\ht0\hss}\box0}}}}
\def\bbbq{{\mathchoice {\setbox0=\hbox{$\displaystyle\rm
Q$}\hbox{\raise
0.15\ht0\hbox to0pt{\kern0.4\wd0\vrule height0.8\ht0\hss}\box0}}
{\setbox0=\hbox{$\textstyle\rm Q$}\hbox{\raise
0.15\ht0\hbox to0pt{\kern0.4\wd0\vrule height0.8\ht0\hss}\box0}}
{\setbox0=\hbox{$\scriptstyle\rm Q$}\hbox{\raise
0.15\ht0\hbox to0pt{\kern0.4\wd0\vrule height0.7\ht0\hss}\box0}}
{\setbox0=\hbox{$\scriptscriptstyle\rm Q$}\hbox{\raise
0.15\ht0\hbox to0pt{\kern0.4\wd0\vrule height0.7\ht0\hss}\box0}}}}
\def\bbbt{{\mathchoice {\setbox0=\hbox{$\displaystyle\rm
T$}\hbox{\hbox to0pt{\kern0.3\wd0\vrule height0.9\ht0\hss}\box0}}
{\setbox0=\hbox{$\textstyle\rm T$}\hbox{\hbox
to0pt{\kern0.3\wd0\vrule height0.9\ht0\hss}\box0}}
{\setbox0=\hbox{$\scriptstyle\rm T$}\hbox{\hbox
to0pt{\kern0.3\wd0\vrule height0.9\ht0\hss}\box0}}
{\setbox0=\hbox{$\scriptscriptstyle\rm T$}\hbox{\hbox
to0pt{\kern0.3\wd0\vrule height0.9\ht0\hss}\box0}}}}
\def\bbbs{{\mathchoice
{\setbox0=\hbox{$\displaystyle     \rm S$}\hbox{\raise0.5\ht0\hbox
to0pt{\kern0.35\wd0\vrule height0.45\ht0\hss}\hbox
to0pt{\kern0.55\wd0\vrule height0.5\ht0\hss}\box0}}
{\setbox0=\hbox{$\textstyle        \rm S$}\hbox{\raise0.5\ht0\hbox
to0pt{\kern0.35\wd0\vrule height0.45\ht0\hss}\hbox
to0pt{\kern0.55\wd0\vrule height0.5\ht0\hss}\box0}}
{\setbox0=\hbox{$\scriptstyle      \rm S$}\hbox{\raise0.5\ht0\hbox
to0pt{\kern0.35\wd0\vrule height0.45\ht0\hss}\raise0.05\ht0\hbox
to0pt{\kern0.5\wd0\vrule height0.45\ht0\hss}\box0}}
{\setbox0=\hbox{$\scriptscriptstyle\rm S$}\hbox{\raise0.5\ht0\hbox
to0pt{\kern0.4\wd0\vrule height0.45\ht0\hss}\raise0.05\ht0\hbox
to0pt{\kern0.55\wd0\vrule height0.45\ht0\hss}\box0}}}}
\def\bbbz{{\mathchoice {\hbox{$\sf\textstyle Z\kern-0.4em Z$}}
{\hbox{$\sf\textstyle Z\kern-0.4em Z$}}
{\hbox{$\sf\scriptstyle Z\kern-0.3em Z$}}
{\hbox{$\sf\scriptscriptstyle Z\kern-0.2em Z$}}}}

% NUMBER THE DESIGN ELEMENTS

\def\Nulle{0} % null element
\def\Afe{1}   % author affiliation
\def\Hae{2}   % heading A
\def\Hbe{3}   % heading B
\def\Hce{4}   % heading C
\def\Hde{5}   % heading D

% TEMPORARY REGISTERS

\newcount\LastMac       \LastMac=\Nulle

\newskip\half      \half=5.5pt plus 1.5pt minus 2.25pt
\newskip\one       \one=11pt plus 3pt minus 5.5pt
\newskip\onehalf   \onehalf=16.5pt plus 5.5pt minus 8.25pt
\newskip\two       \two=22pt plus 5.5pt minus 11pt

\def\Half{\addvspace{\half}}
\def\One{\addvspace{\one}}
\def\OneHalf{\addvspace{\onehalf}}
\def\Two{\addvspace{\two}}

\def\Raggedright{% set lines unjustified
  \rightskip=\z@ plus \hsize\relax
}

\def\Fullout{% set lines justified
  \rightskip=\z@\relax
}

\def\Hang#1#2{% set hanging indentation
  \hangindent=#1%
  \hangafter=#2\relax
}

% Pagestyles

\newif\ifsp@page
\def\pagestyle#1{\csname ps@#1\endcsname}
\def\thispagestyle#1{\global\sp@pagetrue\gdef\sp@type{#1}}

\def\ps@titlepage{%
  \def\@oddhead{\eightpoint\noindent \the\CatchLine
    \ifprod@font\else\qquad Printed\ \today\qquad
      (MN plain \TeX\ macros\ v\@version)\fi \hfil}%
  \let\@evenhead=\@oddhead
  \def\@oddfoot{\eightpoint\copyright\ \@pubyear\ RAS\hfil}%
  \def\@evenfoot{\hfil\eightpoint\noindent\copyright\ \@pubyear\ RAS}%
}

\def\ps@headings{%
  \def\@oddhead{\elevenpoint\it\noindent
    \hfill\the\RightHeader\hskip1.5em\rm\folio}%
  \def\@evenhead{\elevenpoint\noindent
    \folio\hskip1.5em\it\the\LeftHeader\hfill}%
  \def\@oddfoot{\eightpoint\noindent\copyright\ \@pubyear\ RAS,
    MNRAS {\bf \@volume}, \@pagerange\hfil}%
  \def\@evenfoot{\hfil\eightpoint\copyright\ \@pubyear\ RAS,
    MNRAS {\bf \@volume}, \@pagerange}%
}

\def\ps@plate{%
  \def\@oddhead{\eightpoint\noindent\plt@cap\hfil}%
  \def\@evenhead{\eightpoint\noindent\plt@cap\hfil}%
  \def\@oddfoot{\eightpoint\noindent\copyright\ \@pubyear\ RAS,
    MNRAS {\bf \@volume}, \@pagerange\hfil}%
  \def\@evenfoot{\hfil\eightpoint\copyright\ \@pubyear\ RAS,
    MNRAS {\bf \@volume}, \@pagerange}%
}

% DESIGN ELEMENT DEFINITIONS

% Article opening

\def\title#1{% article title
  \bgroup
    \vbox to 8pt{\vss}%
    \seventeenpoint
    \Raggedright
    \noindent \strut{\bf #1}\par
  \egroup
}

\def\author#1{% article author(s)
  \bgroup
    \ifnum\LastMac=\Afe \OneHalf\else \vskip 21pt\fi
    \fourteenpoint
    \Raggedright
    \noindent \strut #1\par
    \vskip 3pt%
  \egroup
}

\def\affiliation#1{% author(s) affiliation
  \bgroup
    \vskip -4pt%
    \eightpoint
    \Raggedright
    \noindent \strut {\it #1}\par
  \egroup
  \LastMac=\Afe\relax
}

\def\acceptedline#1{% acceptance date
  \bgroup
    \Two
    \eightpoint
    \Raggedright
    \noindent \strut #1\par
  \egroup
}

\long\def\abstract#1{%
  \bgroup
    \vskip 20pt%
    \leftskip 11pc\rightskip\z@
    \noindent{\ninebf ABSTRACT}\par
    \tenpoint
    \Fullout
    \noindent #1\par
  \egroup
}

\long\def\keywords#1{% keywords
  \bgroup
    \Half
    \leftskip 11pc\rightskip\z@
    \tenpoint
    \Fullout
    \noindent\hbox{\bf Key words:}\ #1\par
  \egroup
}

% The \maketitle macro ensures that the two spanning material appears
% at the top of the first page, and that it has two lines of space
% underneath it. If you forget this in you input, no output will be produced.
% The \BeginOpening (alias \begintopmatter) macro should be called at the
% very start of the input file, so that it is in force when the document
% starts. This ensures that when \maketitle is called that the group is
% closed, and the material gets printed.

\def\maketitle{%
  \EndOpening
  \ifsinglecol \else \MakePage\fi
}

% Page offset

% Counter setup

\def\@nameuse#1{\csname #1\endcsname}
\def\arabic#1{\@arabic{\@nameuse{#1}}}
\def\alph#1{\@alph{\@nameuse{#1}}}
\def\Alph#1{\@Alph{\@nameuse{#1}}}
\def\@arabic#1{\number #1}
\def\@Alph#1{\ifcase#1\or A\or B\or C\or D\else\@Ialph{#1}\fi}
\def\@Ialph#1{\ifcase#1\or \or \or \or \or E\or F\or G\or H\or I\or J\or
   K\or L\or M\or N\or O\or P\or Q\or R\or S\or T\or U\or V\or W\or X\or
   Y\or Z\else\errmessage{Counter out of range}\fi}
\def\@alph#1{\ifcase#1\or a\or b\or c\or d\else\@ialph{#1}\fi}
\def\@ialph#1{\ifcase#1\or \or \or \or \or e\or f\or g\or h\or i\or j\or
   k\or l\or m\or n\or o\or p\or q\or r\or s\or t\or u\or v\or w\or x\or y\or
   z\else\errmessage{Counter out of range}\fi}

% Equation auto-numbering

\newcount\Eqnno
\newcount\SubEqnno

\def\theeq{\arabic{Eqnno}}
\def\thesubeq{\alph{SubEqnno}}

\def\stepeq{\relax
  \global\SubEqnno \z@
  \global\advance\Eqnno \@ne\relax
  {\rm (\theeq)}%
}

\def\startsubeq{\relax
  \global\SubEqnno \z@
  \global\advance\Eqnno \@ne\relax
  \stepsubeq
}

\def\stepsubeq{\relax
  \global\advance\SubEqnno \@ne\relax
  {\rm (\theeq\thesubeq)}%
}

% Headings

\newcount\Sec        %  heading auto number counters
\newcount\SecSec
\newcount\SecSecSec

\def\thesection{\arabic{Sec}}
\def\thesubsection{\thesection.\arabic{SecSec}}
\def\thesubsubsection{\thesubsection.\arabic{SecSecSec}}

\Sec=\z@

\def\:{\let\@sptoken= } \:  % this makes \@sptoken a space token 
\def\:{\@xifnch} \expandafter\def\: {\futurelet\@tempc\@ifnch}

\def\@ifnextchar#1#2#3{%
  \let\@tempMACe #1%
  \def\@tempMACa{#2}%
  \def\@tempMACb{#3}%
  \futurelet \@tempMACc\@ifnch%
}

\def\@ifnch{%
\ifx \@tempMACc \@sptoken%
  \let\@tempMACd\@xifnch%
\else%
  \ifx \@tempMACc \@tempMACe%
    \let\@tempMACd\@tempMACa%
  \else%
    \let\@tempMACd\@tempMACb%
  \fi%
\fi%
\@tempMACd%
}

\def\@ifstar#1#2{\@ifnextchar *{\def\@tempMACa*{#1}\@tempMACa}{#2}}

\newskip\@tempskipb

\def\addvspace#1{%
  \ifvmode\else \endgraf\fi%
  \ifdim\lastskip=\z@%
    \vskip #1\relax%
  \else%
    \@tempskipb#1\relax\@xaddvskip%
  \fi%
}

\def\@xaddvskip{%
  \ifdim\lastskip<\@tempskipb%
    \vskip-\lastskip%
    \vskip\@tempskipb\relax%
  \else%
    \ifdim\@tempskipb<\z@%
      \ifdim\lastskip<\z@ \else%
        \advance\@tempskipb\lastskip%
        \vskip-\lastskip\vskip\@tempskipb%
      \fi%
    \fi%
  \fi%
}

\newskip\@tmpSKIP

\def\addpen#1{%
  \ifvmode
    \if@nobreak
    \else
      \ifdim\lastskip=\z@
        \penalty#1\relax
      \else
        \@tmpSKIP=\lastskip
        \vskip -\lastskip
        \penalty#1\vskip\@tmpSKIP
      \fi
    \fi
  \fi
}

\newcount\@clubpen   \@clubpen=\clubpenalty
\newif\if@nobreak    \@nobreakfalse

\def\@noafterindent{%
  \global\@nobreaktrue
  \everypar{\if@nobreak
              \global\@nobreakfalse
              \clubpenalty \@M
              {\setbox\z@\lastbox}%
              \LastMac=\Nulle\relax%
            \else
              \clubpenalty \@clubpen
              \everypar{}%
            \fi}%
}

\newcount\gds@cbrk   \gds@cbrk=-300

\def\@nohdbrk{\interlinepenalty \@M\relax}

\let\@par=\par
\def\@restorepar{\def\par{\@par}}

\newif\if@endpe   \@endpefalse
 
\def\@doendpe{\@endpetrue \@nobreakfalse \LastMac=\Nulle\relax%
     \def\par{\@restorepar\everypar{}\par\@endpefalse}%
              \everypar{\setbox\z@\lastbox\everypar{}\@endpefalse}%
}

\def\section{\@ifstar{\@ssection}{\@section}}

\def\@section#1{% heading A (\section{....})
  \if@nobreak
    \everypar{}%
    \ifnum\LastMac=\Hae \addvspace{\half}\fi
  \else
    \addpen{\gds@cbrk}%
    \addvspace{\two}%
  \fi
  \bgroup
    \ninepoint\bf
    \Raggedright
    \global\advance\Sec \@ne
    \ifappendix
      \global\Eqnno=\z@ \global\SubEqnno=\z@\relax
      \def\ch@ck{#1}%
      \ifx\ch@ck\empty \def\c@lon{}\else\def\c@lon{:}\fi
      \noindent\@nohdbrk APPENDIX\ \thesection\c@lon\hskip 0.5em%
        \uppercase{#1}\par
    \else
      \noindent\@nohdbrk\thesection\hskip 1pc \uppercase{#1}\par
    \fi
    \global\SecSec=\z@
  \egroup
  \nobreak
  \vskip\half
  \nobreak
  \@noafterindent
  \LastMac=\Hae\relax
}

\def\@ssection#1{%  main section heading (\section*{....})
  \if@nobreak
    \everypar{}%
    \ifnum\LastMac=\Hae \addvspace{\half}\fi
  \else
    \addpen{\gds@cbrk}%
    \addvspace{\two}%
  \fi
  \bgroup
    \ninepoint\bf
    \Raggedright
%    \ifappendix
%      \global\Eqnno=\z@ \global\SubEqnno=\z@\relax % mh in apps dont reset
%      \noindent\@nohdbrk APPENDIX:\hskip 0.5em%
%        \uppercase{#1}\par
%    \else
    \noindent\@nohdbrk\uppercase{#1}\par
%    \fi
  \egroup
  \nobreak
  \vskip\half
  \nobreak
  \@noafterindent
  \LastMac=\Hae\relax
}

\def\subsection{\@ifstar{\@ssubsection}{\@subsection}}

\def\@subsection#1{% heading B
  \if@nobreak
    \everypar{}%
    \ifnum\LastMac=\Hae \addvspace{1pt plus 1pt minus .5pt}\fi
  \else
    \addpen{\gds@cbrk}%
    \addvspace{\onehalf}%
  \fi
  \bgroup
    \ninepoint\bf
    \Raggedright
    \global\advance\SecSec \@ne
    \noindent\@nohdbrk\thesubsection \hskip 1pc\relax #1\par
    \global\SecSecSec=\z@
  \egroup
  \nobreak
  \vskip\half
  \nobreak
  \@noafterindent
  \LastMac=\Hbe\relax
}

\def\@ssubsection#1{% heading B*
  \if@nobreak
    \everypar{}%
    \ifnum\LastMac=\Hae \addvspace{1pt plus 1pt minus .5pt}\fi
  \else
    \addpen{\gds@cbrk}%
    \addvspace{\onehalf}%
  \fi
  \bgroup
    \ninepoint\bf
    \Raggedright
    \noindent\@nohdbrk #1\par
  \egroup
  \nobreak
  \vskip\half
  \nobreak
  \@noafterindent
  \LastMac=\Hbe\relax
}

\def\subsubsection{\@ifstar{\@ssubsubsection}{\@subsubsection}}

\def\@subsubsection#1{% heading C
  \if@nobreak
    \everypar{}%
    \ifnum\LastMac=\Hbe \addvspace{1pt plus 1pt minus .5pt}\fi
  \else
    \addpen{\gds@cbrk}%
    \addvspace{\onehalf}%
  \fi
  \bgroup
    \ninepoint\it
    \Raggedright
    \global\advance\SecSecSec \@ne
    \noindent\@nohdbrk\thesubsubsection \hskip 1pc\relax #1\par
  \egroup
  \nobreak
  \vskip\half
  \nobreak
  \@noafterindent
  \LastMac=\Hce\relax
}

\def\@ssubsubsection#1{% heading C*
  \if@nobreak
    \everypar{}%
    \ifnum\LastMac=\Hbe \addvspace{1pt plus 1pt minus .5pt}\fi
  \else
    \addpen{\gds@cbrk}%
    \addvspace{\onehalf}%
  \fi
  \bgroup
    \ninepoint\it
    \Raggedright
    \noindent\@nohdbrk #1\par
  \egroup
  \nobreak
  \vskip\half
  \nobreak
  \@noafterindent
  \LastMac=\Hce\relax
}

\def\paragraph#1{% heading D
  \if@nobreak
    \everypar{}%
  \else
    \addpen{\gds@cbrk}%
    \addvspace{\one}%
  \fi%
  \bgroup%
    \ninepoint\it
    \noindent #1\ \nobreak%
  \egroup
  \LastMac=\Hde\relax
  \ignorespaces
}

% Appendix

\newif\ifappendix

\def\appendix{%
  \global\appendixtrue
  \def\thesection{\Alph{Sec}}%
  \def\thesubsection{\thesection\arabic{SecSec}}%
  \def\theeq{\thesection\arabic{Eqnno}}%
  \Sec=\z@ \SecSec=\z@ \SecSecSec=\z@ \Eqnno=\z@ \SubEqnno=\z@\relax
}

% Text

 % provided for backward compatibility

% Lists

\def\beginlist{%
  \par\if@nobreak \else\addvspace{\half}\fi%
  \bgroup%
    \ninepoint
    \let\item=\list@item%
}

\def\list@item{%
  \par\noindent\hskip 1em\relax%
  \ignorespaces%
}

\def\endlist{\par\egroup\addvspace{\half}\@doendpe}

% References

\def\beginrefs{%
  \par
  \bgroup
    \eightpoint
    \Fullout
    \let\bibitem=\bib@item
}

\def\bib@item{%
  \par\parindent=1.5em\Hang{1.5em}{1}%
  \everypar={\Hang{1.5em}{1}\ignorespaces}%
  \noindent\ignorespaces
}

\def\endrefs{\par\egroup\@doendpe}

% Page heads

\newtoks\CatchLine

\def\@journal{Mon.\ Not.\ R.\ Astron.\ Soc.\ }  % The journal title string
\def\@pubyear{1994}        % Assign a default publication year
\def\@pagerange{000--000}  % Assign a default page-range
\def\@volume{000}          % Assign a default volume number
\def\@microfiche{}         %

\def\pubyear#1{\gdef\@pubyear{#1}\@makecatchline}
\def\pagerange#1{\gdef\@pagerange{#1}\@makecatchline}
\def\volume#1{\gdef\@volume{#1}\@makecatchline}
\def\microfiche#1{\gdef\@microfiche{and Microfiche\ #1}\@makecatchline}

\def\@makecatchline{%
  \global\CatchLine{%
    {\rm \@journal {\bf \@volume},\ \@pagerange\ (\@pubyear)\ \@microfiche}}%
}

\@makecatchline % Assign a catchline, using the above defaults

\newtoks\LeftHeader
\def\shortauthor#1{% left page head
  \global\LeftHeader{#1}%
}

\newtoks\RightHeader
\def\shorttitle#1{% right page head
  \global\RightHeader{#1}%
}

\def\PageHead{% recto/verso running heads
  \begingroup
    \ifsp@page
      \csname ps@\sp@type\endcsname
    \fi
    \ifodd\pageno
      \let\the@head=\@oddhead
    \else
      \let\the@head=\@evenhead
    \fi
    \vbox to \z@{\vskip-22.5\p@%
      \hbox to \PageWidth{\vbox to8.5\p@{}%
        \the@head
      }%
    \vss}%
  \endgroup
  \nointerlineskip
}

\gdef\PageFoot{%
  \nointerlineskip%
  \begingroup
  \ifsp@page
    \csname ps@\sp@type\endcsname
    \global\sp@pagefalse
  \fi
  \vbox to 22pt{\vfil%
    \hbox to \PageWidth{%
      \eightpoint\strut\noindent
      \ifodd\pageno
        \@oddfoot
      \else
        \@evenfoot
      \fi
    }%
  }%
  \endgroup
}

\def\today{%
  \number\day\space
  \ifcase\month\or January\or February\or March\or April\or May\or June\or
    July\or August\or September\or October\or November\or December\fi
  \space\number\year%
}

\def\authorcomment#1{%
  \gdef\PageFoot{%
    \nointerlineskip%
    \vbox to 20pt{\vfil%
      \hbox to \PageWidth{\elevenpoint\noindent \hfil #1 \hfil}}%
  }%
}

% Plate pages

\newif\ifplate@page
\newbox\plt@box

\def\beginplatepage{%
  \let\plate=\plate@head
  \let\caption=\fig@caption
  \global\setbox\plt@box=\vbox\bgroup
  \TEMPDIMEN=\PageWidth % For \fig@caption test
  \hsize=\PageWidth\relax
}

\def\endplatepage{\par\egroup\global\plate@pagetrue}
\def\plate@head#1{\gdef\plt@cap{#1}}

% Letters option

\def\letters{%
  \gdef\folio{\ifnum\pageno<\z@ L\romannumeral-\pageno
    \else L\number\pageno \fi}%
}

% Math setup

% The standard math indentation
\newdimen\mathindent

\global\mathindent=\z@
\global\everydisplay{\global\@dspwd=\displaywidth\displaysetup}

% New versions of \displaylines, \eqalign, \eqalignno for
% when non-centered math is in use.

\def\@displaylines#1{% (for non-centered math)
  {}$\displ@y\hbox{\vbox{\halign{$\@lign\hfil\displaystyle##\hfil$\crcr
  #1\crcr}}}${}%
}

\def\@eqalign#1{\null\vcenter{\openup\jot\m@th% (for non-centered math)
  \ialign{\strut\hfil$\displaystyle{##}$&$\displaystyle{{}##}$\hfil
      \crcr#1\crcr}}%
}

\def\@eqalignno#1{% (for non-centered math)
  \global\advance\@dspwd by -\mathindent%
  {}$\displ@y\hbox{\vbox{\halign to\@dspwd%
  {\hfil$\@lign\displaystyle{##}$\tabskip\z@skip
  &$\@lign\displaystyle{{}##}$\hfil\tabskip\centering
  &\llap{$\@lign##$}\tabskip\z@skip\crcr
  #1\crcr}}}${}%
}

% When equations are flushleft ensure, that \displaylines,
% \eqalign, \eqalignno and \leqalignno point to the new versions of
% the macros. Also make \leqalignno act like \eqalignno, otherwise the
% equation text would `crash' into the equation number.

\global\let\displaylines=\@displaylines
\global\let\eqalign=\@eqalign
\global\let\eqalignno=\@eqalignno
\global\let\leqalignno=\@eqalignno

\newdimen\@dspwd   \@dspwd=\z@
\newif\if@eqno
\newif\if@leqno
\newtoks\@eqn
\newtoks\@eq

\def\displaysetup#1$${\displaytest#1\eqno\eqno\displaytest}

\def\displaytest#1\eqno#2\eqno#3\displaytest{%
 \if!#3!\ldisplaytest#1\leqno\leqno\ldisplaytest
 \else\@eqnotrue\@leqnofalse\@eqn={#2}\@eq={#1}\fi
 \generaldisplay$$}

\def\ldisplaytest#1\leqno#2\leqno#3\ldisplaytest{%
\@eq={#1}%
 \if!#3!\@eqnofalse\else\@eqnotrue\@leqnotrue
  \@eqn={#2}\fi}

\def\generaldisplay{%
  \if@eqno
    \if@leqno
      \hbox to \displaywidth{\noindent
        \rlap{$\displaystyle\the\@eqn$}%
        \hskip\mathindent$\displaystyle\the\@eq$\hfil}%
    \else
      \hbox to \displaywidth{\noindent
        \hskip\mathindent
        $\displaystyle\the\@eq$\hfil$\displaystyle\the\@eqn$}%
    \fi
  \else
    \hbox to \displaywidth{\noindent
      \hskip\mathindent$\displaystyle\the\@eq$\hfil}%
  \fi
}

% Finishing notice

\def\@notice{%
  \par\Two%
  \noindent{\b@ls{11pt}\ninerm This paper has been produced using the
    Royal Astronomical Society/Blackwell Science \TeX\ macros.\par}%
}

% redefine \bye to output our identification notice :
\outer\def\bye{\@notice\par\vfill\supereject\end}

% define a sign on :

\def\start@mess{%
  Monthly notices of the RAS journal style (\@typeface)\space
    v\@version,\space \@verdate.%
}

\everyjob{\Warn{\start@mess}}

% Two-column macros

%--------------------------------------------------------%
%                     INITIALISATION                     %
%--------------------------------------------------------%

\newif\if@debug \@debugfalse  %  when false, only warnings displayed

\def\Print#1{\if@debug\immediate\write16{#1}\else \fi}
\def\Warn#1{\immediate\write16{#1}}
\def\wlog#1{}

\newcount\Iteration % temporary loop counter

\def\Single{0} \def\Double{1}                 % ItemSPAN's
\def\Figure{0} \def\Table{1}                  % ItemTYPE's

\def\InStack{0}  % ItemSTATUS
\def\InZoneA{1}
\def\InZoneB{2}
\def\InZoneC{3}

\newcount\TEMPCOUNT % temporary count register
\newdimen\TEMPDIMEN % temporary dimen register
\newbox\TEMPBOX     % temporary box register
\newbox\VOIDBOX     % a box which is permenately void

\newcount\LengthOfStack % number of items currently in stack
\newcount\MaxItems      % maximum number of items allowed in stack
\newcount\StackPointer
\newcount\Point         % used in calculation for generating the
                        % physical address of an item in the stack.
\newcount\NextFigure    % number of next figure to be output
\newcount\NextTable     % number of next table to be output
\newcount\NextItem      % Next item to consider by order in stack

\newcount\StatusStack   % set to point to top of STATUS stack
\newcount\NumStack      % set to point to top of NUMBER stack
\newcount\TypeStack     % set to point to top of TYPE stack
\newcount\SpanStack     % set to point to top of SPAN stack
\newcount\BoxStack      % set to point to top of BOX stack

\newcount\ItemSTATUS    % status of present item
\newcount\ItemNUMBER    % number of present item
\newcount\ItemTYPE      % type of present item
\newcount\ItemSPAN      % span of present item
\newbox\ItemBOX         % box of present item
\newdimen\ItemSIZE      % size of present item
                        % (calculated by GetItemBOX)

\newdimen\PageHeight    % vertical measure of full page
\newdimen\TextLeading   % distance between baselines of body text
\newdimen\Feathering    % amount of interline stretch
\newcount\LinesPerPage  % height of page in text lines
\newdimen\ColumnWidth   % width of 1 column of text
\newdimen\ColumnGap     % size of gap between columns
\newdimen\PageWidth     % = \ColumnWidth * 2 + \ColumnGap
\newdimen\BodgeHeight   % Bodge to align figures and tables with text
\newcount\Leading       % Set to same as \TextLeading above

\newdimen\ZoneBSize  % size of items in ZoneB
\newdimen\TextSize   % size of text in ZoneB
\newbox\ZoneABOX     % contains Zone A material
\newbox\ZoneBBOX     % contains Zone B material
\newbox\ZoneCBOX     % contains Zone C material

\newif\ifFirstSingleItem
\newif\ifFirstZoneA
\newif\ifMakePageInComplete
\newif\ifMoreFigures \MoreFiguresfalse % set true in join stack
\newif\ifMoreTables  \MoreTablesfalse  % set true in join stack

\newif\ifFigInZoneB % used to determine in which zone an item
\newif\ifFigInZoneC % will be placed based on what is in other
\newif\ifTabInZoneB % zones already for a given page.
\newif\ifTabInZoneC

\newif\ifZoneAFullPage

\newbox\MidBOX    % = LeftBOX+gap+RightBOX
\newbox\LeftBOX
\newbox\RightBOX
\newbox\PageBOX   % complete made-up page

\newif\ifLeftCOL  % flags first pass through output routine
\LeftCOLtrue

\newdimen\ZoneBAdjust

\newcount\ItemFits
\def\Yes{1}
\def\No{2}

% Setup file.

\MaxItems=15
\NextFigure=\z@        % used for article opening
\NextTable=\@ne

\BodgeHeight=6pt
\TextLeading=11pt    % baselineskip of body text
\Leading=11
\Feathering=\z@      % amount of interline stretch
\LinesPerPage=61     % number of text lines per full page -1
\topskip=\TextLeading
\ColumnWidth=20pc    % width of text columns
\ColumnGap=2pc       % gap between columns

\newskip\ItemSepamount  % space between floats
\ItemSepamount=\TextLeading plus \TextLeading minus 4pt

\parskip=\z@ plus .1pt
\parindent=18pt
\widowpenalty=\z@
\clubpenalty=10000
\tolerance=1500
\hbadness=1500
\abovedisplayskip=6pt plus 2pt minus 1pt
\belowdisplayskip=6pt plus 2pt minus 1pt
\abovedisplayshortskip=6pt plus 2pt minus 1pt
\belowdisplayshortskip=6pt plus 2pt minus 1pt

\frenchspacing

\ninepoint % start main text size

\PageHeight=682pt
\PageWidth=2\ColumnWidth
\advance\PageWidth by \ColumnGap

\pagestyle{headings}

%--------------------------------------------------------%
%                         STACKS                         %
%--------------------------------------------------------%

% THE ITEM STACK
% The item stack contains contains figures and tables
% in the order in which they appear in the article source
% code.

% allocate stack space

\newcount\DUMMY \StatusStack=\allocationnumber
\newcount\DUMMY \newcount\DUMMY \newcount\DUMMY 
\newcount\DUMMY \newcount\DUMMY \newcount\DUMMY 
\newcount\DUMMY \newcount\DUMMY \newcount\DUMMY
\newcount\DUMMY \newcount\DUMMY \newcount\DUMMY 
\newcount\DUMMY \newcount\DUMMY \newcount\DUMMY

\newcount\DUMMY \NumStack=\allocationnumber
\newcount\DUMMY \newcount\DUMMY \newcount\DUMMY 
\newcount\DUMMY \newcount\DUMMY \newcount\DUMMY 
\newcount\DUMMY \newcount\DUMMY \newcount\DUMMY 
\newcount\DUMMY \newcount\DUMMY \newcount\DUMMY 
\newcount\DUMMY \newcount\DUMMY \newcount\DUMMY

\newcount\DUMMY \TypeStack=\allocationnumber
\newcount\DUMMY \newcount\DUMMY \newcount\DUMMY 
\newcount\DUMMY \newcount\DUMMY \newcount\DUMMY 
\newcount\DUMMY \newcount\DUMMY \newcount\DUMMY 
\newcount\DUMMY \newcount\DUMMY \newcount\DUMMY 
\newcount\DUMMY \newcount\DUMMY \newcount\DUMMY

\newcount\DUMMY \SpanStack=\allocationnumber
\newcount\DUMMY \newcount\DUMMY \newcount\DUMMY 
\newcount\DUMMY \newcount\DUMMY \newcount\DUMMY 
\newcount\DUMMY \newcount\DUMMY \newcount\DUMMY 
\newcount\DUMMY \newcount\DUMMY \newcount\DUMMY 
\newcount\DUMMY \newcount\DUMMY \newcount\DUMMY

\newbox\DUMMY   \BoxStack=\allocationnumber
\newbox\DUMMY   \newbox\DUMMY \newbox\DUMMY 
\newbox\DUMMY   \newbox\DUMMY \newbox\DUMMY 
\newbox\DUMMY   \newbox\DUMMY \newbox\DUMMY 
\newbox\DUMMY   \newbox\DUMMY \newbox\DUMMY 
\newbox\DUMMY   \newbox\DUMMY \newbox\DUMMY

\def\wlog{\immediate\write\m@ne}

% \GetItemSTATUS, \GetItemNUMBER, \GetItemTYPE, \GetItemSPAN,
% \GetItemBox 
% are used to get details of a particular item from the item
% stack. The argument to each of these is the items position
% in the stack (usually \StackPointer)...not the items number.

\def\GetItemAll#1{%
 \GetItemSTATUS{#1}
 \GetItemNUMBER{#1}
 \GetItemTYPE{#1}
 \GetItemSPAN{#1}
 \GetItemBOX{#1}
}

% Note: \LeaveStack uses this routine. Do not destroy \Point
\def\GetItemSTATUS#1{%
 \Point=\StatusStack
 \advance\Point by #1
 \global\ItemSTATUS=\count\Point
}

% Note: \LeaveStack uses this routine. Do not destroy \Point
\def\GetItemNUMBER#1{%
 \Point=\NumStack
 \advance\Point by #1
 \global\ItemNUMBER=\count\Point
}

% Note: \LeaveStack uses this routine. Do not destroy \Point
\def\GetItemTYPE#1{%
 \Point=\TypeStack
 \advance\Point by #1
 \global\ItemTYPE=\count\Point
}

% Note: \LeaveStack uses this routine. Do not destroy \Point
\def\GetItemSPAN#1{%
 \Point\SpanStack
 \advance\Point by #1
 \global\ItemSPAN=\count\Point
}

% Note: \LeaveStack uses this routine. Do not destroy \Point
\def\GetItemBOX#1{%
 \Point=\BoxStack
 \advance\Point by #1
 \global\setbox\ItemBOX=\vbox{\copy\Point}
 \global\ItemSIZE=\ht\ItemBOX
 \global\advance\ItemSIZE by \dp\ItemBOX
 \TEMPCOUNT=\ItemSIZE
 \divide\TEMPCOUNT by \Leading
 \divide\TEMPCOUNT by 65536
 \advance\TEMPCOUNT \@ne
 \ItemSIZE=\TEMPCOUNT pt
 \global\multiply\ItemSIZE by \Leading
}

% item joins stack

\def\JoinStack{%
 \ifnum\LengthOfStack=\MaxItems % stack is full of items
  \Warn{WARNING: Stack is full...some items will be lost!}
 \else
  \Point=\StatusStack
  \advance\Point by \LengthOfStack
  \global\count\Point=\ItemSTATUS
  \Point=\NumStack
  \advance\Point by \LengthOfStack
  \global\count\Point=\ItemNUMBER
  \Point=\TypeStack
  \advance\Point by \LengthOfStack
  \global\count\Point=\ItemTYPE
  \Point\SpanStack
  \advance\Point by \LengthOfStack
  \global\count\Point=\ItemSPAN
  \Point=\BoxStack
  \advance\Point by \LengthOfStack
  \global\setbox\Point=\vbox{\copy\ItemBOX}
  \global\advance\LengthOfStack \@ne
  \ifnum\ItemTYPE=\Figure % used in \MakePage
   \global\MoreFigurestrue
  \else
   \global\MoreTablestrue
  \fi
 \fi
}

% item leaves stack
% #1=physical position of the item to be removed

\def\LeaveStack#1{%
 {\Iteration=#1
 \loop
 \ifnum\Iteration<\LengthOfStack
  \advance\Iteration \@ne
  \GetItemSTATUS{\Iteration}
   \advance\Point by \m@ne
   \global\count\Point=\ItemSTATUS
  \GetItemNUMBER{\Iteration}
   \advance\Point by \m@ne
   \global\count\Point=\ItemNUMBER
  \GetItemTYPE{\Iteration}
   \advance\Point by \m@ne
   \global\count\Point=\ItemTYPE
  \GetItemSPAN{\Iteration}
   \advance\Point by \m@ne
   \global\count\Point=\ItemSPAN
  \GetItemBOX{\Iteration}
   \advance\Point by \m@ne
   \global\setbox\Point=\vbox{\copy\ItemBOX}
 \repeat}
 \global\advance\LengthOfStack by \m@ne
}

% clean stack
% This routine scans through the stack and removes anything
% that does not have STATUS=\InStack.

\newif\ifStackNotClean

\def\CleanStack{%
 \StackNotCleantrue
 {\Iteration=\z@
  \loop
   \ifStackNotClean
    \GetItemSTATUS{\Iteration}
    \ifnum\ItemSTATUS=\InStack
     \advance\Iteration \@ne
     \else
      \LeaveStack{\Iteration}
    \fi
   \ifnum\LengthOfStack<\Iteration
    \StackNotCleanfalse
   \fi
 \repeat}
}

% Find item.
% This macro searches from the top to the bottom of the
% stack for an item of a specified type and number.
% #1=type, #2=number
% If the specified item is found, then \StackPointer is set
% to point to it, else \StackPointer=-1.
% This routine is used to find the next figure or table
% by number.

\def\FindItem#1#2{%
 \global\StackPointer=\m@ne % assume item isn't in stack for now
 {\Iteration=\z@
  \loop
  \ifnum\Iteration<\LengthOfStack
   \GetItemSTATUS{\Iteration}
   \ifnum\ItemSTATUS=\InStack
    \GetItemTYPE{\Iteration}
    \ifnum\ItemTYPE=#1
     \GetItemNUMBER{\Iteration}
     \ifnum\ItemNUMBER=#2
      \global\StackPointer=\Iteration
      \Iteration=\LengthOfStack % terminate loop
     \fi
    \fi
   \fi
  \advance\Iteration \@ne
 \repeat}
}

% Find next type
% This macro searches from the top to the bottom of the stack
% looking for the first item which has STATUS=\InStack.
% If it is a figure then a figure is what will be considered
% next by \MakePage else table.

\def\FindNext{%
 \global\StackPointer=\m@ne % assume stack is empty for now
 {\Iteration=\z@
  \loop
  \ifnum\Iteration<\LengthOfStack
   \GetItemSTATUS{\Iteration}
   \ifnum\ItemSTATUS=\InStack
    \GetItemTYPE{\Iteration}
   \ifnum\ItemTYPE=\Figure
    \ifMoreFigures
      \global\NextItem=\Figure
      \global\StackPointer=\Iteration
      \Iteration=\LengthOfStack % terminate loop
    \fi
   \fi
   \ifnum\ItemTYPE=\Table
    \ifMoreTables
      \global\NextItem=\Table
      \global\StackPointer=\Iteration
      \Iteration=\LengthOfStack % terminate loop
    \fi
   \fi
  \fi
  \advance\Iteration \@ne
 \repeat}
}

% Change status
% Macro to change the status of a specified item in stack.
% #1=item, #2=new status

\def\ChangeStatus#1#2{%
 \Point=\StatusStack
 \advance\Point by #1
 \global\count\Point=#2
}

%--------------------------------------------------------%
%                       MAKEPAGE                         %
%--------------------------------------------------------%

% This macro is called at the start of every new page
% including the first. It scans through the stack picking
% out items which should be placed on this page. It then
% leaves space for the items to be placed later. The routine
% terminates when either there is no room on the page to
% fit the next figure or table, or there are no more items
% in the stack.

\def\Zone{\InZoneA}

\ZoneBAdjust=\z@

\def\MakePage{% allocate space on this page for stack items
 \global\ZoneBSize=\PageHeight
 \global\TextSize=\ZoneBSize
 \global\ZoneAFullPagefalse
 \global\topskip=\TextLeading
 \MakePageInCompletetrue
 \MoreFigurestrue
 \MoreTablestrue
 \FigInZoneBfalse
 \FigInZoneCfalse
 \TabInZoneBfalse
 \TabInZoneCfalse
 \global\FirstSingleItemtrue
 \global\FirstZoneAtrue
 \global\setbox\ZoneABOX=\box\VOIDBOX
 \global\setbox\ZoneBBOX=\box\VOIDBOX
 \global\setbox\ZoneCBOX=\box\VOIDBOX
 \loop
  \ifMakePageInComplete
 \FindNext
 \ifnum\StackPointer=\m@ne
  \NextItem=\m@ne
  \MoreFiguresfalse
  \MoreTablesfalse
 \fi
 \ifnum\NextItem=\Figure
   \FindItem{\Figure}{\NextFigure}
   \ifnum\StackPointer=\m@ne \global\MoreFiguresfalse
   \else
    \GetItemSPAN{\StackPointer}
    \ifnum\ItemSPAN=\Single \def\Zone{\InZoneB}\relax
     \ifFigInZoneC \global\MoreFiguresfalse\fi
    \else
     \def\Zone{\InZoneA}
     \ifFigInZoneB \def\Zone{\InZoneC}\fi
    \fi
   \fi
   \ifMoreFigures\Print{}\FigureItems\fi
 \fi
\ifnum\NextItem=\Table
   \FindItem{\Table}{\NextTable}
   \ifnum\StackPointer=\m@ne \global\MoreTablesfalse
   \else
    \GetItemSPAN{\StackPointer}
    \ifnum\ItemSPAN=\Single\relax
     \ifTabInZoneC \global\MoreTablesfalse\fi
    \else
     \def\Zone{\InZoneA}
     \ifTabInZoneB \def\Zone{\InZoneC}\fi
    \fi
   \fi
   \ifMoreTables\Print{}\TableItems\fi
 \fi
   \MakePageInCompletefalse % assume page is complete
   \ifMoreFigures\MakePageInCompletetrue\fi
   \ifMoreTables\MakePageInCompletetrue\fi
 \repeat
%\Print{TextSize=\the\TextSize}
%\Print{ZoneBSize=\the\ZoneBSize}
 \ifZoneAFullPage
  \global\TextSize=\z@
  \global\ZoneBSize=\z@
  \global\vsize=\z@\relax
  \global\topskip=\z@\relax
  \vbox to \z@{\vss}
  \eject
 \else
 \global\advance\ZoneBSize by -\ZoneBAdjust
 \global\vsize=\ZoneBSize
 \global\hsize=\ColumnWidth
 \global\ZoneBAdjust=\z@
 \ifdim\TextSize<23pt
 \Warn{}
 \Warn{* Making column fall short: TextSize=\the\TextSize *}
 \vskip-\lastskip\eject\fi
 \fi
}

\def\MakeRightCol{% allocate space for the right column of text
 \global\TextSize=\ZoneBSize
 \MakePageInCompletetrue
 \MoreFigurestrue
 \MoreTablestrue
 \global\FirstSingleItemtrue
 \global\setbox\ZoneBBOX=\box\VOIDBOX
 \def\Zone{\InZoneB}
 \loop
  \ifMakePageInComplete
 \FindNext
 \ifnum\StackPointer=\m@ne
  \NextItem=\m@ne
  \MoreFiguresfalse
  \MoreTablesfalse
 \fi
 \ifnum\NextItem=\Figure
   \FindItem{\Figure}{\NextFigure}
   \ifnum\StackPointer=\m@ne \MoreFiguresfalse
   \else
    \GetItemSPAN{\StackPointer}
    \ifnum\ItemSPAN=\Double\relax
     \MoreFiguresfalse\fi
   \fi
   \ifMoreFigures\Print{}\FigureItems\fi
 \fi
 \ifnum\NextItem=\Table
   \FindItem{\Table}{\NextTable}
   \ifnum\StackPointer=\m@ne \MoreTablesfalse
   \else
    \GetItemSPAN{\StackPointer}
    \ifnum\ItemSPAN=\Double\relax
     \MoreTablesfalse\fi
   \fi
   \ifMoreTables\Print{}\TableItems\fi
 \fi
   \MakePageInCompletefalse % assume page is complete
   \ifMoreFigures\MakePageInCompletetrue\fi
   \ifMoreTables\MakePageInCompletetrue\fi
 \repeat
 \ifZoneAFullPage
  \global\TextSize=\z@
  \global\ZoneBSize=\z@
  \global\vsize=\z@\relax
  \global\topskip=\z@\relax
  \vbox to \z@{\vss}
  \eject
 \else
 \global\vsize=\ZoneBSize
 \global\hsize=\ColumnWidth
 \ifdim\TextSize<23pt
 \Warn{}
 \Warn{* Making column fall short: TextSize=\the\TextSize *}
 \vskip-\lastskip\eject\fi
\fi
}

\def\FigureItems{% Stack pointer points to next figure
 \Print{Considering...}
 \ShowItem{\StackPointer}
 \GetItemBOX{\StackPointer} % auto calculates ItemSIZE
 \GetItemSPAN{\StackPointer}
  \CheckFitInZone % check to see if item fits
  \ifnum\ItemFits=\Yes
   \ifnum\ItemSPAN=\Single
     \ChangeStatus{\StackPointer}{\InZoneB} % flag to be output
     \global\FigInZoneBtrue
     \ifFirstSingleItem
      \hbox{}\vskip-\BodgeHeight
     \global\advance\ItemSIZE by \TextLeading
     \fi
     \unvbox\ItemBOX\ItemSep
     \global\FirstSingleItemfalse
     \global\advance\TextSize by -\ItemSIZE% allocate space
     \global\advance\TextSize by -\TextLeading
   \else
    \ifFirstZoneA
     \global\advance\ItemSIZE by \TextLeading
     \global\FirstZoneAfalse\fi
    \global\advance\TextSize by -\ItemSIZE
    \global\advance\TextSize by -\TextLeading
    \global\advance\ZoneBSize by -\ItemSIZE
    \global\advance\ZoneBSize by -\TextLeading
    \ifFigInZoneB\relax
     \else
     \ifdim\TextSize<3\TextLeading
     \global\ZoneAFullPagetrue
     \fi
    \fi
    \ChangeStatus{\StackPointer}{\Zone}
    \ifnum\Zone=\InZoneC \global\FigInZoneCtrue\fi
  \fi
   \Print{TextSize=\the\TextSize}
   \Print{ZoneBSize=\the\ZoneBSize}
  \global\advance\NextFigure \@ne
   \Print{This figure has been placed.}
  \else
   \Print{No space available for this figure...holding over.}
   \Print{}
   \global\MoreFiguresfalse
  \fi
}

\def\TableItems{% Stack pointer points to next table
 \Print{Considering...}
 \ShowItem{\StackPointer}
 \GetItemBOX{\StackPointer} % auto calculates ItemSIZE
 \GetItemSPAN{\StackPointer}
  \CheckFitInZone % check to see of item fits in Zone
  \ifnum\ItemFits=\Yes
   \ifnum\ItemSPAN=\Single
    \ChangeStatus{\StackPointer}{\InZoneB}
     \global\TabInZoneBtrue
     \ifFirstSingleItem
      \hbox{}\vskip-\BodgeHeight
     \global\advance\ItemSIZE by \TextLeading
     \fi
     \unvbox\ItemBOX\ItemSep
     \global\FirstSingleItemfalse
     \global\advance\TextSize by -\ItemSIZE
     \global\advance\TextSize by -\TextLeading
   \else
    \ifFirstZoneA
    \global\advance\ItemSIZE by \TextLeading
    \global\FirstZoneAfalse\fi
    \global\advance\TextSize by -\ItemSIZE
    \global\advance\TextSize by -\TextLeading
    \global\advance\ZoneBSize by -\ItemSIZE
    \global\advance\ZoneBSize by -\TextLeading
    \ifFigInZoneB\relax
     \else
     \ifdim\TextSize<3\TextLeading
     \global\ZoneAFullPagetrue
     \fi
    \fi
    \ChangeStatus{\StackPointer}{\Zone}
    \ifnum\Zone=\InZoneC \global\TabInZoneCtrue\fi
   \fi
%   \Print{TextSize=\the\TextSize}
%   \Print{ZoneBSize=\the\ZoneBSize}
  \global\advance\NextTable \@ne
   \Print{This table has been placed.}
  \else
  \Print{No space available for this table...holding over.}
   \Print{}
   \global\MoreTablesfalse
  \fi
}

% These macros check to see if an item of ItemSIZE will
% fit in a particular zone. If it will, then ItemFits
% will be set true else false.

\def\CheckFitInZone{%
{\advance\TextSize by -\ItemSIZE
 \advance\TextSize by -\TextLeading
 \ifFirstSingleItem
  \advance\TextSize by \TextLeading
 \fi
 \ifnum\Zone=\InZoneA\relax
  \else \advance\TextSize by -\ZoneBAdjust
 \fi
 \ifdim\TextSize<3\TextLeading \global\ItemFits=\No
 \else \global\ItemFits=\Yes\fi}
}

\def\BeginOpening{%
  % start 9pt a.s.a.p. so that \New.. commands get a chance to init.
  \ninepoint
  \thispagestyle{titlepage}%
  \global\setbox\ItemBOX=\vbox\bgroup%
    \hsize=\PageWidth%
    \hrule height \z@
    \ifsinglecol\vskip 6pt\fi % Bodge, to get same pos. as two-column!
}

\let\begintopmatter=\BeginOpening  %  alias for \BeginOpening

\def\EndOpening{%
  \One%  1 line fixed space below opening
  \egroup
  \ifsinglecol
    \box\ItemBOX%
    \vskip\TextLeading plus 2\TextLeading% var. space: min 1, max 3 lines
    \@noafterindent
  \else
    \ItemNUMBER=\z@%
    \ItemTYPE=\Figure
    \ItemSPAN=\Double
    \ItemSTATUS=\InStack
    \JoinStack
  \fi
}

% Figures

\newif\if@here  \@herefalse

\def\no@float{\global\@heretrue}
\let\nofloat=\relax % only enabled for one column

\def\beginfigure{%
  \@ifstar{\global\@dfloattrue \@bfigure}{\global\@dfloatfalse \@bfigure}%
}

\def\@bfigure#1{%
  \par
  \if@dfloat
    \ItemSPAN=\Double
    \TEMPDIMEN=\PageWidth
  \else
    \ItemSPAN=\Single
    \TEMPDIMEN=\ColumnWidth
  \fi
  \ifsinglecol
    \TEMPDIMEN=\PageWidth
  \else
    \ItemSTATUS=\InStack
    \ItemNUMBER=#1%
    \ItemTYPE=\Figure
  \fi
  \bgroup
    \hsize=\TEMPDIMEN
    \global\setbox\ItemBOX=\vbox\bgroup
      \eightpoint\nostb@ls{10pt}%
      \let\caption=\fig@caption
      \ifsinglecol \let\nofloat=\no@float\fi
}

\def\fig@caption#1{%
  \vskip 5.5pt plus 6pt%
  \bgroup % grouping and size change needed for plate pages
    \eightpoint\nostb@ls{10pt}%
    \setbox\TEMPBOX=\hbox{#1}%
    \ifdim\wd\TEMPBOX>\TEMPDIMEN
      \noindent \unhbox\TEMPBOX\par
    \else
      \hbox to \hsize{\hfil\unhbox\TEMPBOX\hfil}%
    \fi
  \egroup
}

\def\endfigure{%
  \par\egroup % end \vbox
  \egroup
  \ifsinglecol
    \if@here \midinsert\global\@herefalse\else \topinsert\fi
      \unvbox\ItemBOX
    \endinsert
  \else
    \JoinStack
    \Print{Processing source for figure \the\ItemNUMBER}%
  \fi
}

% Tables

\newbox\tab@cap@box
\def\tab@caption#1{\global\setbox\tab@cap@box=\hbox{#1\par}}

\newtoks\tab@txt@toks
\long\def\tab@txt#1{\global\tab@txt@toks={#1}\global\table@txttrue}

\newif\iftable@txt  \table@txtfalse
\newif\if@dfloat    \@dfloatfalse

\def\begintable{%
  \@ifstar{\global\@dfloattrue \@btable}{\global\@dfloatfalse \@btable}%
}

\def\@btable#1{%
  \par
  \if@dfloat
    \ItemSPAN=\Double
    \TEMPDIMEN=\PageWidth
  \else
    \ItemSPAN=\Single
    \TEMPDIMEN=\ColumnWidth
  \fi
  \ifsinglecol
    \TEMPDIMEN=\PageWidth
  \else
    \ItemSTATUS=\InStack
    \ItemNUMBER=#1%
    \ItemTYPE=\Table
  \fi
  \bgroup
    \eightpoint\nostb@ls{10pt}%
    \global\setbox\ItemBOX=\vbox\bgroup
      \let\caption=\tab@caption
      \let\tabletext=\tab@txt
      \ifsinglecol \let\nofloat=\no@float\fi
}

\def\endtable{%
  \par\egroup % end \vbox
  \egroup
  \setbox\TEMPBOX=\hbox to \TEMPDIMEN{%
    \eightpoint\nostb@ls{10pt}%
    \hss
    \vbox{%
      \hsize=\wd\ItemBOX
      \ifvoid\tab@cap@box
      \else
        \noindent\unhbox\tab@cap@box
        \vskip 5.5pt plus 6pt%
      \fi
      \box\ItemBOX
      \iftable@txt
        \vskip 10pt%
        \noindent\the\tab@txt@toks
        \global\table@txtfalse
      \fi
    }%
    \hss
  }%
  \ifsinglecol
    \if@here \midinsert\global\@herefalse\else \topinsert\fi
      \box\TEMPBOX
    \endinsert
  \else
    \global\setbox\ItemBOX=\box\TEMPBOX
    \JoinStack
    \Print{Processing source for table \the\ItemNUMBER}%
  \fi
}

\def\UnloadZoneA{%
\FirstZoneAtrue
 \Iteration=\z@
  \loop
   \ifnum\Iteration<\LengthOfStack
    \GetItemSTATUS{\Iteration}
    \ifnum\ItemSTATUS=\InZoneA
     \GetItemBOX{\Iteration}
     \ifFirstZoneA \vbox to \BodgeHeight{\vfil}%
     \FirstZoneAfalse\fi
     \unvbox\ItemBOX\ItemSep
     \LeaveStack{\Iteration}
     \else
     \advance\Iteration \@ne
   \fi
 \repeat
}

\def\UnloadZoneC{%
\Iteration=\z@
  \loop
   \ifnum\Iteration<\LengthOfStack
    \GetItemSTATUS{\Iteration}
    \ifnum\ItemSTATUS=\InZoneC
     \GetItemBOX{\Iteration}
     \ItemSep\unvbox\ItemBOX
     \LeaveStack{\Iteration}
     \else
     \advance\Iteration \@ne
   \fi
 \repeat
}

%--------------------------------------------------------%
%                         DIAGNOSTICS                    %
%--------------------------------------------------------%

\def\ShowItem#1{% Show details of on item entry in stack
  {\GetItemAll{#1}
  \Print{\the#1:
  {TYPE=\ifnum\ItemTYPE=\Figure Figure\else Table\fi}
  {NUMBER=\the\ItemNUMBER}
  {SPAN=\ifnum\ItemSPAN=\Single Single\else Double\fi}
  {SIZE=\the\ItemSIZE}}}
}

\def\ShowStack{% 
 \Print{}
 \Print{LengthOfStack = \the\LengthOfStack}
 \ifnum\LengthOfStack=\z@ \Print{Stack is empty}\fi
 \Iteration=\z@
 \loop
 \ifnum\Iteration<\LengthOfStack
  \ShowItem{\Iteration}
  \advance\Iteration \@ne
 \repeat
}

\def\B#1#2{%
\hbox{\vrule\kern-0.4pt\vbox to #2{%
\hrule width #1\vfill\hrule}\kern-0.4pt\vrule}
}

%-------------------------------------------------------%
%             SINGLE COLUMN OUTPUT ROUTINE              %
%-------------------------------------------------------%

\newif\ifsinglecol   \singlecolfalse

\def\onecolumn{%
  \global\output={\singlecoloutput}%
  \global\hsize=\PageWidth
  \global\vsize=\PageHeight
  \global\ColumnWidth=\hsize
  \global\TextLeading=12pt
  \global\Leading=12
  \global\singlecoltrue
  \global\let\onecolumn=\relax%         stop them using \onecolumn again
  \global\let\footnote=\sing@footnote%  enable footnotes
  \global\let\vfootnote=\sing@vfootnote
  \ninepoint % reset \baselineskip after leading change
  \message{(Single column)}%
}

\def\singlecoloutput{%
  \shipout\vbox{\PageHead\vbox to \PageHeight{\pagebody\vss}\PageFoot}%
  \advancepageno
  \ifplate@page
    \shipout\vbox{%
      \sp@pagetrue
      \def\sp@type{plate}%
      \global\plate@pagefalse
      \PageHead\vbox to \PageHeight{\unvbox\plt@box\vfil}\PageFoot%
    }%
    \message{[plate]}%
    \advancepageno
  \fi
  \ifnum\outputpenalty>-\@MM \else\dosupereject\fi%
}

\def\ItemSep{\vskip\ItemSepamount\relax}

\def\ItemSepbreak{\par\ifdim\lastskip<\ItemSepamount
  \removelastskip\penalty-200\ItemSep\fi%
}

% Modify plain's \endinsert so that the mn's spacing is used

\let\@@endinsert=\endinsert % save plain's original \endinsert

\def\endinsert{\egroup % finish the \vbox
  \if@mid \dimen@\ht\z@ \advance\dimen@\dp\z@ \advance\dimen@12\p@
    \advance\dimen@\pagetotal \advance\dimen@-\pageshrink
    \ifdim\dimen@>\pagegoal\@midfalse\p@gefalse\fi\fi
  \if@mid \ItemSep\box\z@\ItemSepbreak
  \else\insert\topins{\penalty100 % floating insertion
    \splittopskip\z@skip
    \splitmaxdepth\maxdimen \floatingpenalty\z@
    \ifp@ge \dimen@\dp\z@
    \vbox to\vsize{\unvbox\z@\kern-\dimen@}% depth is zero
    \else \box\z@\nobreak\ItemSep\fi}\fi\endgroup%
}

% Footnotes (only enabled in single column)

\def\gobbleone#1{}
\def\gobbletwo#1#2{}
\let\footnote=\gobbletwo % Gobble footnote's unless enabled by \onecolumn
\let\vfootnote=\gobbleone

\def\sing@footnote#1{\let\@sf\empty % parameter #2 (the text) is read later
  \ifhmode\edef\@sf{\spacefactor\the\spacefactor}\/\fi
  \hbox{$^{\hbox{\eightpoint #1}}$}\@sf\sing@vfootnote{#1}%
}

\def\sing@vfootnote#1{\insert\footins\bgroup\eightpoint\b@ls{9pt}%
  \interlinepenalty\interfootnotelinepenalty
  \splittopskip\ht\strutbox % top baseline for broken footnotes
  \splitmaxdepth\dp\strutbox \floatingpenalty\@MM
  \leftskip\z@skip \rightskip\z@skip \spaceskip\z@skip \xspaceskip\z@skip
  \noindent $^{\scriptstyle\hbox{#1}}$\hskip 4pt%
    \footstrut\futurelet\next\fo@t%
}

% Kill footnote rule
\def\footnoterule{\kern-3\p@ \hrule height \z@ \kern 3\p@}

\skip\footins=19.5pt plus 12pt minus 1pt
\count\footins=1000
\dimen\footins=\maxdimen

% for footnotes in double column: use \note{$\star$}{footnote}
\def\note#1#2{%
  \let\@sf=\empty \ifhmode\edef\@sf{\spacefactor\the\spacefactor}\/\fi
  #1\insert\footins\bgroup
    \eightpoint\b@ls{10pt}\rm
    \interlinepenalty\interfootnotelinepenalty
%    \splittopskip\ht\strutbox % top baseline for broken footnotes
    \splitmaxdepth\dp\strutbox \floatingpenalty\@MM
    \leftskip\z@skip \rightskip\z@skip \spaceskip\z@skip \xspaceskip\z@skip
    \noindent\footstrut #1$\,$#2\strut\par
  \egroup
  \@sf\relax}

% Landscape

\def\landscape{%
  \global\TEMPDIMEN=\PageWidth
  \global\PageWidth=\PageHeight
  \global\PageHeight=\TEMPDIMEN
  \global\let\landscape=\relax%         stop them using \landscape again.
  \onecolumn
  \message{(landscape)}%
  \raggedbottom
}

%-------------------------------------------------------%
%               TWO COLUMN OUTPUT ROUTINE               %
%-------------------------------------------------------%

% Very slight redefinition of the \output routine of mn.tex, to allow footnotes.
\output{%
  \ifLeftCOL
    \global\setbox\LeftBOX=\vbox to \ZoneBSize{\box255\unvbox\ZoneBBOX
      \ifvoid\footins\else
        \vskip\skip\footins\unvbox\footins\fi
    }%
    \global\LeftCOLfalse
    \MakeRightCol
  \else
    \setbox\RightBOX=\vbox to \ZoneBSize{\box255\unvbox\ZoneBBOX
      \ifvoid\footins\else
        \vskip\skip\footins\unvbox\footins\fi
    }%
    \setbox\MidBOX=\hbox{\box\LeftBOX\hskip\ColumnGap\box\RightBOX}%
    \setbox\PageBOX=\vbox to \PageHeight{%
      \UnloadZoneA\box\MidBOX\UnloadZoneC}%
    \shipout\vbox{\PageHead\vbox to \PageHeight{\box\PageBOX\vss}\PageFoot}%
    \advancepageno
    \ifplate@page
      \shipout\vbox{%
        \sp@pagetrue
        \def\sp@type{plate}%
        \global\plate@pagefalse
        \PageHead\vbox to \PageHeight{\unvbox\plt@box\vfil}\PageFoot%
      }%
      \message{[plate]}%
      \advancepageno
    \fi
    \global\LeftCOLtrue
    \CleanStack
    \MakePage
  \fi
}

% Startup message

\Warn{\start@mess}

\newif\ifCUPmtplainloaded % for use in documents
\ifprod@font
  \global\CUPmtplainloadedtrue
\fi

\def\mnmacrosloaded{} % so articles can see if a format file has been used.

\catcode `\@=12 % @ signs are non-letters

% \dump

% end of mn.tex

\fi
%
%\input mn.tex
%
% If your system has the AMS fonts version 2.0 installed, MN.tex can be
% made to use them by uncommenting the line: %\AMStwofontstrue
%
% By doing this, you will be able to obtain upright Greek characters.
% e.g. \umu, \upi etc.  See the section on "Upright Greek characters" in
% this guide for further information.

\newif\ifAMStwofonts
%\AMStwofontstrue

\ifCUPmtplainloaded \else
  \NewTextAlphabet{textbfit} {cmbxti10} {}
  \NewTextAlphabet{textbfss} {cmssbx10} {}
  \NewMathAlphabet{mathbfit} {cmbxti10} {} % for math mode
  \NewMathAlphabet{mathbfss} {cmssbx10} {} %  "   "    "
  \ifAMStwofonts
    \NewSymbolFont{upmath} {eurm10}
    \NewSymbolFont{AMSa} {msam10}
    \NewMathSymbol{\upi}     {0}{upmath}{19}
    \NewMathSymbol{\umu}     {0}{upmath}{16}
    \NewMathSymbol{\upartial}{0}{upmath}{40}
    \NewMathSymbol{\leqslant}{3}{AMSa}{36}
    \NewMathSymbol{\geqslant}{3}{AMSa}{3E}

    \let\leq=\leqslant \let\le=\leqslant
    \let\geq=\geqslant 
  \else
    \def\umu{\mu}
    \def\upi{\pi}
    \def\upartial{\partial}
  \fi
\fi

% Marginal adjustments using \pageoffset maybe required when printing
% proofs on a Laserprinter (this is usually not needed).
% Syntax: \pageoffset{ +/- hor. offset}{ +/- vert. offset}
% e.g.    \pageoffset{-3pc}{-4pc}

%\pageoffset{-2.5pc}{0pc}

\loadboldmathnames

%\Referee   %  uncomment this for referee mode (double spaced)

% \pagerange, \pubyear and \volume are defined at the Journals office and
% not by an author.

\onecolumn        % enable one column mode
% \letters          % for `letters' articles
%\pagerange{1--7}    % `letters' articles should use \pagerange{Ln--Ln}
%\pubyear{1989}
%\volume{226}
% \microfiche{}     % for articles with microfiche
% \authorcomment{}  % author comment for footline

% macros
%
% some definitions for the references
%
%
%
\def\PBvp #1 #2{ #1, #2}
\def\PBa #1:#2 #3 #4 {#1,#2, {A\&A} \PBvp #3 #4}
\def\PBapj #1:#2 #3 #4 {#1,#2, {ApJ} \PBvp #3 #4}
\def\PBasupl #1:#2 #3 #4 {#1,#2, {A\&AS} \PBvp #3 #4}
\def\PBapjsupl #1:#2 #3 #4 {#1,#2, {ApJS} \PBvp #3 #4}
\def\PBpasp #1:#2 #3 #4 {#1,#2, { PASP} \PBvp #3 #4}
\def\PBpaspc #1:#2 #3 #4 {#1,#2, { PASPC } \PBvp #3 #4}
\def\PBmn #1:#2 #3 #4 {#1,#2, {MNRAS} \PBvp #3 #4}
\def\PBmsait #1:#2 #3 #4 {#1,#2, {Mem. S.A.It.} \PBvp #3 #4}
\def\PBnat #1:#2 #3 #4 {#1,#2, {Nat} \PBvp #3 #4}
\def\PBaj #1:#2 #3 #4 {#1,#2, {AJ} \PBvp #3 #4}
\def\PBjaa #1:#2 #3 #4 {#1,#2, {JA\& A} \PBvp #3 #4}
\def\PBaspsc #1:#2 #3 #4 {#1,#2, {Ap\&SS} \PBvp #3 #4}
\def\PBanrev #1:#2 #3 #4 {#1,#2, {ARA\&A} \PBvp #3 #4}
\def\PBrevmex #1:#2 #3 #4 {#1,#2, {Rev. Mex. de Astron. y Astrof.} \PBvp #3 #4}
\def\PBscie #1:#2 #3 #4 {#1,#2, {Sci} \PBvp #3 #4}
\def\PBesomsg #1:#2 #3 #4 {#1,#2, {The Messenger} \PBvp #3 #4}
\def\PBrmp #1:#2 #3 #4 {#1,#2, {Rev. Mod. Phys.} \PBvp #3 #4}
\def\PBans #1:#2 #3 #4 {#1,#2, {Ann. Rev. of Nucl. Sci.} \PBvp #3 #4}
\def\PBphrev #1:#2 #3 #4 {#1,#2, {Phys. Rev.} \PBvp #3 #4}
\def\PBphreva #1:#2 #3 #4 {#1,#2, {Phys. Rev. A} \PBvp #3 #4}
\def\PBphs #1:#2 #3 #4 {#1,#2, {Physica Scripta} \PBvp #3 #4}
\def\PBjqsrt #1:#2 #3 #4 {#1,#2, {J. Quant. Spectrosc. Radiat.
       Transfer} \PBvp #3 #4}
\def\PBcjp #1:#2 #3 #4 {#1,#2, {Can. J. Phys. } \PBvp #3 #4}
\def\PBjphb #1:#2 #3 #4 {#1,#2, {J. Phys. B} \PBvp #3 #4}
\def\PBapop #1:#2 #3 #4 {#1,#2, {Appl. Opt.} \PBvp #3 #4}
\def\PBgca #1:#2 #3 #4 {#1,#2, {Geochim. Cosmochim. Acta}\PBvp #3 #4}

\def\PBs{\phantom{n}}
%
% some shorthand
%
\def\PBkms{$\rm km s^{-1}$}

\def\PBt{${T}_{\rm eff}~$}
\def\PBg{$\rm \log g$}
\def\PBal{{ et al.~}}
\def\PBew {equivalent width}

\def\PBfeh{[Fe/H]~}

\def\PBcogs{curves of growth}
%
% letters with accents
%
\chardef\PBii="10
\def\PBi{\'\PBii}
%
%
%
% table  numbering
%
\newcount\PBtn
\def\PBcleartn{\global\PBtn=0}
\def\PBtbl #1{\global\advance\PBtn by 1
\begintable{\the\PBtn}
\caption{{\bf Table \the\PBtn .} #1}
}
\def\PBtbltwo #1{\global\advance\PBtn by 1
\begintable*{\the\PBtn}
\caption{{\bf Table \the\PBtn .} #1}
}

\begintopmatter  %  start the two spanning material

\title{The primordial lithium abundance}
\author{
 P. Bonifacio and P. Molaro}
\affiliation{Osservatorio Astronomico di Trieste, 
Via G.B. Tiepolo 11 34131, Trieste -- Italy}

\shortauthor{P. Bonifacio and P. Molaro }
\shorttitle{The primordial lithium abundance}

% \acceptedline is to be defined at the Journals office and not
% by an author.

%\acceptedline{ }
\PBcleartn

\abstract {Lithium abundances in a
selected sample of halo stars have been
revised
by using the  new accurate IRFM effective temperatures
by Alonso, Arribas \& Mart\PBi nez-Roger  (1996a). From 41
plateau stars (\PBt $>$ 5700 and [Fe/H] $\le$ -1.5) we found no
evidence for intrinsic dispersion,
a tiny trend with T$_{eff}$ and no  trend with [Fe/H].
The  trend with the \PBt is fully consistent with the standard
Li isochrones of Deliyannis, Demarque \& Kawaler
(1990) implying a primordial
value for Li of
A(Li)$=2.238 \pm 0.012_{1 \sigma}  \pm 0.05_{sys}$ .
The present results argue against any kind of depletion
predicted by
diffusion, rotational mixing or stellar winds. Therefore the Li observed
in Pop II
stars provides a direct and reliable estimate of the baryonic density
that can  rival other baryonic indicators such as the deuterium in high
redshift systems. The present upwards revision of primordial Li
in the framework of SBBN gives at $1\sigma$
two  solutions for the baryonic density:
$ \Omega_{B}h^2  = 0.0062^{+0.0018}_{-0.0011}$
or  $\Omega_{B}h^2  = 0.0146^{+0.0029}_{-0.0033} $.
}

\keywords {Stars: abundances -- Stars: Population II --
Stars: fundamental parameters -- Galaxy: halo -- Cosmology: observations}

\maketitle  %  finish the two spanning material

\section{Introduction}

In a seminal paper Spite \& Spite (1982) found that the
warm Pop II dwarfs (\PBt $>$ 5700 K)
share a unique Li abundance, at variance
with their Pop I counterparts, where a large spread
in Li abundances is observed.
Since no sign  of stellar depletion
was evident on the first dozen stars the Spite plateau
was readily interpreted as a signature of pristine Li
abundance directly related to primordial
nucleosynthesis.
The original finding has
been later confirmed by a number of investigations
(Spite, Maillard \& Spite  1984,
Spite \& Spite 1986,
Hobbs \& Duncan 1987,
Rebolo, Molaro \& Beckmann 1988).
The  tight connection between Pop II Li abundance and
BBN is of far reaching cosmological
importance since in the standard BBN
the primordial Li abundance is a function of $\eta\equiv n_b/n_\gamma$, the
baryon to photon ratio, which is the only free parameter
now left in the standard model.
The information provided by Li is especially interesting 
in view of  the recently conflicting  observations of
 deuterium
in high redshift absorption systems (Songaila \PBal 1994,
Carswell \PBal
1996, Wampler \PBal 1996, Rugers \& Hogan 1996,
Tytler, Fan \& Burles 1996, Burles \& Tytler 1996) .
\par
In the recent years the existence of a true plateau in the Li
abundances of the warm halo dwarfs has  been debated.
Deliyannis, Pinsonneault \& Duncan (1993)  claimed the existence
of a rather small dispersion of 10\%, at 1 $\sigma$ level, in Li abundances
on the Spite plateau.
Thorburn (1994) found an even larger
dispersion of Li abundance
on the plateau and  found  a trend
both with \PBt ~ and [Fe/H].
Norris, Ryan \&
Stringfellow (1994) and
Ryan \PBal (1996) also support the existence of  a trend
of Li abundance with both \PBt and \PBfeh .
The existence of dispersion and  trends on the plateau could argue in favour of
depletion and/or Galactic enrichment in halo dwarfs. In particular,
stellar  models have been proposed
where Li is considerably depleted by rotational mixing, diffusion
or stellar winds (Vauclair \& Charbonnel 1995).
If this is the case then the abundance of Li in warm Pop II
dwarfs does not reflect the primordial Li abundance.

On the other hand the existence of the Spite
plateau has been reasserted by
Molaro, Primas \& Bonifacio
(1995a) from the analysis   of a subsample of stars
with  uniform \PBt  derived from
Balmer line profiles by Fuhrmann, Axer \& Gehren  (1994).
Their result
was that there is no dispersion on the plateau above what is
expected by observational errors and that there is no slope
in either \PBt or \PBfeh.
Spite \PBal (1996) tackled the problem in a similar fashion
using three samples of stars
for which  \PBt was determined from either excitation
equilibria, Balmer line profiles, or $(b-y)_0$, and
concluded  that there is no evidence of dispersion or trends.
So it appears
that the existence of the Spite plateau
is real when small samples of stars
with homogeneous \PBt are examined, but it is blurred when
the whole set of stars is considered.

In this paper we use a new \PBt
scale based on the semi-direct
Infrared Flux Method (IRFM, Blackwell \PBal 1990)
applied to a large sample of stars by Alonso
\PBal (1996a), to investigate
the distribution of
Li abundance with \PBfeh and \PBt
for a statistically significant fraction of the stars with Li measured.
Preliminary results
have been already presented by Bonifacio \& Molaro (1996).

\section{A new sample for L\lowercase{i} }

Li is present mainly in the form of the
singly ionized  ion in the atmospheres
of cool dwarfs and
a large ionization correction is required  when deriving the Li
abundance from the Li I 6707 \AA ~
resonance doublet. This correction is strongly dependent on the adopted
\PBt  making  the determination of \PBt
crucial for the discussion of possible trends and/or
dispersion on the Spite plateau.
Other stellar parameters such as
surface gravity,
metallicity or microturbulence are much less
important for the Li abundance.
In this paper we consider a new sample of stars with Li observations
available and with accurate and internally coherent $T_{\rm eff}$'s.

Effective temperatures are  usually determined either from the
continuum spectrum (colours, flux distribution) or
from the line spectrum (Balmer line profiles, excitation
equilibria).  These methods are calibrated on the few stars for
which the direct measure of \PBt is obtained  from the bolometric
fluxes and
angular diameter measurements.
The Sun is the only star cooler than F5V for which the angular diameter
is known determining an intrinsic uncertainty in the
scale of effective temperatures in the low main sequence. In particular,
for Pop II stars there are no stars with
measured
angular diameter which can be used as primary calibrators;
therefore,
one is forced to rely on theoretical models to derive the metallicity
dependence of his favourite \PBt indicator.

 In the impossibility of obtaining a direct measure of \PBt,
the semi--direct Infrared
Flux Method (IRFM,
Blackwell \PBal 1990)
is the best  alternative to derive the
temperature of F and G stars.
The IRFM  relies only weakly on theoretical models
and its main uncertainty
is related to the absolute flux calibrations of the infrared magnitudes.
It has been extensively applied to different samples of Pop I stars
(Saxner \& Hammarback 1985, Blackwell \PBal 1990).
  The works of Magain (1987)
and  Arribas \& Mart\PBi nez-Roger (1989)  applied
the IRFM to  limited samples
of metal poor stars.
 Alonso \PBal (1996a) have recently determined IRFM temperatures
for an enlarged
sample of 475 dwarfs
of spectral types from F5 to K0, with a wide range of metallicities.
The effective temperatures are reddening corrected.
Alonso \PBal (1996a) derived three effective temperatures for the
J,H and K fluxes showing a good internal consistency.
They provided a final \PBt as the mean of the three,  weighted with
the inverse of their errors.
The zero point of the IRFM \PBt by Alonso \PBal (1996a)
differ by    112, 0 and -56 K  from the zero point of
the IRFM \PBt derived
by  Magain (1987), Saxner \& Hammarback (1985) and Bell \&
Gustafsson (1989) respectively,
which may be explained by several improvements in the Alonso \PBal (1996a)
analysis.
The  mean accuracy of the \PBt estimated by Alonso \PBal (1996a)
is about 1.5 \%,
which includes both systematic and random errors.
\par

Out of  the sample of Alonso \PBal (1996a),  64 stars have
been observed for Li,
and they  form  our
sample  for the study of the Spite plateau.  The star names together with
the stellar parameters and Li equivalent widths are reported in Table 1.
The Li EWs for the 64 stars have been taken from the literature.
For multiple
measurements  we adopted the weighted average.
The errors in the EW have been taken from the original papers,
when available, or from Ryan \PBal (1996), which estimated the errors
 according to the Cayrel (1988)
prescriptions.
Following Ryan \PBal (1996) we kept a 1 m\AA ~ as the highest precision
claimed in the observations, with the only exception of the
few stars observed
by Deliyannis, Boesgaard \& King  (1995)
with the Keck telescope, at very high S/N.
Multiple observations are in general in excellent agreement and
an accurate statistical analysis of the errors for multiple observations
of the same stars has been performed by Ryan \PBal (1996) confirming that
in the  majority of  cases the scatter between
multiple observations is consistent with the stated errors.
The measurement of EWs may be occasionally affected by
non gaussian noise, such as that arising from
cosmic rays hits or scattered light in the spectrograph.
This event is rare but has to be considered in case of claims
of intrinsic dispersion of
isolated stars for which only single observations are available.

\par

The frequency of binarity among halo dwarfs is not well established
but it may be between  20-30\% up to  50\% as it is for Pop I
stars. An investigation on the impact of binarity was done
by Molaro (1991) showing that known binaries were not significantly different
from other stars. However,
the influence of  binaries may be important in assessing the
presence of a {\it small} dispersion.
In Molaro \PBal (1995a) the A(Li) in
three stars was consistent with the  mean value only considering the full
error. One of these, HD 116064, has been later found to be a spectroscopic
binary by Spite \PBal (1996) and the  compensation for the veiling, which
is of  20 \% in the EW, move the A(Li) closer to the mean value.
This
is a clear example of how binarity may affect the dispersion analysis
on the plateau. Another  example is G020-008
found to be double by Fouts (1987),
which, in the small
sample of stars
considered by Spite \PBal (1996), contributes to increase the dispersion.

Several binaries are  present in our sample.
HD 219617, BD+20 3603, HD 3567, HD 16031, BD+01 2341p,
HD 84937, HD 132475, G206-034, HD 188510,
BD+38 4955
have been identified  as binaries or suspected radial velocity variables
by Carney (1983).
  By means
of spectroscopic analysis over a four-year
temporal baseline,  Stryker \PBal (1985) classified as
certain binaries:
HD  84937, HD 94028, HD 201891
BD+26 2606, G090-025, HD 188510; as  possible binaries: BD+28 2137,
BD+34 2476, BD+13 3683, BD+ 26 3578;  near significance criterion:
BD+01 2341p, BD+29 2091, BD+38 4955, BD+21 607, HD 108177, G090-003.
 Lu \PBal (1987) by means of the speckle technique identified: BD+17 4708.
 G020-008, G020-024, HD 16031 have been found
to be binaries by Fouts (1987).
Thus in total
there are 25 known or suspected binaries
out of 64 stars, which gives a binarity fraction of at
least  39\% in our sample.

\section{Model atmospheres and L\lowercase{i} abundances}

The different
atmospheric models  adopted
by different authors to derive
lithium abundances are known to be generally
consistent with one another, but  in some cases
they may be   responsible for systematic
differences in lithium abundances.
As shown by Molaro \PBal (1995a) and
Molaro, Bonifacio \& Primas (1995b), the Kurucz (1993) models
which include overshooting are hotter than the Bell \& Gustafsson
models in the Li doublet forming region. As a consequence,
the Thorburn (1994) Li abundances, derived  by using
Kurucz  models with overshooting, are systematically
$\approx $ 0.1 dex
higher than those
of other researchers. On the other hand the Kurucz (1993) models
with the overshooting  switched off are
in good agreement  with the others.
Castelli, Gratton \& Kurucz  (1996) have shown that for
all F and G stars, except the Sun,
the Kurucz models computed without overshooting
are more consistent with the observations than models with
overshooting.
Consistently with these findings,
in this investigation  we use model atmospheres
computed with the ATLAS9 code (Kurucz 1993)
with the overshooting option switched off.

% the two plots should be placed side by side
%
\beginfigure*{1}
\vskip 7cm
\caption{{\bf Figure 1.}
a) $c1,(b-y)$ diagram  for our sample of Pop II stars;
b) $c0,(b-y)_0$ diagram.
 The solid line represents the locus of points
with log g = 3.5 for [Fe/H]=-1.5. The dashed line
is the same but for [Fe/H]=-3.0. The region above these lines
is populated by subgiants and giants.
 The crosses are the stars with [Fe/H] $> -1.5$
 while the $\times$ symbols are stars with [Fe/H] $\le -1.5$.
 }
 \endfigure
\par
For the opacities we used the ODFs with enhanced $\alpha$ elements,
which provide a more
realistic chemical composition for Pop II stars
than the solar scaled ODFs.
The microturbulent velocity was assumed to be 1 \PBkms. However,
the precise value of the microturbulent velocity is relatively unimportant
for the Li abundance in halo stars, since
0.5 \PBkms change in the microturbulent velocity induce a change of
0.005 dex in Li.

We  checked  our models with those used in Ryan \PBal (1996)  by comparing
our  curve of growth for a model with \PBt = 6500 , \PBg = 4.0,
\PBfeh -2.0 and microturbulence of 1 \PBkms  with
the corresponding curve of growth, based on Bell-Gustafsson models,
published by Ryan \PBal (1996). The  difference in Li abundance
is zero for an \PBew ~ of 14 m\AA ~ (A(Li)$\approx 2.1$)
and 0.008 dex for an
\PBew ~  of 24 m\AA ~(A(Li)$\approx 2.4$), with our
models giving abundances higher than the Bell--Gustafsson ones.
\par
 Surface
 gravities have been
redetermined  for each star to identify
possible subgiants. As already mentioned,
 the Li doublet \PBew ~ depends
little on surface gravity with  a 0.02 dex
change in Li abundance for a 0.7 dex change in \PBg, but
the identification of subgiants in the sample is important because their
Li abundance is expected to be lower due to dilution (Deliyannis
\PBal ~ 1990).
As a gravity indicator we used the c0 index of Str\"omgren photometry,
which measures the Balmer discontinuity.
We derived surface gravities from c0 in the same way as described
in Molaro \PBal (1996).
For all our stars, except G201-005, we took the photometry from
the General Catalogue of Photometric Data
(Mermillod, Hauck \& Mermillod 1996),
this was dereddened using the Schuster \& Nissen (1989) intrinsic
colour calibration. For BD +71 31, for which there is no $\beta$ available,
we assumed a zero reddening.
 For some stars we could not find any solution;  this
happened if the observed
c0 value was smaller than any of the theoretical values
for the given temperature and metallicity. This was the case for
G246-038, HD25329, G090-025 and G190-015. For these  stars, as well as for
G201-005, we have taken the
surface gravity given by Alonso \PBal (1996a).
From our model grid, described in
Molaro \PBal (1996), we computed fluxes and both Johnson 
and Str\"omgren photometry.
In Fig. 1
 the $c1,(b-y)$ and $c0,(b-y)_0$ diagrams for our sample of Pop II
stars are shown. The solid line represents the locus of points
with log g = 3.5 for [Fe/H]=-1.5, from
our synthetic Str\"omgren photometry, and the  dashed line
is the same but for [Fe/H]=-3.0. The region above these lines
is populated by subgiants and giants.
In our sample of 41 plateau stars only 3 have a surface gravity less
than or equal to 3.6, namely G090-003, G126-062 and BD+71 31, with
\PBg ~ 3.55, 3.58 and 3.58, respectively.

\par
A model atmosphere was computed for each star  with the appropriate
gravity and metallicity.
We then computed a synthetic profile of the Li doublet with the SYNTHE
code (Kurucz 1993) and obtained the \PBew ~ of the doublet by
integrating this profile. The process was iterated, changing the Li
abundance, until the computed \PBew ~ matched the observed one.
This approach is different from that used by Thorburn (1994), Molaro
\PBal (1995a) and Spite \PBal (1996),
who treated the doublet
as a single line. The differences are very small, because for our stars
the stronger Li line is never saturated, however the explicit computation
of the doublet is physically more realistic.
\par
The present analysis is carried out strictly under the LTE
assumption.  NLTE effects have been  studied by
Carlsson \PBal (1994) and Pavlenko \& Magazz\`u (1996) and
were shown to be minimal for halo dwarfs.  The corrections
provided by Carlsson \PBal (1994) have  been applied to the
LTE abundances and considered in the discussion of the results.

\beginfigure{2}
\vskip 7cm
\caption{{\bf Figure 2.}
The A(Li)-- [Fe/H] diagram for our sample of stars.
 The circles are stars with \PBt $>$ 5700 K, while
 the squares are those with \PBt $\le$ 5700 K.
 Subgiants are shown with crossed symbols.
Three upper limits are shown as vertical downward arrows.
}
\endfigure

\par

The errors reported in  Table 1 are those following from the
error in EW and in \PBt. The two errors are summed 
under quadrature as it is generally done. 
The errors given do not consider a possible error originated by
uncertaintes in the gravity and
in the microturbulence
which are  of the order of 0.01 dex each.
These are also  independent sources of errors
and should be added to the above
quoted errors.
We found  that  ignoring their contribution to the total error
does not affect our results.

\par

\PBtbltwo{Li data}
\halign to \hsize{\tabskip=0ptplus12ptminus15pt
$\rm #$\hfill &
# \hfill&
# \PBs&
# &
\hfill # &
# &
# &
\hfill# &
# &
# &
# &
# &
# &
# &
# &
#\cr
\multispan{16}{\hrulefill}\cr
\hfill BD \hfill & \hfill HD \hfill &
\hfill G \# \hfill & \hfill \PBt \hfill & $\sigma T_{\rm eff}$ \hfill &
\hfill \PBg \hfill & \hfill \PBfeh \hfill & \hfill EW \hfill &
\hfill $\sigma\rm EW$ \hfill  & \hfill A(Li)\phantom{$_1$}
\hfill & \hfill A(Li)$_1$
\hfill & \hfill A(Li)$_2$ \hfill & \hfill A(Li)$_3$ \hfill &
\hfill $\sigma _T $\phantom{$_W$}
\hfill & $\sigma _{EW}$  \hfill &
\hfill $\sigma_{tot}^\ast$  \hfill \cr
\multispan{16}{\hrulefill}\cr
 BD -09~122^b   & HD 3567     & G270-023   &  5858 &   63 &   3.70 &  -1.34 &
   45.0 &   6.0 &   2.28 &   2.30 &   2.32 &   2.33 &   0.05 &   0.08 &
   0.10 \cr
 BD +71~31    &            &            &  6026 &   94 &   3.58 &  -1.91 &
   31.0 &   2.3 &   2.19 &   2.20 &   2.21 &   2.22 &   0.07 &   0.04 &
   0.08 \cr
 BD +29~366   &            & G074-005   &  5685 &   85 &   4.48 &  -1.05 &
   14.0 &   2.9 &   1.55 &   1.57 &   1.64 &   1.66 &   0.07 &   0.11 &
   0.13 \cr
 BD -13~482^b
 & HD 16031    &            &  6114 &   72 &   4.00 &  -2.09 &
   28.0 &   2.2 &   2.20 &   2.21 &   2.21 &   2.23 &   0.05 &   0.04 &
   0.06 \cr
 BD +09~352   &            & G076-021   &  5931 &   85 &   4.03 &  -2.15 &
   34.0 &   1.5 &   2.16 &   2.18 &   2.21 &   2.22 &   0.06 &   0.03 &
   0.07 \cr
             &            & G004-037   &  6337 &   92 &   4.13 &  -3.31 &
   20.0 &   2.2 &   2.15 &   2.16 &   2.15 &   2.16 &   0.06 &   0.05 &
   0.08 \cr
 BD +25~495   & HD 19445    & G037-026   &  6050 &  109 &   4.45 &  -2.15 &
   34.0 &   0.5 &   2.26 &   2.27 &   2.28 &   2.30 &   0.08 &   0.01 &
   0.08 \cr
 BD +66~268   &            & G246-038   &  5279 &   81 &   4.50 &  -2.94 &
   10.0 &   7.8 &   1.09 &   1.11 &   1.74 &   1.76 &   0.08 &   0.65 &
   0.66 \cr
 BD -04~680   & HD 24289    & G080-028   &  5866 &   87 &   3.73 &  -2.07 &
   47.0 &   2.1 &   2.29 &   2.31 &   2.35 &   2.36 &   0.07 &   0.03 &
   0.08 \cr
             & HD 25329   &            &  4842 &   97 &   4.50 &  -1.64 &
   $<$
   3.0 &    --  &   0.00 &    --  &    --  &    --  &    --  &    --  &
    --  \cr
 BD +21~607^b
 & HD 284248   & G008-016   &  6034 &   74 &   4.05 &  -1.61 &
   25.0 &   2.2 &   2.11 &   2.12 &   2.12 &   2.14 &   0.05 &   0.05 &
   0.07 \cr
 BD +03~740   &            & G084-029   &  6110 &   82 &   3.73 &  -2.01 &
   21.0 &   1.0 &   2.06 &   2.07 &   2.07 &   2.09 &   0.06 &   0.02 &
   0.06 \cr
 BD +37~1458  &            & G098-058   &  5510 &   56 &   3.80 &  -1.86 &
   14.0 &   1.4 &   1.41 &   1.43 &   1.65 &   1.67 &   0.05 &   0.05 &
   0.07 \cr
             &            & G088-010   &  5856 &   74 &   3.85 &  -2.20 &
   30.0 &   2.9 &   2.05 &   2.07 &   2.12 &   2.13 &   0.06 &   0.05 &
   0.08 \cr
   \phantom{BD +00 0000}^b
         &            & G090-003   &  5786 &  132 &   3.55 &  -1.99 &
   44.0 &   1.9 &   2.19 &   2.21 &   2.27 &   2.29 &   0.10 &   0.03 &
   0.10 \cr
 BD +24~1676  &            & G088-032   &  6208 &   67 &   3.78 &  -2.59 &
   27.0 &   1.8 &   2.23 &   2.24 &   2.25 &   2.26 &   0.05 &   0.04 &
   0.06 \cr
 BD +31~1684^b
 & HD 64090    & G090-025   &  5441 &   68 &   4.50 &  -1.82 &
   12.0 &   1.0 &   1.30 &   1.32 &   1.61 &   1.63 &   0.07 &   0.04 &
   0.08 \cr
 BD -15~2546  & HD 74000    &            &  6224 &   96 &   4.08 &  -2.00 &
   25.0 &   3.4 &   2.22 &   2.23 &   2.23 &   2.24 &   0.07 &   0.07 &
   0.10 \cr
 BD -15~2656  & HD 76932    &            &  5727 &   77 &   3.78 &  -1.05 &
   28.7 &   1.0 &   1.94 &   1.96 &   2.01 &   2.03 &   0.06 &   0.02 &
   0.06 \cr
             &            & G115-049   &  5950 &   72 &   5.00 &  -1.86 &
   38.0 &   3.4 &   2.26 &   2.28 &   2.29 &   2.30 &   0.05 &   0.05 &
   0.07 \cr
 BD +09~2190  &            & G041-041   &  6333 &   89 &   3.83 &  -2.89 &
   19.0 &   1.4 &   2.14 &   2.15 &   2.14 &   2.16 &   0.06 &   0.04 &
   0.07 \cr
 BD +01~2341p^b
 & HD 83769C$^1$   & G048-029   &  6313 &   80 &   3.90 &  -2.81 &
   22.0 &   2.3 &   2.20 &   2.21 &   2.21 &   2.22 &   0.06 &   0.05 &
   0.08 \cr
 BD +14~2151^b
 & HD 84937    & G043-003   &  6330 &   83 &   3.90 &  -2.49 &
   24.0 &   1.2 &   2.26 &   2.27 &   2.26 &   2.28 &   0.06 &   0.03 &
   0.07 \cr
 BD +29~2091^b
 &            & G119-032   &  5815 &  119 &   4.80 &  -1.86 &
   45.0 &   6.9 &   2.24 &   2.26 &   2.30 &   2.32 &   0.09 &   0.09 &
   0.12 \cr
   BD +21~2247^b
         & HD 94028    & G058-025   &  6001 &  131 &   4.15 &  -1.50 &
   35.0 &   1.6 &   2.25 &   2.26 &   2.27 &   2.28 &   0.10 &   0.02 &
   0.10 \cr
 BD +02~2406  & HD 97916    &            &  6393 &  155 &   3.98 &  -1.22 &
   $<$
   3.0 &    --  &   1.28 &   1.29 &   1.32 &   1.33 &   0.10 &    --  &
    --  \cr
             &            & G011-044   &  5943 &  108 &   4.08 &  -1.80 &
   29.0 &   3.1 &   2.10 &   2.12 &   2.13 &   2.14 &   0.08 &   0.06 &
   0.10 \cr
 BD +02~2538^b
 & HD 108177   & G013-035   &  6097 &   94 &   4.03 &  -1.98 &
   32.0 &   1.3 &   2.26 &   2.27 &   2.27 &   2.29 &   0.07 &   0.02 &
   0.07 \cr
             &            & G059-024   &  6108 &   83 &   4.28 &  -2.69 &
   36.0 &   2.1 &   2.31 &   2.32 &   2.34 &   2.35 &   0.06 &   0.03 &
   0.07 \cr
 BD +28~2137^b
 &            & G059-027   &  6107 &  127 &   3.85 &  -2.34 &
   27.0 &   2.2 &   2.17 &   2.18 &   2.19 &   2.20 &   0.09 &   0.05 &
   0.10 \cr
             &            & G064-012   &  6468 &   87 &   4.10 &  -3.35 &
   27.0 &   1.8 &   2.39 &   2.40 &   2.36 &   2.37 &   0.06 &   0.03 &
   0.07 \cr
 BD +34~2476^b
 &            & G165-039   &  6282 &   72 &   3.88 &  -2.41 &
   23.0 &   1.5 &   2.21 &   2.22 &   2.22 &   2.23 &   0.05 &   0.04 &
   0.06 \cr
             &            & G064-037   &  6432 &   70 &   4.15 &  -2.51 &
   15.0 &   1.0 &   2.10 &   2.11 &   2.10 &   2.11 &   0.05 &   0.03 &
   0.06 \cr
             &            & G201-005   &  6328 &  111 &   4.50 &  -2.64 &
   24.0 &   1.7 &   2.26 &   2.27 &   2.26 &   2.28 &   0.08 &   0.04 &
   0.09 \cr
 BD +26~2606^b
 &            & G166-045   &  6037 &   73 &   3.93 &  -2.33 &
   30.0 &   1.3 &   2.16 &   2.17 &   2.19 &   2.20 &   0.05 &   0.03 &
   0.06 \cr
 BD -21~4009^b  & HD 132475   &            &  5788 &   68 &   3.65 &  -1.53 &
   51.5 &   1.4 &   2.29 &   2.31 &   2.35 &   2.37 &   0.05 &   0.02 &
   0.05 \cr
 BD +04~2969  & HD 134169   &            &  5870 &   84 &   3.83 &  -1.15 &
   44.4 &   1.7 &   2.29 &   2.31 &   2.33 &   2.34 &   0.07 &   0.03 &
   0.07 \cr
             &            & G152-035   &  5491 &   63 &   5.10 &  -1.61 &
   53.7 &   4.5 &   2.07 &   2.09 &   2.31 &   2.33 &   0.05 &   0.05 &
   0.07 \cr
 BD -10~4149  & HD 140283   &            &  5691 &   69 &   3.35 &  -2.37 &
   47.5 &   0.6 &   2.14 &   2.16 &   2.29 &   2.31 &   0.05 &   0.01 &
   0.05 \cr
 BD +42~2667  &            & G180-024   &  6059 &  128 &   4.10 &  -1.67 &
   28.0 &   4.8 &   2.18 &   2.19 &   2.19 &   2.21 &   0.09 &   0.09 &
   0.13 \cr
 BD +02~3375^b
 &            & G020-008   &  5891 &   75 &   3.83 &  -2.59 &
   33.0 &   1.3 &   2.10 &   2.12 &   2.18 &   2.19 &   0.06 &   0.02 &
   0.06 \cr
 BD -08~4501  &            & G020-015   &  5682 &   63 &   3.50 &  -2.79 &
   29.4 &   1.8 &   1.87 &   1.89 &   2.06 &   2.08 &   0.05 &   0.03 &
   0.06 \cr
 BD +20~3603^b  &            & G183-011   &  6441 &   76 &   4.30 &  -2.05 &
   27.0 &   2.2 &   2.40 &   2.41 &   2.40 &   2.41 &   0.05 &   0.04 &
   0.07 \cr
 BD +01~3597^b
 &            & G020-024   &  6464 &   77 &   4.30 &   -.91 &
   20.0 &   3.4 &   2.31 &   2.32 &   2.38 &   2.39 &   0.05 &   0.09 &
   0.10 \cr
 BD +13~3683^b
 &            & G141-019   &  5138 &   77 &   3.20 &  -2.39 &
   31.8 &   1.7 &   1.44 &   1.46 &   2.27 &   2.29 &   0.08 &   0.03 &
   0.09 \cr
 \phantom{BD +00 0000}^b
 &            & G206-034   &  6191 &  109 &   4.18 &  -3.10 &
   27.0 &   2.5 &   2.21 &   2.22 &   2.24 &   2.25 &   0.08 &   0.05 &
   0.10 \cr
             &            & G021-022   &  5869 &   82 &   3.93 &  -1.63 &
   37.0 &   1.9 &   2.18 &   2.20 &   2.22 &   2.24 &   0.06 &   0.03 &
   0.07 \cr
 BD +26~3578^b
 & HD 338529   &            &  6310 &   81 &   3.88 &  -2.58 &
   23.0 &   0.7 &   2.23 &   2.24 &   2.24 &   2.25 &   0.06 &   0.02 &
   0.06 \cr
 BD +10~4091^b
 & HD 188510   & G143-017   &  5564 &  121 &   4.88 &  -1.80 &
   18.0 &   3.4 &   1.60 &   1.62 &   1.79 &   1.81 &   0.10 &   0.10 &
   0.14 \cr
 BD -12~5613  & HD 189558   &            &  5663 &   87 &   3.68 &  -1.30 &
   42.0 &   1.6 &   2.08 &   2.10 &   2.18 &   2.20 &   0.07 &   0.02 &
   0.07 \cr
             &            & G024-003   &  5859 &   89 &   3.88 &  -1.78 &
   29.0 &   2.6 &   2.03 &   2.05 &   2.08 &   2.09 &   0.07 &   0.05 &
   0.08 \cr
 BD +23~3912  & HD 345957   &            &  5788 &   65 &   3.73 &  -1.45 &
   77.6 &   0.5 &   2.53 &   2.55 &   2.59 &   2.60 &   0.05 &   0.01 &
   0.05 \cr
 BD -21~5703  & HD 193901   &            &  5750 &   90 &   4.40 &  -1.13 &
   30.0 &   3.4 &   1.98 &   2.00 &   2.05 &   2.06 &   0.07 &   0.06 &
   0.09 \cr
 BD +04~4551  &            &            &  5819 &   91 &   3.78 &  -1.80 &
   29.0 &   6.9 &   2.00 &   2.02 &   2.06 &   2.08 &   0.07 &   0.14 &
   0.15 \cr
 BD +09~4529  & HD 194598   & G024-015   &  6017 &   76 &   4.15 &  -1.37 &
   28.0 &   0.9 &   2.15 &   2.16 &   2.17 &   2.18 &   0.06 &   0.02 &
   0.06 \cr
 BD +17~4519^b
 & HD 201891   &            &  5909 &   70 &   4.18 &  -1.22 &
   24.0 &   0.8 &   2.00 &   2.02 &   2.03 &   2.04 &   0.05 &   0.02 &
   0.05 \cr
 BD +23~4264  & HD 201889   &            &  5635 &   84 &   3.80 &  -1.10 &
   $<$
   4.0 &    --  &   0.85 &   0.87 &   0.97 &   0.98 &   0.08 &    --  &
    --  \cr
 BD +17~4708^b
 &            & G126-062   &  5941 &  106 &   3.58 &  -1.80 &
   25.0 &   1.4 &   2.02 &   2.04 &   2.05 &   2.06 &   0.08 &   0.03 &
   0.09 \cr
 BD +07~4841  &            & G018-039   &  5976 &   92 &   4.00 &  -1.34 &
   37.0 &   7.4 &   2.26 &   2.28 &   2.28 &   2.29 &   0.07 &   0.11 &
   0.13 \cr
             &            & G018-054   &  5762 &   86 &   3.75 &  -1.52 &
   35.0 &   2.6 &   2.07 &   2.09 &   2.14 &   2.16 &   0.07 &   0.04 &
   0.08 \cr
 BD +38~4955^b
 &            & G190-015   &  5115 &   92 &   5.00 &  -3.00 &
   8.0 &   0.3 &   0.95 &   0.97 &   1.92 &   1.94 &   0.10 &   0.02 &
   0.10 \cr
 BD -14~6437^b
 & HD 219617   & G273-001   &  6012 &   89 &   4.30 &  -1.63 &
   40.0 &   1.3 &   2.33 &   2.34 &   2.35 &   2.36 &   0.07 &   0.02 &
   0.07 \cr
\multispan{16}{\hrulefill}\cr
   }
   \endtable
\advance\PBtn by -1
\PBtbltwo{Li data ({\it continued})}
\halign to \hsize{\tabskip=0ptplus12ptminus15pt
$\rm #$\hfill &
# \hfill&
# \PBs&
# &
\hfill # &
# &
# &
\hfill# &
# &
# &
# &
# &
# &
# &
# &
#\cr
\multispan{16}{\hrulefill}\cr
\hfill BD \hfill & \hfill HD \hfill &
\hfill G \# \hfill & \hfill \PBt \hfill & $\sigma T_{\rm eff}$ \hfill &
\hfill \PBg \hfill & \hfill \PBfeh \hfill & \hfill EW \hfill &
\hfill $\sigma\rm EW$ \hfill  & \hfill A(Li)\phantom{$_1$}
\hfill & \hfill A(Li)$_1$
\hfill & \hfill A(Li)$_2$ \hfill & \hfill A(Li)$_3$ \hfill &
\hfill $\sigma _T $\phantom{$_W$}
\hfill & $\sigma _{EW}$  \hfill &
\hfill $\sigma_{tot}$  \hfill \cr
\multispan{16}{\hrulefill}\cr
 BD +02~4651  &            & G029-023   &  6102 &   90 &   3.80 &  -2.02 &
   27.0 &   2.2 &   2.18 &   2.19 &   2.19 &   2.21 &   0.07 &   0.05 &
   0.09 \cr
 BD +59~2723  &            & G217-008   &  6134 &   70 &   4.25 &  -1.91 &
   29.0 &   2.1 &   2.23 &   2.24 &   2.24 &   2.25 &   0.05 &   0.04 &
   0.06 \cr
\multispan{16}{\hrulefill}\cr
}
\tabletext
{\noindent A(Li)$_1$ is the NLTE corrected  value according to Carlsson
\PBal (1995)\par
\noindent A(Li)$_2$ is the value corrected for the depletion predicted
by the standard Li isochrones of Deliyannis \PBal (1990)\par
\noindent A(Li)$_3$ is the value corected both for NLTE and standard
depletion\par
\noindent
$^\ast$ the sum under qaudrature has been made taking into account three
significant figures and then rounded off.
\par
\noindent
$^b$ binaries
\par\noindent
$^1$ this star is LP 608-62=Ross  889, it is indicated
as HD 83769C by Olsen (1993) }
\endtable

\section{Results}

\subsection{L\lowercase{i} -- \PBt,\PBfeh correlations}

The derived Li abundances
are given in Table 1 (where ${\rm A(Li)}=\log(N_{\rm Li}/N_H)+12$).
The corresponding A(Li) -- \PBt and  A(Li) -- \PBfeh  diagrams are given in
figures 2 and 3. We considered plateau stars those
with \PBt $> 5700$
and  with \PBfeh $\le -1.5$; this extracts a sample of  41 stars.
\beginfigure{3}
\vskip 7cm
\caption{{\bf Figure 3.}
 The A(Li) --  \PBt diagram for our sample of stars.
 The circles are stars with [Fe/H] $\le -1.5$ , while
 the squares are those with [Fe/H] $> -1.5$.
 Subgiants are shown with crossed symbols.
 Three upper limits
 are shown with vertical arrows pointing downwards.
 }
 \endfigure

\par
 To investigate the existence of  trends  in the
A(Li) -- \PBt and A(Li) -- [Fe/H]
planes, we first performed univariate fits by means
of four different  estimators:
(i)
the Bivariate Correlated
Errors and intrinsic Scatter (BCES) of Akritas \& Bershady (1996);
(ii) BCES
simulations  bootstrap based on 10000 samples;
(iii) least squares with errors in the independent variable only
 (as implemented in the routine {\tt fitxy} of Press \PBal (1992));
(iv) least squares with errors in both variables
 (as implemented in the routine {\tt fitexy} of Press \PBal (1992)).
In addition,  the presence of a
slope was searched by means of a  rank correlation
analysis using Kendall's
$\tau$.
As a second step  we  performed bivariate fits
using both \PBt and \PBfeh as independent variables.
 The need for such bivariate fits follow from the
 known trend
of \PBt with \PBfeh,
 which   is  an observational
 bias, due to the fact
 that  the more metal--poor stars are, on average, the more
 distant and  therefore also the more luminous and hotter
 members of the class.
 So one could suspect that if there is, indeed, a trend
 of A(Li) with \PBt {\em and \ } \PBfeh, the trend of \PBt with
 \PBfeh could be exactly tuned so as to suppress the trends
 when performing univariate fits.
For the bivariate fits we used an ordinary least squares
method taking into account only errors in the independent
variable.
\par
\PBtbl{ Univariate fits in the A(Li)--\PBt plane.}
\halign to\hsize{
\tabskip=24ptplus54ptminus24pt
\hfill $#$ \hfill&
 \phantom{Method} #\hfill\cr
\multispan2{\hrulefill}\cr
 & Method\cr
\multispan2{\hrulefill}\cr
{\rm A(Li)}
= 0.766(\pm 0.635)+0.0235(\pm 0.0104)\times (T_{eff}/100) & BCES\cr
{\rm A(Li)} = 0.774(\pm 0.644)+0.0233(\pm0.0106)\times (T_{eff}/100)
& BCES bootstrap\cr
{\rm A(Li)} =1.346(\pm 0.353)+0.0140(\pm 0.0058)\times (T_{eff}/100)&
{\tt fitxy\phantom{e}} $\chi^2 = 56.59$ ; P = 0.034\cr
{\rm A(Li)} =1.153(\pm 4.056 )+0.0171(\pm 0.0511)\times (T_{eff}/100)&
{\tt fitexy} $\chi^2 = 0.4291$ ; P = 1.000\cr
\multispan2{\hrulefill}\cr
\multispan2{\hfill
after the theoretical
correction for the standard depletion \hfill}\cr
\multispan2{\hrulefill}\cr
{\rm A(Li)}
= 1.63
\phantom{0}(\pm 0.637)+0.0097(\pm 0.0104)\times (T_{eff}/100) & BCES\cr
{\rm A(Li)} 
= 1.64\phantom{0}(\pm 0.646)+0.0095(\pm0.0106)\times (T_{eff}/100)
& BCES bootstrap\cr
{\rm A(Li)} =2.009(\pm 0.353)+0.0035(\pm 0.0058)\times (T_{eff}/100)&
{\tt fitxy\phantom{e}} $\chi^2 = 56.47$ ; P = 0.035\cr
{\rm A(Li)} =1.827(\pm 4.053 )+0.0064(\pm 0.0444)\times (T_{eff}/100)&
{\tt fitexy} $\chi^2 = 0.4250$ ; P = 1.000\cr
\multispan2{\hrulefill}\cr
\multispan2{\hfill
 after the theoretical
 correction for the NLTE effect \hfill}\cr
\multispan2{\hrulefill}\cr
{\rm A(Li)}
= 0.858(\pm 0.635)+0.0222(\pm 0.0104)\times (T_{eff}/100) & BCES\cr
{\rm A(Li)} = 0.866(\pm 0.644)+0.0220(\pm 0.0106)\times (T_{eff}/100)
& BCES bootstrap\cr
{\rm A(Li)} =1.421(\pm 0.353)+0.0130(\pm 0.0058)\times (T_{eff}/100)&
{\tt fitxy\phantom{e}} $\chi^2 = 56.59$ ; P = 0.034\cr
{\rm A(Li)} =1.230(\pm 4.056 )+0.0160(\pm 0.0503)\times (T_{eff}/100)&
{\tt fitexy} $\chi^2 = 0.4291$ ; P = 1.000\cr
\multispan2{\hrulefill}\cr
\multispan2{\hfill
after the theoretical
corrections for the standard depletion and NLTE\hfill}\cr
\multispan2{\hrulefill}\cr
{\rm A(Li)}
= 1.72\phantom{0}(\pm 0.637)+0.0084(\pm 0.0105)\times (T_{eff}/100) & BCES\cr
{\rm A(Li)} 
= 1.73\phantom{0}(\pm 0.646)+0.0083(\pm0.0106)\times (T_{eff}/100)
& BCES bootstrap\cr
{\rm A(Li)} =2.084(\pm 0.353)+0.0025(\pm 0.0058)\times (T_{eff}/100)&
{\tt fitxy\phantom{e}} $\chi^2 = 56.47$ ; P = 0.035\cr
{\rm A(Li)} =1.902(\pm 4.053 )+0.0054(\pm 0.0441)\times (T_{eff}/100)&
{\tt fitexy} $\chi^2 = 0.4250$ ; P = 1.000\cr
\multispan2{\hrulefill}\cr
}
\endtable
 The results of the statistical 
 analysis are  reported  in Tables 2 -- 4.
 Inspection of
Table 2 shows that a
small slope of the order 0.02($\pm$ 0.01)/100~K is marginally detected
using BCES and {\tt fitxy},  but
not  using {\tt fitexy}.
 We used Kendall's $\tau$ to check the detection of the slope
 and we found a $\tau= 0.2112$ with a two--sided significance
 level of 0.05, consistent with a positive detection of the slope.
 Note that  the slope found is
about one half of
those  found by Thorburn (1994) of 0.034($\pm$ 0.006)/100~K
and Ryan \PBal (1996)  of 0.0408($\pm$ 0.0052)/ 100K.
One may also note that the goodness-of-fit implied by {\tt fitxy}
is quite small, i.e. of the order of 0.035. This value means that the
fit may be believable if the errors are nonnormal {\it or}
if they have been slightly underestimated (Press \PBal  1992).
To check this interpretation we performed the same fits, but using
the slightly larger errors which you obtain by summing linearly,
rather than under quadrature, the errors due to \PBt ~ and EW.
While the value of the fitting parameters are little sensitive to
this assumption, the goodness-of-fit boosts up to values of the order
of 0.7. This comment applies also to the results of Table 3, obtained from
{\tt fitxy}, and to those of Table 4.
\par
\beginfigure{4}
\vskip 7cm
\caption{{\bf Figure 4.}
 Zoom of the A(Li) --  \PBt diagram for our sample of stars.
 The solid line is the ZAMS isochrone of Deliyannis \PBal
 (1990) corresponding to \PBfeh $=-1.3$ and an age
 of 16 Gyrs, the dashed line is the same but for \PBfeh $=-2.3$.
 Both isochrones have been shifted upwards by 0.1 dex to better
 match the observations.
 }
 \endfigure
Table 3 shows the results of the investigations of
  trends of A(Li) with [Fe/H].
 Irrespective of the method used
 the slope is not  detected,
Kendall's $\tau$ is  -0.857 with a two--sided significance
level of 0.43, i.e.  no  slope.
 Note that in almost all cases the slope
 with \PBfeh is {\em negative}, contrary to the findings
 of Thorburn (1994) and Ryan \PBal (1996).
 A negative slope is physically plausible since it
 could be present if astration is important.
An increase of Li with the metallicity is expected when Li synthesis by
GCR begins to become significant. Using $^9$Be as an indirect indicator,
from the $^9$Be-metallicity correlation found by
Molaro \PBal (1996),
the contribution to Li by energetic galactic cosmic rays is of only
0.03 dex
at [Fe/H]=-1.5 and of 0.1 dex at [Fe/H]=-1.0, which is below our
detectability threshold.

Table 4 displays the results for the bivariate fits,
 which show that the values of the slopes found
 by performing bivariate fits are not significantly
 different from the values found performing univariate
 fits.
 \par
\beginfigure{5}
\vskip 7cm
\caption{{\bf Figure 5.}
Zoom of the plateau in the A(Li) -- \PBt plane using the Li abundances
corrected for standard depletion. The solid line is our estimate
of the primordial Li, the two dashed lines are the $1\sigma $ errors.
}
\endfigure
According to the Deliyannis \PBal (1990) standard ZAMS
Li isochrones,
a very little depletion is indeed expected in the \PBt range
between 5700 K and 6400 K. In Fig. 4
one can see the Li
isochrones of Deliyannis \PBal (1990) corresponding to [Fe/H]=$-2.3$
and [Fe/H]=$-1.3$, and an age of 16 Gyr, which
have been shifted upwards by 0.1 dex
to better match the observations.
Using these isochrones we corrected the observed
Li abundances  so that each star was
assigned the A(Li) it should have if its temperature
were 6400 K and its metallicity -2.3.
In Fig. 5 we show the corrected data and no trend is evident.
In fact the formal statistical analysis of this corrected data shows
no detectable trend   irrespective of    the estimators used.
Kendall's $\tau$ is  0.0804  with a two--sided
significance level of 0.458,  consistent with a non--detection
of the slope. Thus  the small slope with \PBt , which
is marginally detected using our sample, is consistent
with a constant Li abundance on the plateau with minimal  depletion
predicted by the standard Li isochrones of Deliyiannis \PBal (1990).
Also for the case of bivariate fits  the small slope,
marginally detected using the uncorrected Li, disappears
after accounting for the Li depletion predicted
by standard isochrones.

 \par
 NLTE correction are even smaller, but again
they act in the sense of decreasing the slope. Thus when  both
corrections are applied no slope is detected.

\par

\PBtbl{ Univariate fits in the A(Li)--\PBfeh plane.}
\halign to\hsize{
\tabskip=24ptplus54ptminus24pt
\hfill $#$ \hfill&
 \phantom{Method} #\hfill\cr
\multispan2{\hrulefill}\cr
 & Method\cr
\multispan2{\hrulefill}\cr
\rm A(Li) = 2.09(\pm 0.09)-0.05(\pm 0.04)\times[Fe/H] & BCES\cr
\rm A(Li) = 2.10(\pm 0.09)-0.04(\pm 0.04)\times[Fe/H] &BCES bootstrap\cr
\rm A(Li)= 2.15(\pm 0.06)-0.02(\pm 0.03)\times[Fe/H] &{\tt fitxy
\phantom{e}} $\chi^2 = 61.49$ ; P = 0.012\cr
\rm A(Li) = 2.11(\pm 0.15)-0.04(\pm 0.06)\times[Fe/H] & {\tt fitexy}
$\chi^2 = 8.50$ ; P = 1.000\cr
\multispan2{\hrulefill}\cr
\multispan2{\hfill
after the theoretical
correction for the standard depletion \hfill}\cr
\multispan2{\hrulefill}\cr
\rm A(Li) = 2.17(\pm 0.08)-0.02(\pm 0.04)\times[Fe/H] & BCES\cr
\rm A(Li) = 2.18(\pm 0.08)-0.02(\pm 0.03)\times[Fe/H] &BCES bootstrap\cr
\rm A(Li)= 2.22(\pm 0.06)-0.00(\pm 0.03)\times[Fe/H] &{\tt fitxy
\phantom{e}} $\chi^2 = 56.84$ ; P = 0.032\cr
\rm A(Li) = 2.18(\pm 0.15)-0.02(\pm 0.06)\times[Fe/H] & {\tt fitexy}
$\chi^2 = 7.67$ ; P = 1.000\cr
\multispan2{\hrulefill}\cr
\multispan2{\hfill
after the theoretical
correction for the NLTE effect \hfill}\cr
\multispan2{\hrulefill}\cr
\rm A(Li) = 2.11(\pm 0.09)-0.04(\pm 0.04)\times[Fe/H] & BCES\cr
\rm A(Li) = 2.12(\pm 0.08)-0.04(\pm 0.04)\times[Fe/H] &BCES bootstrap\cr
\rm A(Li)= 2.17(\pm 0.06)-0.02(\pm 0.02)\times[Fe/H] &{\tt fitxy
\phantom{e}} $\chi^2 = 60.90$ ; P = 0.014\cr
\rm A(Li) = 2.13(\pm 0.15)-0.03(\pm 0.06)\times[Fe/H] & {\tt fitexy}
$\chi^2 = 8.41$ ; P = 1.000\cr
\multispan2{\hrulefill}\cr
\multispan2{\hfill
after the theoretical
corrections for the standard depletion and NLTE\hfill}\cr
\multispan2{\hrulefill}\cr
\rm A(Li) = 2.19(\pm 0.08)-0.02(\pm 0.04)\times[Fe/H] & BCES\cr
\rm A(Li) = 2.20(\pm 0.08)-0.02(\pm 0.03)\times[Fe/H] &BCES bootstrap\cr
\rm A(Li)= 2.24(\pm 0.06)+0.00(\pm 0.02)\times[Fe/H] &{\tt fitxy
\phantom{e}} $\chi^2 = 56.66$ ; P = 0.033\cr
\rm A(Li) = 2.20(\pm 0.15)-0.01(\pm 0.06)\times[Fe/H] & {\tt fitexy}
$\chi^2 = 7.64$ ; P = 1.000\cr
\multispan2{\hrulefill}\cr
}
\endtable
\PBtbl{ Bivariate fits }
\halign to\hsize{
\tabskip=24ptplus24ptminus24pt
\hfill $#$ \hfill&
 \phantom{Li correction} #\hfill& # &#\cr
\multispan2{\hrulefill}\cr
 & $\chi^2$& P & \hbox{Li correction}\hfill\cr
  1.10(\pm 0.43)+0.034(\pm 0.034)\times {\rm [Fe/H]}+
 0.019(\pm 0.008)\times (T_{eff}/100) &60.21 & 0.012 & none \cr
  1.88(\pm 0.43)+0.018(\pm 0.034)\times {\rm [Fe/H]}+
 0.006(\pm 0.008)\times (T_{eff}/100)
 &61.03 & 0.010 & \hbox{standard depletion}\hfill \cr
  1.17(\pm 0.43)+0.034(\pm 0.034)\times {\rm [Fe/H]}+
 0.018(\pm 0.008)\times (T_{eff}/100)
 &60.57 & 0.011 & NLTE \cr
  1.95(\pm 0.43)+0.018(\pm 0.034)\times {\rm [Fe/H]}+
 0.005(\pm 0.008)\times (T_{eff}/100)
 &61.14 & 0.010 & standard depletion and NLTE \cr
\multispan2{\hrulefill}\cr
}
\endtable

As far as the slope is concerned,
a possible explanation of the different results may lie  in the different
\PBt adopted.
 Thorburn (1994)  derived a common photometric temperature scale
by adapting the temperature calibrations between \PBt-(V-K); ({\it b-y})
of Carney (1983) and Carney \PBal (1987) into a \PBt-(B-V) scale,
which is assumed to be valid
for all metal poor stars.
There are 24 stars in common with Thorburn (1994) and
in Fig. 6a the temperature difference between Alonso \PBal (1996a)
and Thorburn (1994) shows a systematic trend with \PBt, with
the \PBt of Alonso \PBal (1996a)  higher at the hot edge
of the plateau and lower at the cool edge.
This systematic difference  in the \PBt
produces the slope in Li versus \PBt  which is observable
in Fig.  6b, where the differences between our A(Li) values
and those of Thorburn (1994) are shown as a function of \PBt .
In addition to the trend one may notice an offset, which is due
to the inclusion of overshooting in the ATLAS9  models
used by Thorburn.
All our stars, but one, are in common with the sample
of Ryan \PBal (1996) and
in Fig. 6c and 6d we  compare
our results with those of Ryan \PBal (1996) in the same way
we did for the Thorburn data. Ryan \PBal (1996)
adopted the average
of the ({\it b-y})-\PBt from Magain (1987), the (R-I)-\PBt
from Buser \& Kurucz (1992) and a (B-V)-\PBt worked out by
interpolation from a grid obtained from the literature.
Also in this case,
as can be seen in Fig. 6d,
a temperature scale difference, 
which is responsible for the differences in the slope
A(Li)--\PBt found in the two analyses,
is clear.
Although
 the \PBt scales used by
Thorburn (1994) and Ryan \PBal (1996)
may be sufficient  for a general abundance
analysis, they may be  not adequate
to discuss dispersion and/or trends at the  levels claimed, since
the indices  used ($b-y$, $B-V$ and $R-I$) are all metallicity
dependent and the first two depend also on gravity.
To illustrate the effect of metallicity we divided the 40 stars
in common with Ryan \PBal (1996) in four metallicity bins
($\rm -2.0 < [Fe/H] \le -1.5$, $\rm -2.5 < [Fe/H] \le -2.0$ ,
$\rm -3.0 < [Fe/H] \le -2.5$ and $\rm  [Fe/H] \le -3.0 $),
the bins have 16, 13, 8 and 3 stars respectively.
For each bin we compute the mean of $\rm T_{Alonso}- T_{Ryan}$,
the results (rounded to the nearest degree) are: +10 K,
+50 K, +96 K and +132 K.

\beginfigure*{6}
\vskip 22.5cm
\caption{{\bf Figure 6.}
Temperature and A(Li) differences between the
present paper and those of Thorburn (1994)(panels a) and b)), 
Ryan \PBal (1996) (panels c) and d)) and
Molaro \PBal (1995a)(panels e) and f)), for the stars in common.
}\endfigure

The same analysis with the 16 stars we have  in
common with  Molaro  \PBal
 (1995a) 
 shows a considerable scatter, but no obvious trend.
 They
used the \PBt obtained by
Fuhrmann \PBal (1994) by fits of the
Balmer lines, i.e with a totally different approach.
Balmer line
temperatures have the considerable
advantage that they are reddening independent.
A trend  with 
temperature cannot be excluded, but
the number is too small to draw any firm conclusion.
Ryan \PBal (1996), 
reanalysed the Molaro \PBal (1995a) sample, after
excluding the two subgiants and outliers, and found 
a trend of $\approx$ 0.04/100 K. However, no trends are found
using the whole original sample. 
\par
We thus conclude that the presence or absence of trends of lithium
abundance with \PBt ~ 
appears to be strongly dependent on the temperature scale adopted.

\beginfigure*{7}
\vskip 7cm
\caption{{\bf Figure 7.}
The temperatures of Thorburn (1994), panel a), Ryan \PBal (1996),
panel b), and Fuhrmann \PBal (1994), panel c), are plotted
as a function of the temperatures of Alonso \PBal (1996).
The bisector is shown as a solid line.
}
\endfigure

\subsection{Dispersion on the plateau?}

The straight average value of A(Li) for the selected stars is
 A(Li) = 2.19 with a dispersion around the
mean value of $\pm $0.094.
 This dispersion  is compatible with the root mean square of the
 estimated observational errors (0.083 dex)
 showing no
 clear evidence for intrinsic dispersion.
In fact 29 stars
($\approx 71 \%$) are within 1 $\sigma$ of the mean A(Li),
which is compatible with a normal distribution, such as would be expected
for observational errors.
The dispersion around the mean value further decreases down to
0.088 if the Li abundances
corrected for the theoretical depletion  and the NLTE effects are considered.
In Fig. 5  the corrected Li abundances on the plateau are zoomed
illustrating that the data are consistent with a
unique Li abundance within the errors.
The absence of intrinsic scatter is remarkable
in the light of the high percentage of binaries or suspected
binaries in our sample.
\par
\beginfigure*{8}
\vskip 7cm
\caption{{\bf Figure 8.}
Equivalent widths as a function of effective temperature. 
The solid lines are the theoretical loci for A(Li)=2.20 and
\PBfeh $=-2.00$; the dashed lines are the same but
for A(Li)$=2.20\pm 0.1$. The dotted line in panel b) is
the fit to the data of a line of slope $-0.0623$ log(m\AA)/100 K,
which is the slope of the theoretical lines in the same panel.
}
\endfigure
Thorburn (1994) analysed the presence of scatter
in the \PBew ~ -- \PBt plane, claiming that the null hypothesis of
no intrinsic dispersion is rejected at the $6\sigma$ confidence level.
In Fig. 8a we show such a diagram for  our plateau sample, together
with  the theoretical \PBew ~-- \PBt locus obtained from
our \PBcogs ~ corresponding to three Li abundances ($2.20\pm 0.1 $dex)
for a metallicity of [Fe/H]=-2.0.
Note that this locus is not a
straight line as assumed both by Thorburn (1994)
and Deliyannis \PBal (1993), but
is better approximated by an exponential, as can be seen
by the linear relation displayed in
the $\log(EW)$--\PBt ~~ diagram shown in Fig. 8b.
The slope of the theoretical lines is $\approx - 0.0623$  log(m\AA)/ 100 K.
Performing a simple least squares fit with a straight line of
this slope (i.e. fitting only one free parameter) we find the dotted line
shown in Fig 8b. The fit has
a $\chi^2= 0.393$, corresponding to a goodness-of-fit of practically 1.00, and
an R.M.S = 0.066 log(m\AA). 
 The fitting line almost
coincides with the theoretical line corresponding to A(Li) = 2.20.
This shows that, even neglecting
the metallicity effects on the curve of growth, the data are consistent
with the hypothesis of a constant Li, around 2.20,  and a scatter due
to observational errors only.

The  disagreement with
Thorburn (1994) on this point can be explained by the
fact that she   fits  an exponential with a straight line. She
attributes the very  low goodness of fit  ($\approx 10^{-10}$)
to either an intrinsic scatter in the Li abundances  or to an
underestimation of the errors
 by about 50 \%, while this can be explained by  the use of an incorrect
fitting function.
 In addition,
as pointed out  by Spite \PBal (1996), the temperature errors of
 Thorburn (1994) appear to be
somewhat underestimated.

 Deliyannis \PBal (1993)
discussed the dispersion on the Spite plateau
in the $ W_\lambda - (b-y)_0$
plane which is strictly observational and avoided the
 problems implied in the
determination of \PBt .
Note, however, that like Thorburn (1994) they
also assumed  a linear relation
between   $ W_\lambda $ and $ (b-y)_0$ to look for dispersion.
Moreover,  there
is not a one-to-one mapping from $ (b-y)_0$ to \PBt
since this index depends both on gravity and metallicity.
A sample of halo stars with different gravities and metallicities
will show a scatter in the $ W_\lambda - (b-y)_0$ plane even if there
is no scatter in the
Li abundances, simply because when comparing stars of
the same $(b-y)_0$ we compare stars with different \PBt.

Ryan \PBal (1996)
identified a triplet of stars (G064-012, G064-037,
and CD -33 1173) with similar metallicity and
temperature, but significantly different
Li abundances.
Only the first two belong to our sample and in fact show
one of the maximum differences in the A(Li)
values in the whole sample, namely
2.39$\pm 0.07$ and 2.10 $\pm 0.06$ , which agree
at the 2.23 $\sigma$ level. For comparison  the values of Ryan \PBal (1996)
 are 2.29$\pm$0.05 and 2.01 $\pm$0.04.
Our errors  associated
with the \PBt are  in general larger
than those of Ryan \PBal (1996).
The temperature errors of Ryan \PBal (1996) are
rather small
($\sigma(B-V)= 50 $K,
$\sigma (b-y)=35 $ K and $\sigma (R-I)=90 $ K).
If we compare them with the
 standard deviations of  the
colour--temperature calibrations  derived by  Alonso
Arribas \& Mart\PBi nez-Roger (1996b)
($\sigma(B-V)= 130 $K,
$\sigma (b-y)= 110 $ K and $\sigma (R-I)= 135 $ K)
we see that the latter are larger
by  almost a factor of three.
The calibrations of Alonso \PBal (1996b)
are based  on
over 400  stars of which roughly
50\% are Pop II, while those of Magain (1987), on which
Ryan \PBal (1996)
partially rely,  
 are based on a total of 44 stars 
out of which only 11 belong to  Pop II.
This comparison gives rise to the suspicion that the temperature
errors of Ryan \PBal (1996) are underestimated.

The absence of intrinsic
dispersion  found by us is in agreement with the results
 of Molaro \PBal (1995a) and
Spite \PBal (1996). The latter authors
 have  considered three small samples of stars
with accurate \PBt derived either from excitation temperature or
from the dereddened colour ({\it b-y})$_0$ or from
H$\alpha$ wings.
They found that the dispersion of the Li abundances in
these small
subsamples is significantly reduced, with an   R.M.S.  of the order
of 0.06-0.08 dex,  and they  argue that it is fully explained by the errors
in the \PBt and EWs. It is interesting that
the  presence of the binary  G020-008, with a A(Li)
lower than the average, is found to
contribute considerably to the scatter.

The absence of dispersion on the plateau is
remarkable considering
the number of processes that might
produce it, such as localized Galactic enrichment,
chromospheric Li
enrichment, different amounts of
accretion or astration, binarity,
anomalous reddening,
mistaken subgiants, pre main-sequence depletion and others.
This implies that
such possibilities are either  intrinsically or statistically
irrelevant.
\par

\section{Discussion}

\subsection {Li  depletion?}

We have found that in our sample of 41 stars on the plateau
with the IRFM \PBt derived by Alonso \PBal (1996a) there is
no evidence for intrinsic dispersion in excess of what  can be
expected  from the measurement uncertainties.
In addition, there is no evidence
of trends with the metallicity, although our
stars span more than two orders of
magnitude in [Fe/H]. The tiny trend
with \PBt , which has been detected,
can be entirely explained by the standard Li isochrones
of Deliyannis  \PBal (1990).

The view that the Li abundance on the Spite plateau is the
primordial abundance requires that Li has  not been destroyed,
in appreciable amounts, in the atmospheres of halo dwarfs.
Simple models of  Li evolution,  referred  to as
{\it standard models},  predict little or no depletion
of Li on the MS, thus supporting the primordial nature of Li in
halo dwarfs. However, it has been suggested that
 three different mechanisms  are  able to deplete
 Li in halo dwarfs in significant amounts. They are
diffusion (Michaud, Fontaine \& Beaudet 1984),
rotational mixing (Pinsonneault, Deliyannis \& Demarque  1992)
and  stellar
winds (Vauclair \& Charbonnel 1995).
 More elaborate scenarios with
some combination of the three have been also considered.
To reproduce the observations such
depletion models must
be able to
deplete Li uniformly in a way which is  independent from the
metallicity and  from the  mass (and therefore \PBt) of the star.
The detection of sizeable  dispersion on the plateau would
be a strong evidence in favour of  some Li depletion
during the life of halo stars.
\par
Diffusion has been proposed to be responsible for abundance anomalies
in Ap and Am
stars which are somewhat hotter than halo dwarfs, have non convective
atmospheres and the Ap are strongly magnetic.
In halo stars there are no  abundance anomalies  which
have been ascribed to diffusion. In fact Li is
the only element for which diffusion has been claimed to
be effective in halo dwarfs.
First computations of the effects of diffusion
on the Li abundances in halo stars have
been performed by Michaud \PBal (1984).
Diffusion causes Li to sink  below the photosphere
and  Michaud \PBal (1984) found that
the process is more  efficient at the
hot end of the plateau producing a pronounced downturn of the Li abundance
for \PBt $>$ 6000 K. The depth of the
surface convection zone
is sensitive
to opacities and to the mixing length, so that some model dependence
is present. However the downturn persists in the various assumptions
(Chaboyer \& Demarque 1994, Vauclair \& Charbonnel 1995).
From an initial A(Li)=2.5, at \PBt = 6500 K the
predicted Li abundance
becomes A(Li)=1.5, while our  observed abundance is  0.72 dex higher.
The total absence of
any bend at the warmer edge of the plateau  makes  uninhibited diffusion
unlikely in depleting lithium.
Moreover, an age spread of only 3 Gyrs would produce a dispersion of 0.2 dex
on the plateau, which, again, has   not been observed.

Vauclair (1988) suggested rotation induced turbulence as the agent
leading to  a nuclear destruction of
Li from an abundance as high as A(Li)=3.0 down to the observed
plateau value,
 still
approximately preserving
the plateau shape.
Similar computations with a different formulation
of the rotation induced turbulence have been performed by
Pinsonneault \PBal (1992) and Chaboyer \& Demarque (1994).
These models predict a bending of the
Li abundance on the warmer stars very similar to that predicted by
 diffusive models. In these models a considerable
 dispersion is expected depending
on the stellar initial angular momenta. Only assuming
an identical
initial rotational velocity it is possible to deplete Li in a uniform way
 down to the observed value.
Moreover, the initial rotational velocity has to be rather high,
which then requires an efficient breaking mechanism
to slow down the stars to the low
rotational velocities observed.
Vauclair \& Charbonnel (1995) started
from an initial value of 100 \PBkms ~
going down to the 2 \PBkms ~ presently observed.
From an initial value of A(Li)=2.7 their model
predicts  at \PBt = 6500 K a A(Li)= 1.6,
with a negative slope on the plateau resembling the one obtained by
pure diffusive models.
These authors found   it difficult to account for lithium dispersion
less than about
20 to 50 \% in the plateau with such  models. It is also to
note that an age spread of 3 Gyrs alone would produce a dispersion  of
0.3 dex on the plateau.
\par
Li depletion induced by
winds has been recently proposed by Vauclair \& Charbonnel (1995).
A mass loss flux between 10$^{-13}$ and 10$^{-12}$  M\sun yr$^{-1}$
can deplete Li  producing a positive slope on the plateau as a function of
the stellar mass (i.e.  \PBt).
Higher mass loss rates would destroy Li entirely, while smaller rates
would deplete less  Li.
A stellar wind of
10$^{-13}$ M\sun yr$^{-1}$ produces an almost unnoticeable
Li depletion, while a
wind of 10$^{-12}$ M\sun yr$^{-1}$ produces a slope of 0.4 dex along
$\approx$ 1000 K; that is  a slope of the same order of that
detected by Thorburn (1994) or
Ryan \PBal  (1996). As a comparison the solar wind is of
10$^{-14}$ M\sun yr$^{-1}$ , but
there are no
measurements of winds in Pop II dwarfs.
The absence of such a slope as well as the absence of any
detectable dispersion
is  arguing against the mass loss model.
If the winds were of different
rates in different stars
they would  produce  a dispersion on the Li abundances, and
an age spread
of 3 Gyr would produce a dispersion of 0.25 dex.
\par
 In summary, all depletion mechanisms predict features which
are not observed in the present data.
 The absence of a downturn in Li abundance at the hottest edge
of the plateau
and the absence of dispersion on the plateau itself
argue strongly against significant depletion by
diffusion
or  rotational mixing
(Vauclair 1988, Pinsonneault \PBal 1992);
the absence of a significant slope with \PBt
and the absence of intrinsic dispersion
rule out  stellar winds as a possible  source for Li depletion
(Vauclair \& Charbonnel 1995).
 Thus we consider that our results strongly support the
view that the
observed A(Li) in Pop II stars coincides with the
primordial value.

\subsection{ Primordial Li}

The absence of dispersion and of significant trends in our analysis,
in particular when the small trend with \PBt ,
expected from standard models,
is considered,
allows  to take the mean value of Li of the 41 stars on the plateau
as the value of the primordial Li.
The weighted mean, in which  each abundance is
 weighted  inversely by its own
variance,  is
 A(Li) = 2.20 with an uncertainty  of the mean value of only $\pm 0.012$.
  When we take the
values corrected for the
mild depletion predicted by standard models
and NLTE effects the weighted mean  is:
\bigskip

\centerline{     A(Li)$_p$ =  2.238$\pm$ 0.012}

\bigskip\noindent
This   value is  in excellent agreement
with the NLTE corrected value of A(Li)=2.224$\pm 0.013$ by Molaro
\PBal (1995a).
It is significant that the two samples, which make use
of a  spectroscopically determined  \PBt and of the semi-direct method of the
IRFM,  give the same average Li abundance.

Any   offset in the zero point of the different \PBt scales
would affect the
absolute value of the plateau,
while different slopes in the \PBt scales
may introduce (or suppress)  trends with \PBt.
The Alonso \PBal (1996a)
and Fuhrmann \PBal (1994) scales are on average
hotter than photometric scales and this explains why old mean values for the Li
on the plateau were close to A(Li)=2.08
(Molaro  1991).
As already mentioned in section 2, the  zero-point of \PBt scales in the low
main sequence is difficult to establish. The Alonso \PBal (1996a) scale
is able to reproduce both the Procyon and Sun
\PBt, within the measurement errors,
and therefore no significant offsets are expected.
Alonso \PBal (1996a) estimate in their \PBt a systematic error
of $\approx$ 1.25\% at \PBt = 6000 K, related to the uncertainty
of the calibrations of J,H and K magnitudes.
This     implies
a systematic error for  Li of  $\pm$ 0.05 dex,
which we have considered separately since it is not eliminated
in the averaging process.
The determination of the
value of primordial Li is  therefore dominated by   systematic
rather than random
errors.
Still larger systematic errors can be hidden in our ultimate ability of
modelling  stellar atmospheres (Kurucz 1995).
Our estimate for the primordial lithium is:
\bigskip

\centerline{     A(Li)$_p$ = 2.238$\pm 0.012_{1\sigma} \pm 0.05_{sys}$}
\bigskip\noindent

\beginfigure{9}
\vskip 7cm
\caption{{\bf Figure 9.}
The theoretical SBBN $^7$Li yields as a function of $\eta$, following the
parametrization of Sarkar (1996), are compared to our estimate
of primordial Li. The thick solid line corresponds to A(Li)=2.238,
the thin solid lines correspond to $1\sigma$ errors, while
the dotted lines correspond to the $1\sigma$ plus systematic error.
The thick solid curve is the prediction of SBBN, thin solid
curves are $1\sigma $ errors related to uncertainties in
the nuclear reaction rates.
}
\endfigure

In Fig. 9  the   SBBN  theoretical  Li yields are shown, computed
by using   the parametrization provided by
Sarkar (1996).
The Li Pop II value  corresponds to
two possible values for $\eta = n_{B}/n_{\gamma}$:
$ \eta_{10}
= 1.7^{+0.5}_{-0.3}$
%(1.42 - 2.25)
or $ \eta_{10} =4.0^{+0.8}_{-0.9} $,
%(3.05 - 4.81)
where $\eta_{10} =
10^{10}\eta$.
The minimum of the theoretical
Li yield is excluded when one considers $1\sigma$ errors
in the theoretical predictions and $1\sigma$ errors,
plus the full systematic error,
in the Li abundances.
However, considering  the $2\sigma$ errors, both in the
theoretical yields and in the observations ($2\sigma_{stat}+
1\sigma_syst$),
the minimum of the Li curve
is allowed,
increasing considerably the allowed $\eta$ range to $1.3 < \eta_{10} < 5.4$ .

Our low  $\eta$  intercept coincides
with  the $\eta_{10}=1.8\pm 0.3$ obtained from
high D/H ($\approx$ 10$^{-4}$) values (Songaila \PBal 1994, Carswell \PBal
1996, Wampler \PBal 1996, Rugers \& Hogan 1996) and
is in good agreement with the $\eta$ derived from the primordial He,
$\rm Y_p =0.228 \pm 0.005$ (Pagel \PBal 1992), with
three neutrino flavours.
The  high  $\eta$  intercept
is more  consistent with the $\eta_{10} \approx 3$
implied by the primordial helium Yp=0.241 $\pm 0.003$ obtained by
Izotov, Thuan \& Lipovetsky  (1994) by
using  revised
recombination coefficients for the He I lines.
\par
Our low  $\eta$ value   is inconsistent with the values
$\eta_{10}= 6.4
^{+0.9}_{-0.7}$ from
the low D/H=2$\times 10^{-5}$ observed in high redshift absorption systems
(Tytler \PBal 1996, Burles \& Tytler 1996; but see Wampler 1996 ).
To achieve  consistency
between the low D/H value and Li  a 0.5 dex of depletion
in Li is required, which is not  supported by the present analysis.
However, considering the $2\sigma$ errors, the total allowed window
for $\eta$ overlaps with the $\eta$ window implied by
the low D/H observations.
\par
Turner \PBal (1996) use the presolar value of
D+$^3$He 
from the recent measurement of $^3$He in the local ISM
of Gloeckler \& Geiss (1996) to deduce
the value of $\eta$ for different assumptions on the evolution
of $^3$He in low-mass stars and of metal ejection by
massive stars.
Their allowed window for $\eta$ ($2\le\eta_{10}\le 7$)
overlaps with ours, although our low $\eta$ value
is close to their lower limit. 
Comparison of
figure 2 of Turner \PBal (1996) with
our $\eta$ range seems to suggest that $^3$He is
preserved but not produced by low-mass stars.

\par
The two intercepts on the Li theoretical curve correspond to two
values for the baryonic  density. From the relation
 $ \Omega_{B} h^{2} = 0.00366 \eta _{10}$ we have:
\bigskip

\centerline{$ \Omega_{B}h^2  = 0.0062^{+0.0018}_{-0.0011}
~~~~~~~~~or~~~~~~~~~ \Omega_{B}h^2  = 0.0146^{+0.0029}_{-0.0033} $}

\bigskip

\noindent where the Hubble parameter
is H$_{0}$ = 100$h$ $\rm kms^{-1} Mpc^{-1}$.
Considering that  the luminous matter has a contribution to $\Omega$ of
$\Omega_{LUM}  = 0.004+0.0007h ^{-3/2}$
(Persic \& Salucci 1996), we have that over the entire range of allowed
H$_{0}$  the baryonic density  from SBBN remains always greater
than $\Omega_{LUM}$
suggesting the presence of
dark baryons.
\par

\section*{Acknowledgments}

We wish to thank  several people who helped us along
this work: A. Alonso and S. Arribas for providing
us with an electronic version of their data,
M.A. Bershady for sending us his codes,
M. Mermillod for clarifying  the
cross--identifications of BD +01 2341p, R.L. Kurucz
for helpful discussions on the determination of Li abundances
and C. Morossi for useful comments on an early draft of this paper.
This research has made use of the Simbad database, operated at CDS,
Strasbourg, France.


\section*{References}

\beginrefs
\bibitem\PBapj Akritas M.G., Bershady M.A.:1996 470 {astro-ph/960502}
\bibitem\PBasupl  Alonso A., Arribas S., Mart\PBi nez-Roger C.:1996a 117 227
\bibitem\PBa  Alonso A., Arribas S., Mart\PBi nez-Roger C.:1996b {in~press}
  {IAC preprint}
\bibitem\PBa Arribas S., Mart\PBi nez-Roger C.:1989 215 305
\bibitem\PBmn Bell R. A., Gustafsson B.:1989 236 653
\bibitem\PBa Blackwell D.E., Petford A.D.,  Arribas S., Haddock D.J., 
  Selby M.J.:1990 232 396
\bibitem Bonifacio P., Molaro, P., 1996  in
Persic M., Salucci P., eds,  
Dark and Visible Matter in Galaxies
ASP Conf. Ser., 
in press, astro-ph/9610051
\bibitem\PBscie  Burles S., Tytler D.:1996 {submitted} {astro-ph~9603069}
\bibitem\PBa Buser R., Kurucz R. L.:1992 264 557
\bibitem\PBa Carlsson M., Rutten R.J., Bruls J.H.M.J.,
Shchukina N.G.:1994 288 860
\bibitem\PBaj Carney B.W.:1983 88 623
\bibitem\PBaj Carney B.W., Laird J.B., Latham D. W. , Kurucz R. L.:1987 94 1066
\bibitem\PBmn Carswell R. F.,
Webb J. K., Lanzetta K. M., Baldwin J. A., Cooke A. J., Williger G. M.,
Rauch M., Irwin M. J., Robertson J. G., Shaver P. A.:1996   278 506
\bibitem\PBa Castelli F., Gratton R.G., Kurucz R.L.:1996 {in~press} ~
\bibitem  Cayrel R., 1988 in
Cayrel de Strobel G., Spite M., eds, Proc. IAU Symp. 132, 
The Impact of Very High S/N Spectroscopy
on Stellar Physics. Kluwer,
Dordrecht, p. 345
\bibitem\PBapj Chaboyer B., Demarque P.:1994 433 510
%\bibitem\PBaj Crawford D.L.:80 955
\bibitem\PBapjsupl Deliyannis C. P., Demarque P., Kawaler S. D.:1990 73 21
\bibitem\PBapj  Deliyannis C. P., Pinsonneault M. H., Duncan, D. K.:1993
 414 740
\bibitem\PBapj Deliyannis C. P., Boesgaard A. M., King J. R.:1995 452 L13
\bibitem\PBpasp Fouts G.:1987 99 986
\bibitem\PBa Fuhrmann K., Axer M, Gehren T.:1994 285 585
\bibitem\PBnat Gloeckler G., Geiss J.:1996 381 210
\bibitem\PBapj Hobbs L. M., Duncan D. K.:1987 317 796
\bibitem\PBapj Izotov Y. I., Thuan T. X., Lipovetsky V. A.:1994  435 647
\bibitem Kurucz R.L., 1993, CD-ROM No. 13, 18
\bibitem\PBapj Kurucz R. L.:1995 452 102
\bibitem\PBaj Lu P. K., Demarque P., van Altena W., McAlister H.,
Hartkopf W.:1987 94 1318
\bibitem\PBa Magain P.:1987 181 323
\bibitem\PBasupl Mermillod J.C., Hauck B., Mermillod M.:1996 {in~preparation}
{http://obswww.unige.ch/gcpd/gcpd.html}
\bibitem\PBapj Michaud G., Fontaine G., Beaudet G.:1984 282 206
\bibitem\PBmsait Molaro P.:1991 62 17
\bibitem\PBa Molaro P., Primas F., Bonifacio P.,:1995a 295 L47
\bibitem\PBmsait Molaro P., Bonifacio P., Primas F.:1995b  66 323
\bibitem\PBa Molaro P., Bonifacio P., Castelli F., Pasquini L.:1996 {in~press}
{astro-ph~9608011}
\bibitem\PBapj Norris J. E., Ryan S. G., Stringfellow G. S.:1994 423 386
\bibitem\PBasupl Olsen E.H.:1993 102 89
\bibitem\PBa Pavlenko Ya.V., Magazz\`u A.:1996 {in~press} {astro-ph 9606090}
\bibitem\PBmn Pagel B. E. J.,
Simonson E. A., Terlevich R. J., Edmunds M. G.:1992  255 325
\bibitem\PBmn Persic M.,  Salucci P.:1996  {submitted} ~
\bibitem\PBapjsupl Pinsonneault M. H., Deliyannis C. P., Demarque P.:1992 78 179
\bibitem Press W. H., Teukolsky A. A., Vetterling W. T., Flannery B. P.,
1992, Numerical Recipes, 2nd edn., Cambridge Univ. Press, Cambridge
\bibitem\PBa Rebolo R., Molaro P., Beckman J.:1988   192 192
\bibitem\PBapj Rugers M., and Hogan C. J.:1996  459 L1
\bibitem\PBapj Ryan S.G., Beers T.C.,
Deliyannis C.P., Thorburn J.A.:1996 458 543
\bibitem\PBa  Saxner M.,  Hammarback G.:1985 151 372
\bibitem Sarkar S., 1996 Reports on Progress in Physics,
submitted, hep-ph 9602260
\bibitem\PBa Schuster W.J., Nissen P.E.:1989 221 65
\bibitem\PBnat Songaila A., Cowie L.L., Hogan C.J., Rugers M.:1994  368 599
\bibitem\PBa  Spite F., Spite M.:1982  115 357
\bibitem\PBa  Spite F., Spite M.:1986  163  140
\bibitem\PBa Spite M., Maillard J.P., Spite F.:1984 141 56
\bibitem\PBa Spite M., Francois P., Nissen P.E., Spite F.:1996 307 172
\bibitem\PBpasp Stryker L. L., Hesser J. E., Hill G., Garlick G. S.,
O'Keefe L. M.:1986 97 247
\bibitem\PBapj Thorburn J.A.:1994 421 318
\bibitem\PBapj Turner M. S., Truran J. W., Schramm D. N., 
Copi C. J.:1996 466 L59
\bibitem\PBnat Tytler  D., Fan X.M,  Burles S.:1996  381 207
\bibitem\PBapj Vauclair S.:1988  335 971
\bibitem\PBa  Vauclair S., Charbonnel C.:1995 295 715
\bibitem\PBnat Wampler E. J.:1996 383 308
\bibitem\PBa Wampler E. J., Willinger, G. M., Baldwin J.A.,
Carswell R. F., Hazard C., McMahon R. G.:1996 {in~press} {astro-ph~9512084}
\endrefs
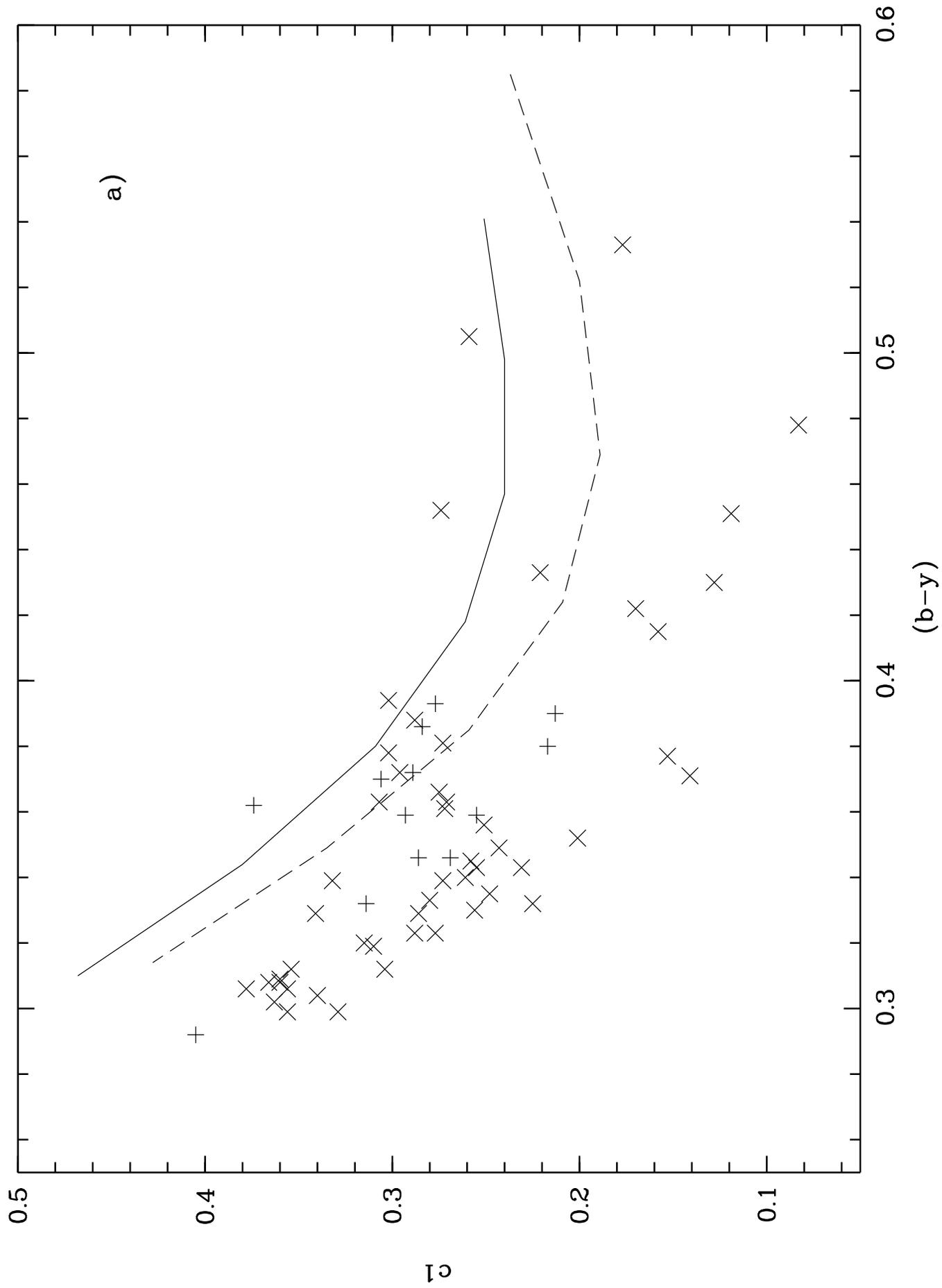
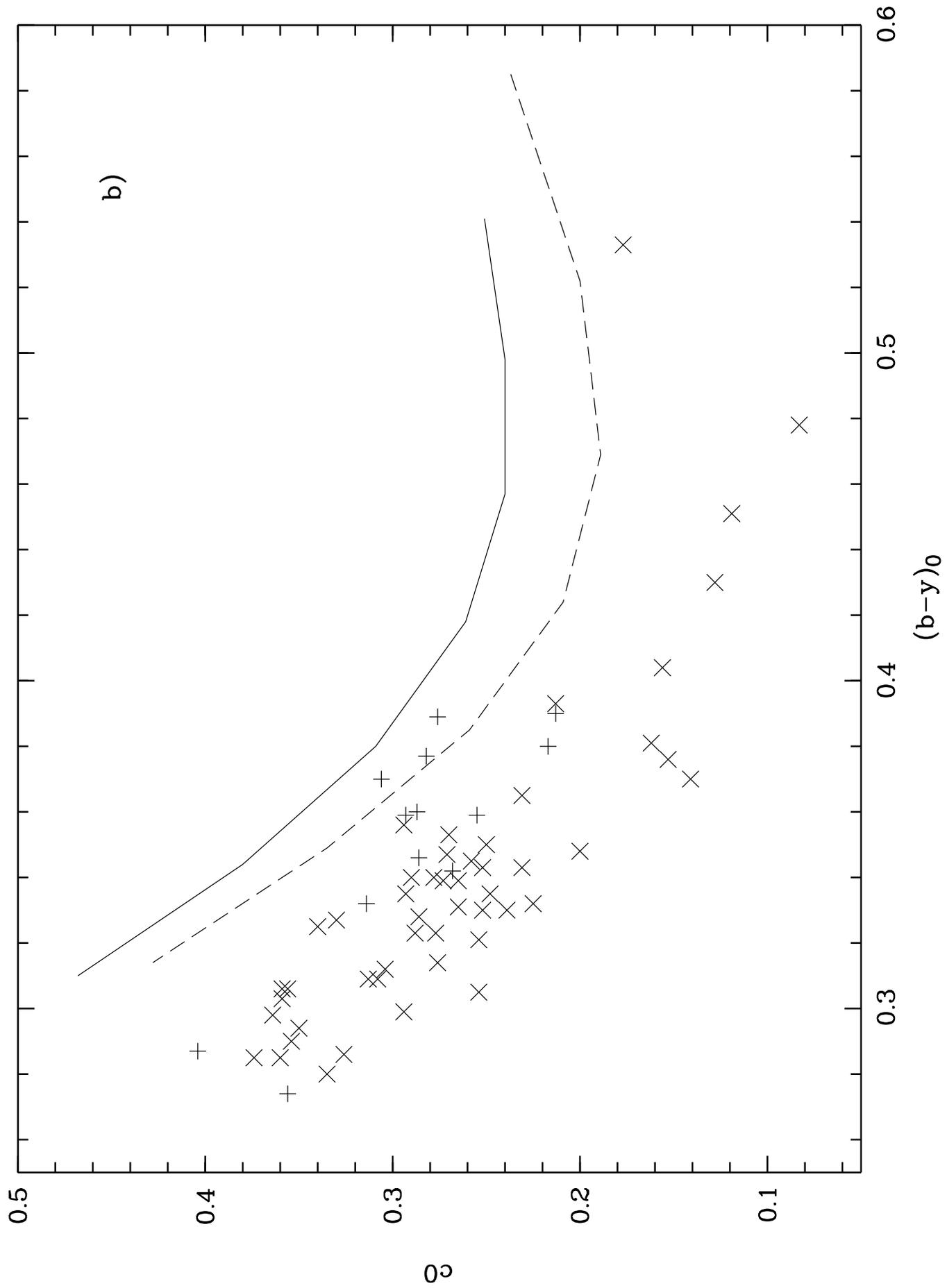
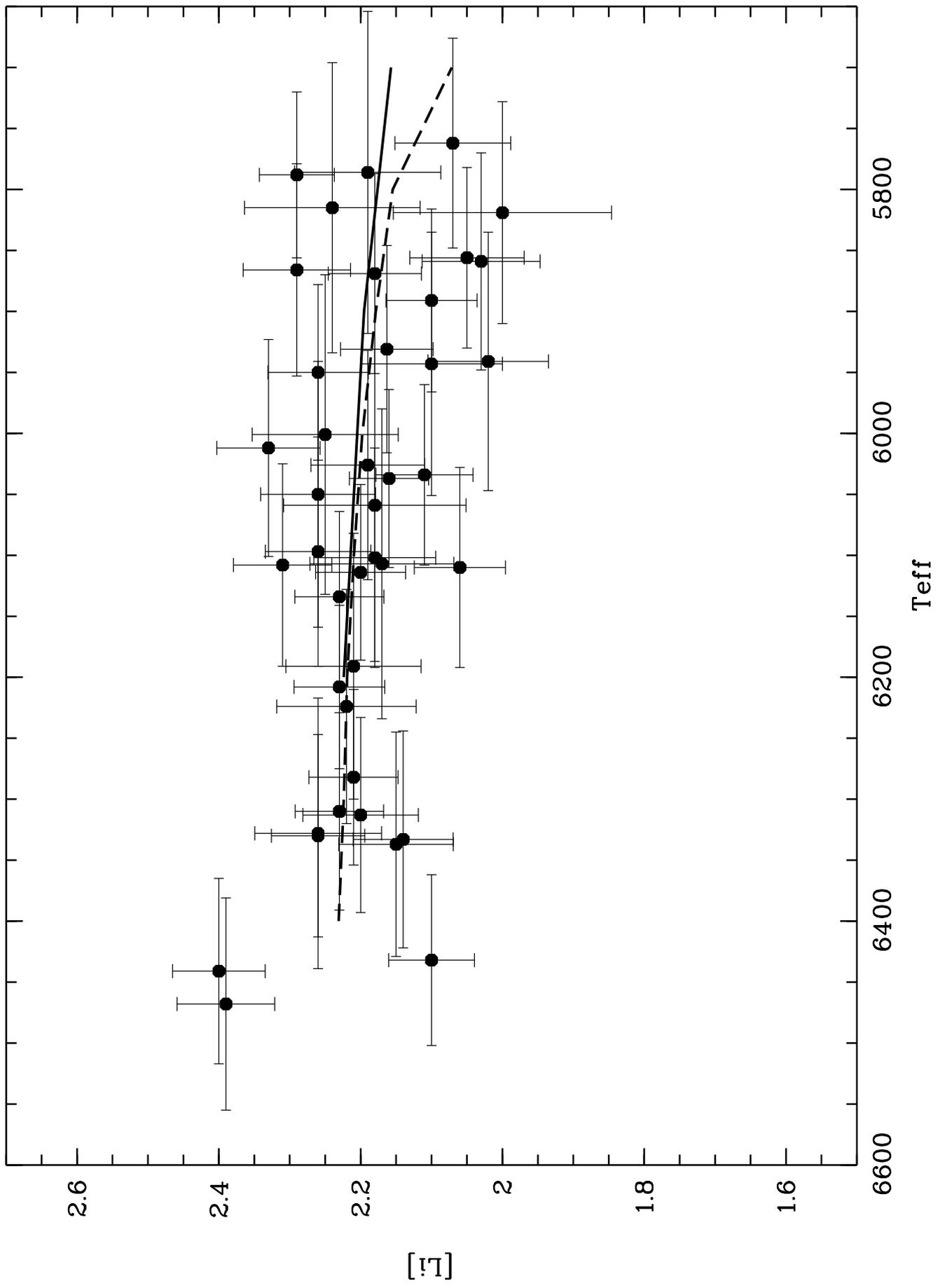
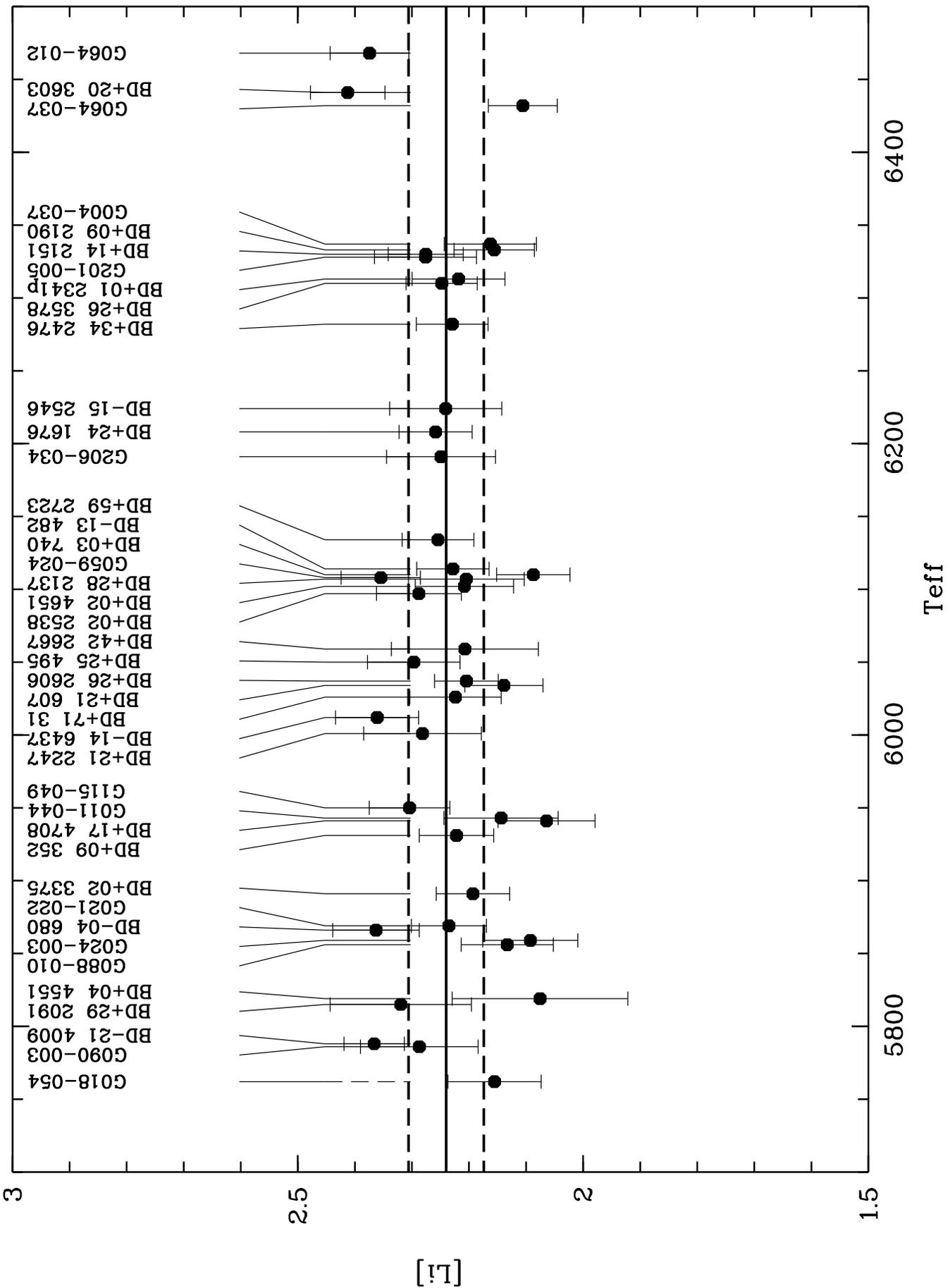
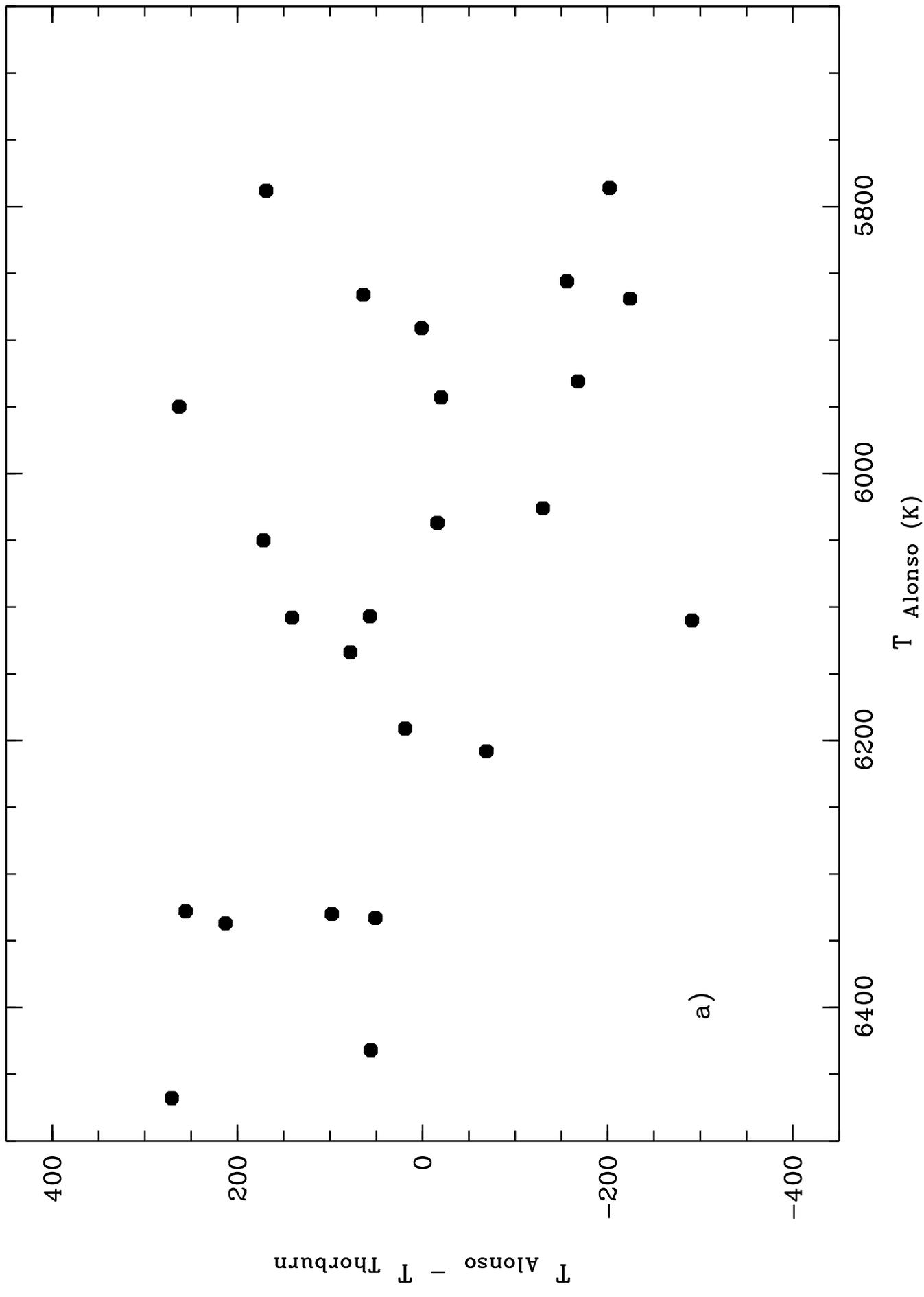
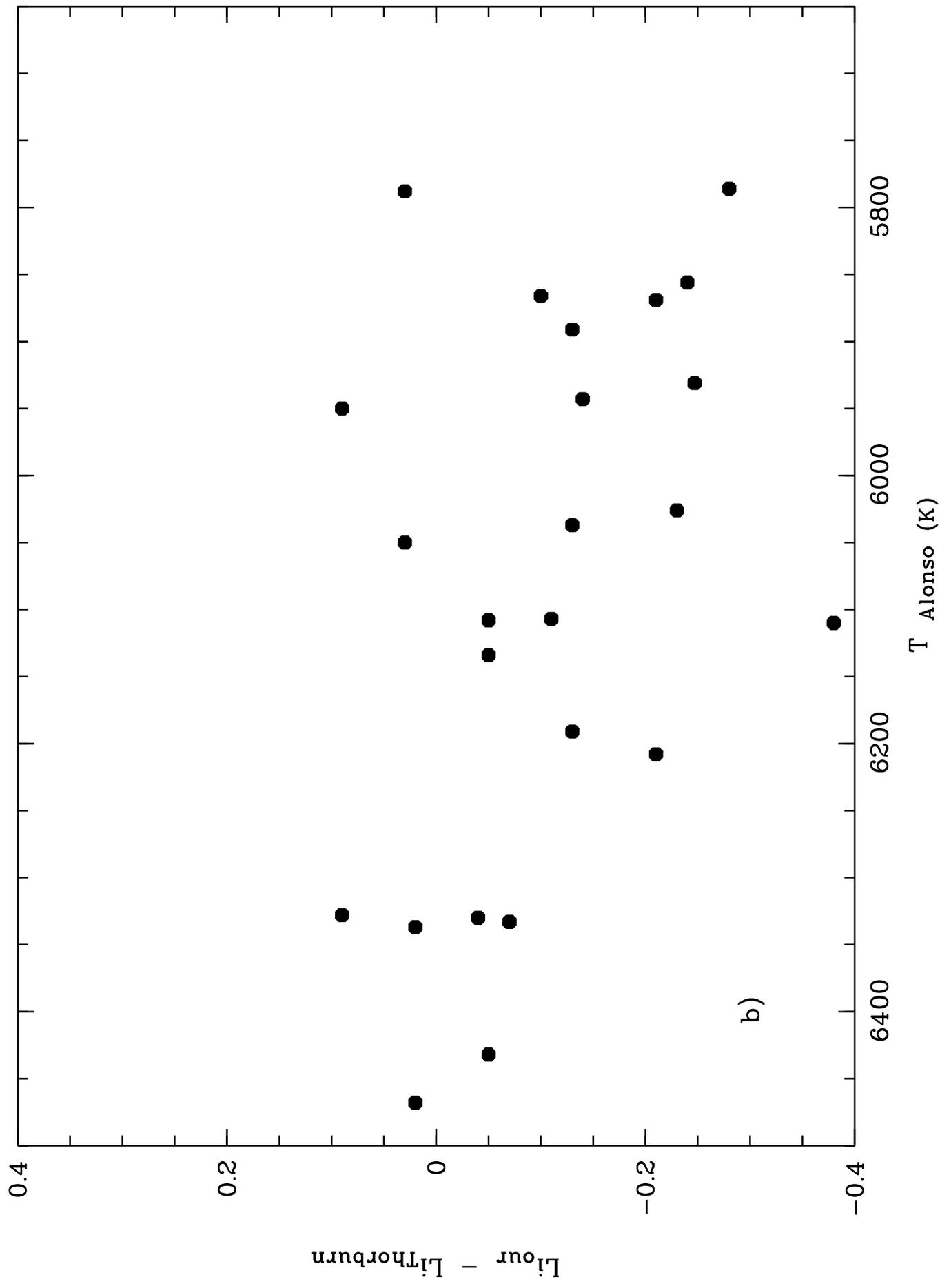
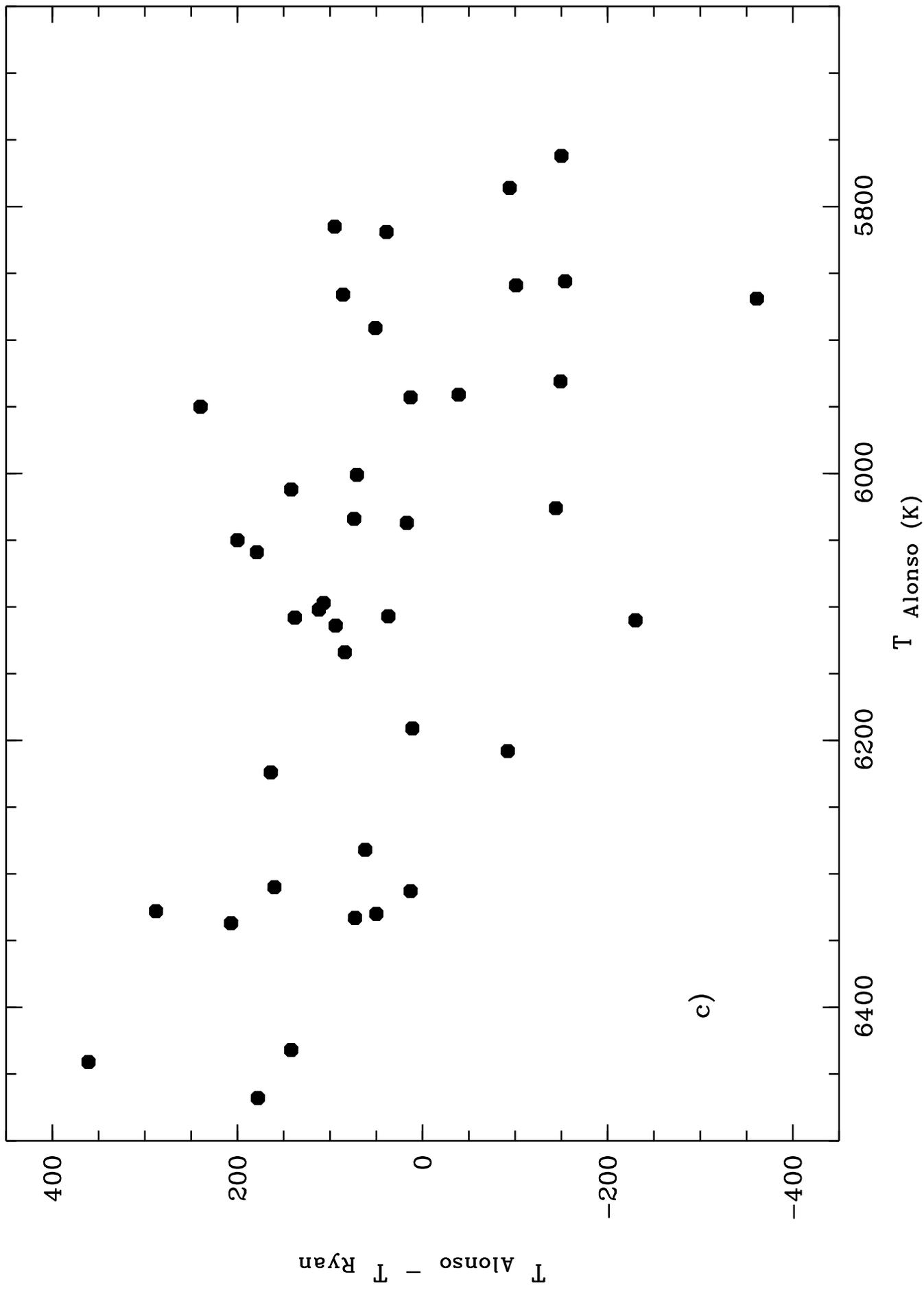
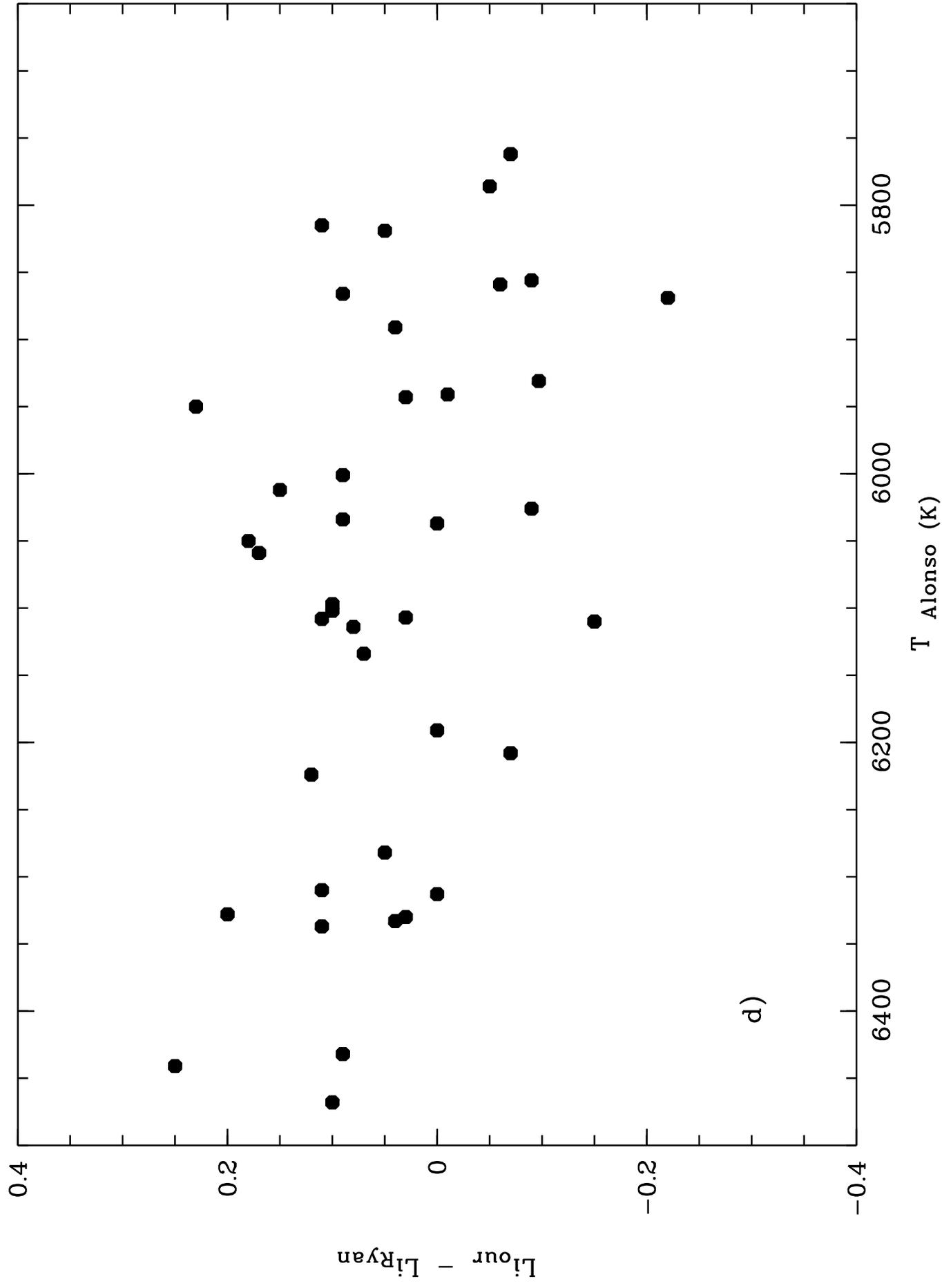
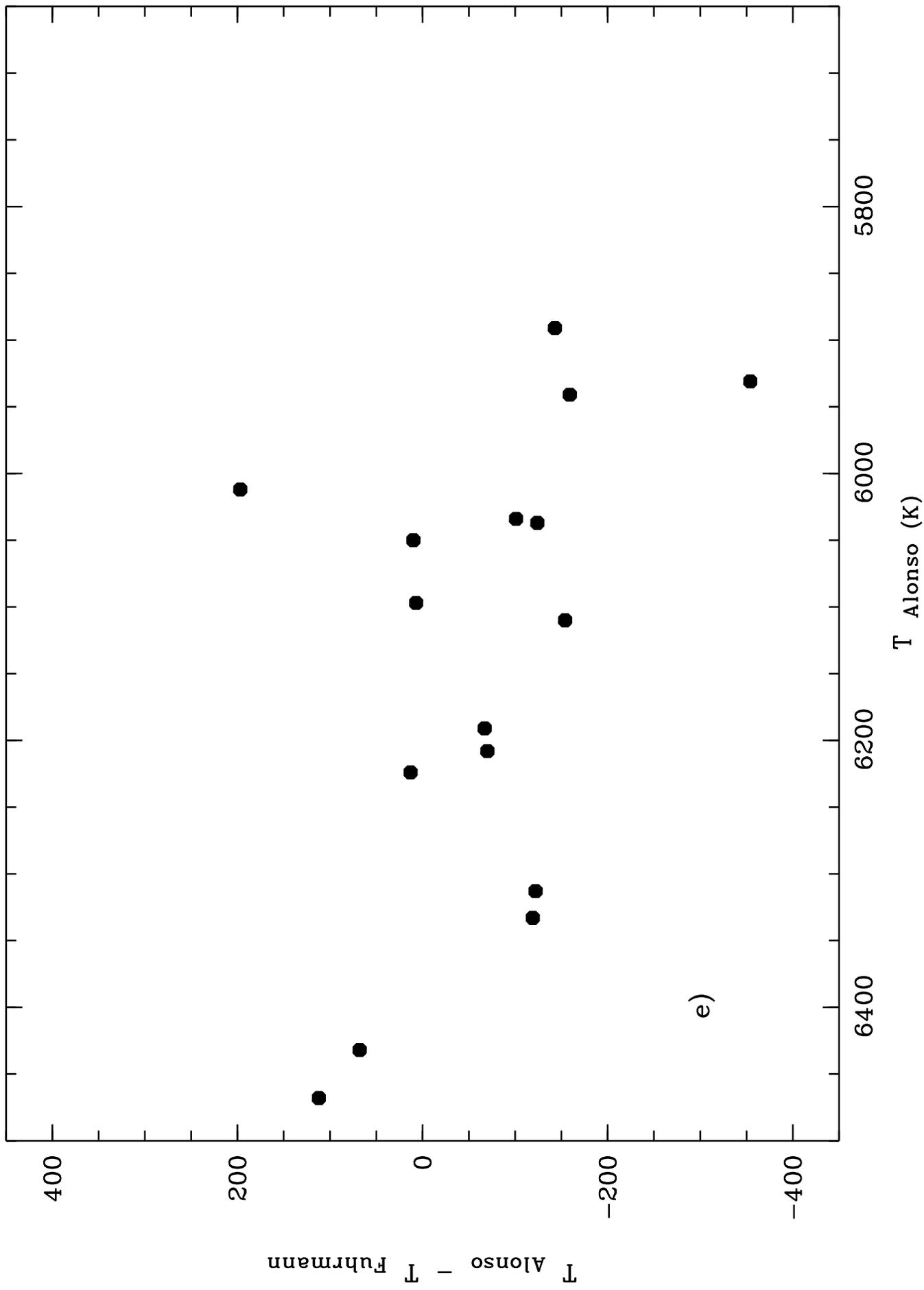
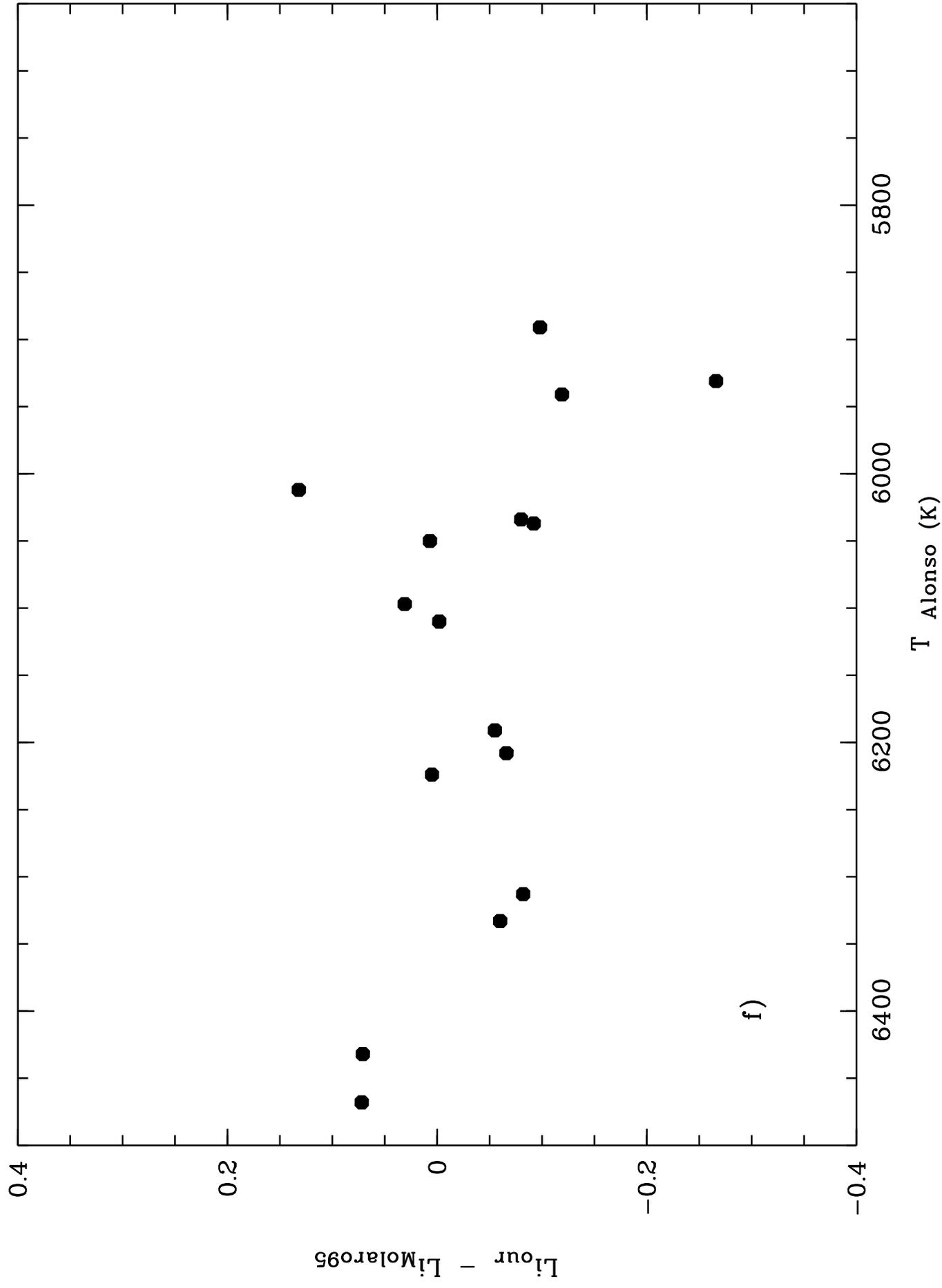
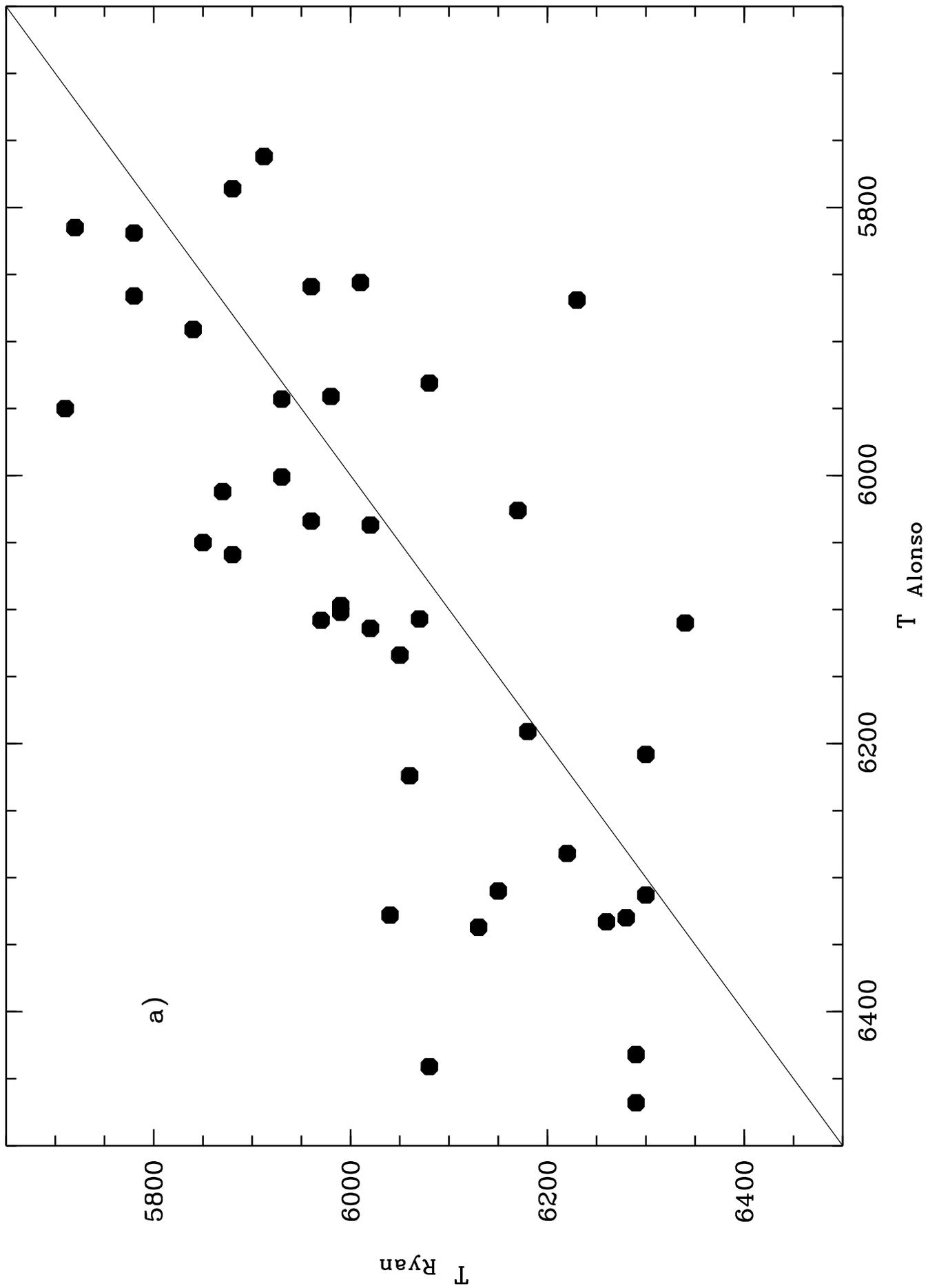
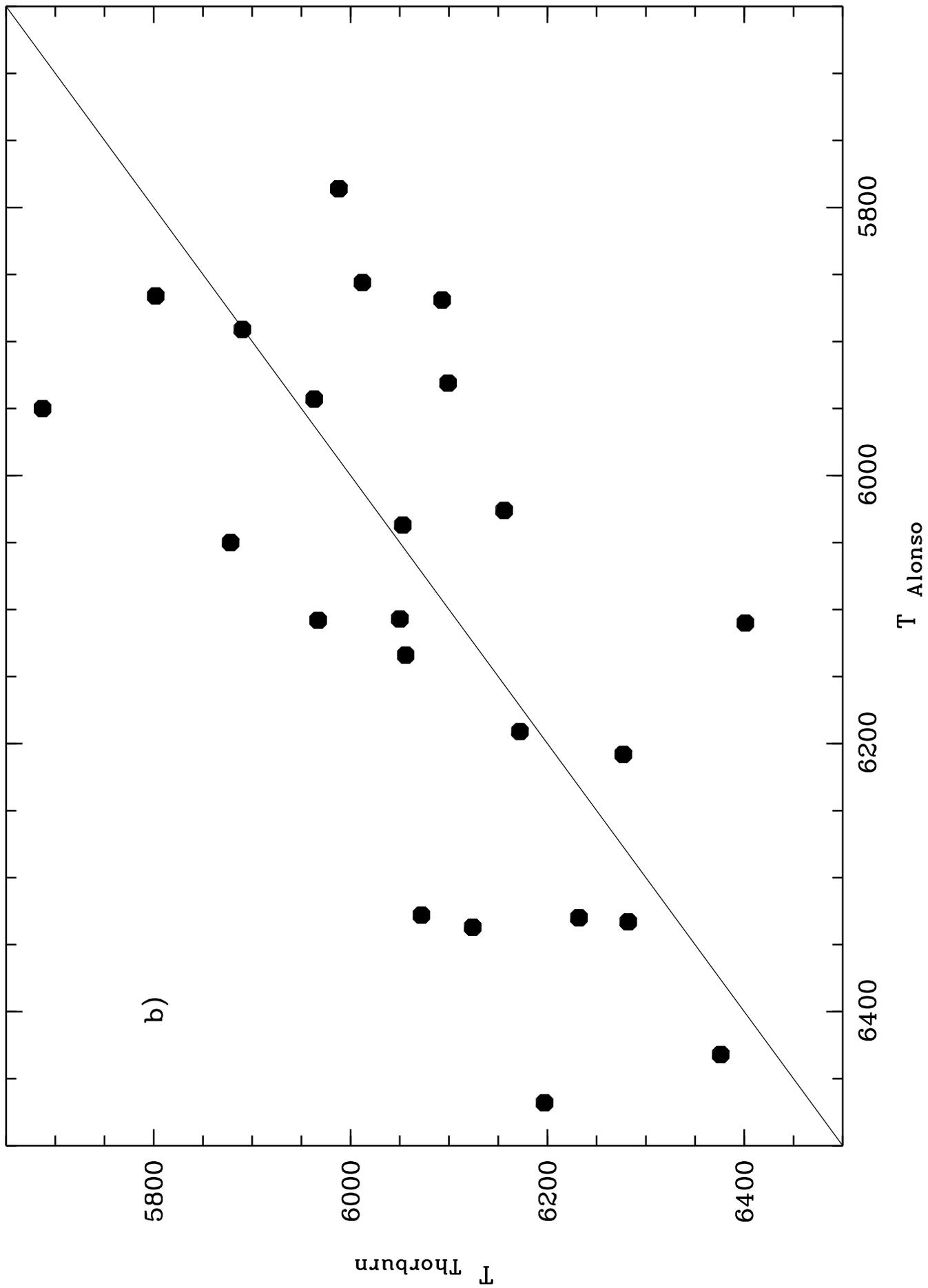
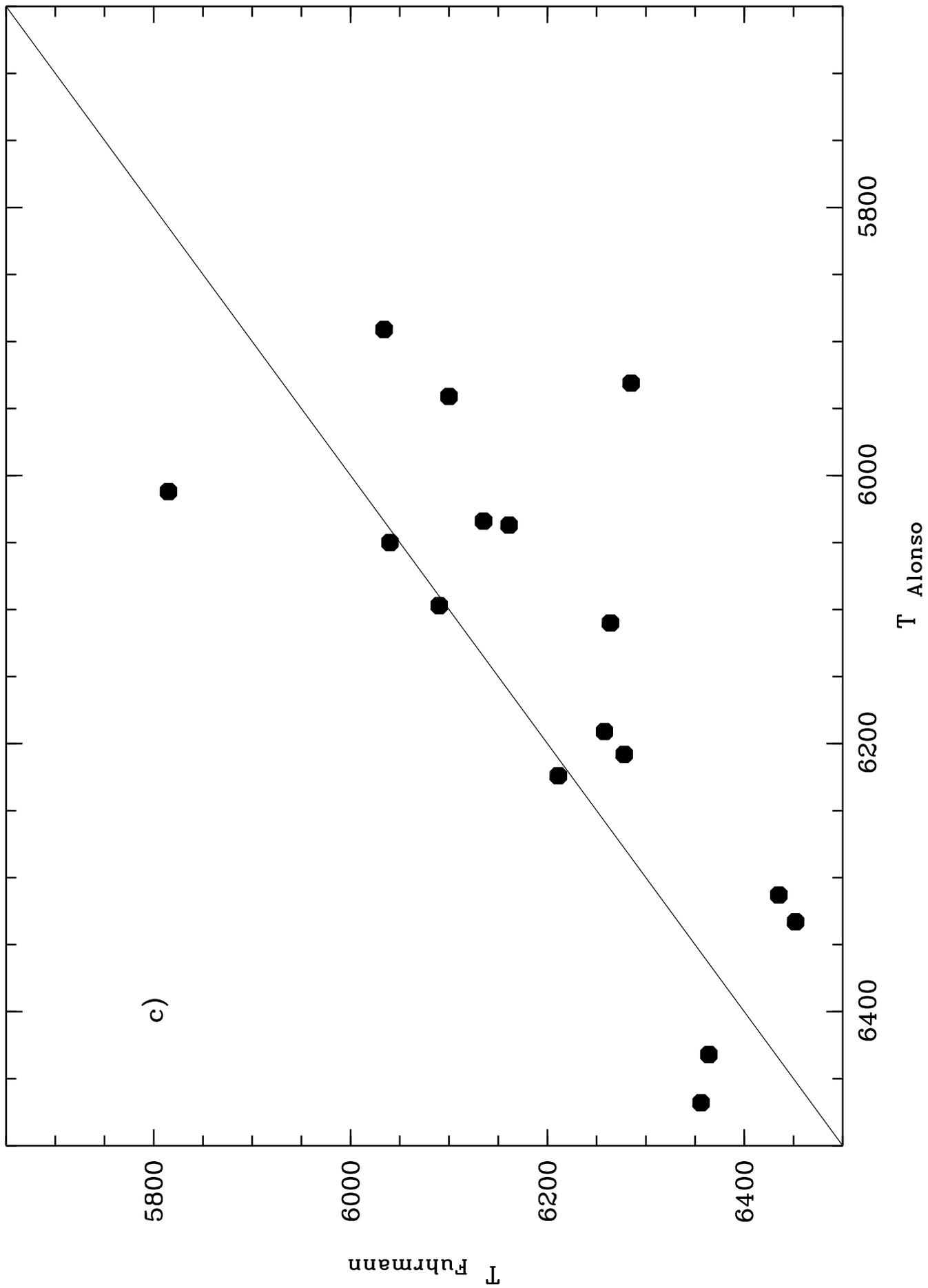
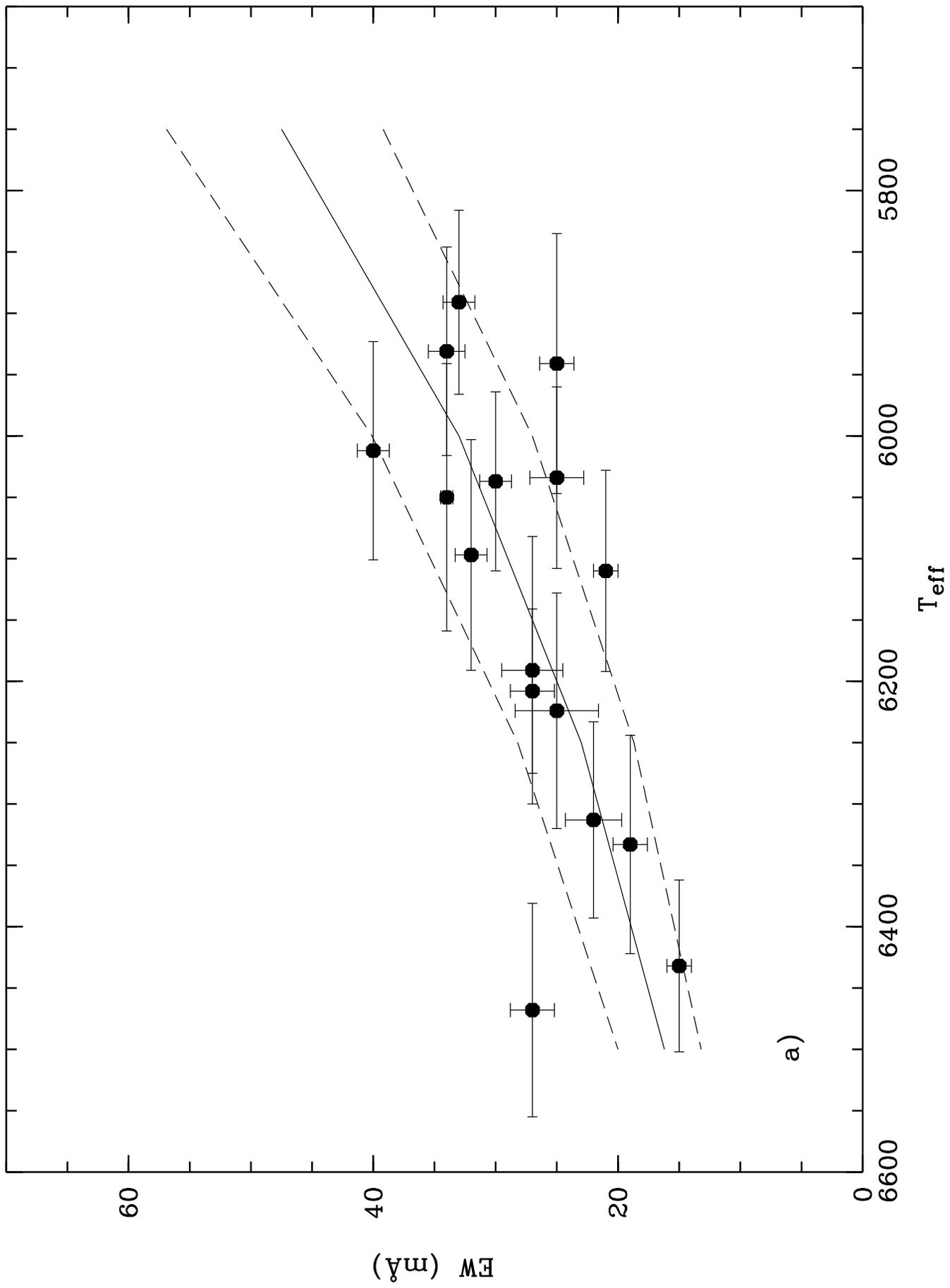
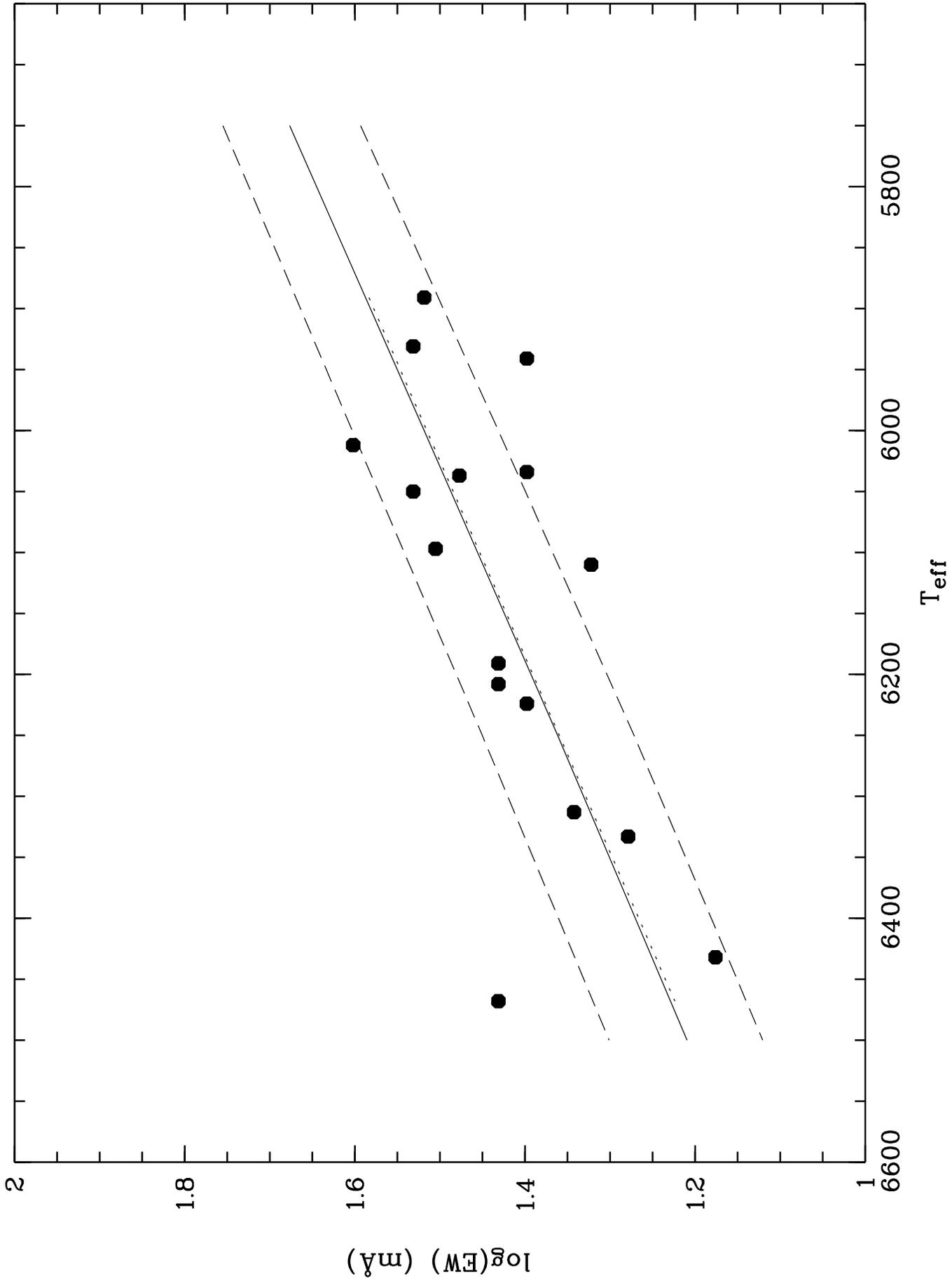
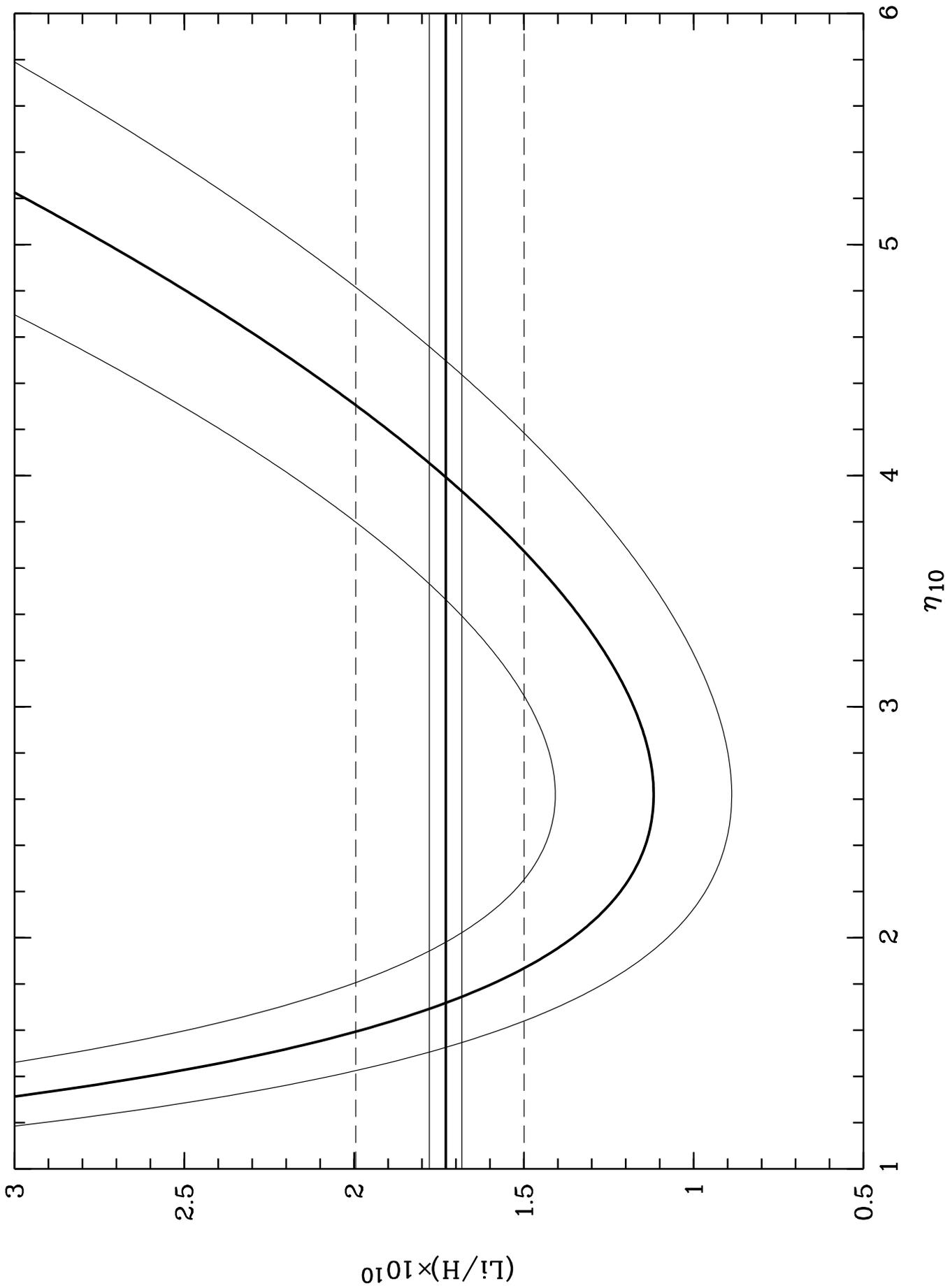
\bye